\documentstyle[12pt,psfig,aps]{revtex}

\begin{document}
\pagenumbering{arabic}
\title{\bf  Relativistic quantum transport theory of 
hadronic matter: the coupled nucleon, delta and pion system}
\author{Guangjun Mao, L.~Neise, H.~St\"{o}cker, and W.~Greiner}  
\address{Institut f\"{u}r Theoretische Physik der
J. W. Goethe-Universit\"{a}t \\
Postfach 11 19 32,  D-60054 Frankfurt am Main, Germany}
\date{\today}
\maketitle
\begin{abstract}
\begin{sloppypar}
We derive  the
 relativistic quantum  transport
equation for the pion distribution function based on an effective Lagrangian
of the QHD-II model. 
The closed time-path Green's function technique, 
the semi-classical, quasi-particle and Born approximation
are employed in the derivation. Both the mean field and collision term are 
derived from the same Lagrangian and presented analytically. The dynamical 
equation for the pions is consistent with that for the nucleons and deltas which
we developed before. Thus, we obtain a   
 relativistic transport model which 
describes the hadronic matter with $N$, $\Delta$ and $\pi$ degrees of freedom
simultaneously. Within this approach, we investigate the medium effects on
the pion dispersion relation as well as the  pion absorption and pion 
production channels in
cold nuclear matter. In contrast to the results of the non-relativistic model,
the pion dispersion relation becomes harder at low momenta and softer at
high momenta as compared to the free one, which is mainly caused by the
relativistic kinetics. The theoretically predicted free
$\pi N \rightarrow \Delta$ cross section is in  agreement with the
experimental data. Medium effects on the $\pi N \rightarrow \Delta$ cross 
section and momentum-dependent $\Delta$-decay width are shown to be substantial.

\end{sloppypar}
\bigskip
\noindent {\bf PACS} number(s): 24.10.Cn; 13.75.Cs; 21.65.+f; 25.70.-z
\end{abstract}
\newcounter{cms}
\setlength{\unitlength}{1mm}
\newpage
\begin{center}
{\bf I. INTRODUCTION}
\end{center}
\begin{sloppypar}
Pion physics is an  important topic in nuclear physics. Recently,
it received renewed interests in relativistic heavy-ion collisions because
pions are the most abundantly produced particles at relativistic energies.
Studies of pionic many-body degrees of freedom in high-energy nucleus-nucleus
collisions were initiated by Gyulassy and Greiner \cite{Gyu77} and by Migdal
\cite{Mig78}. Since then, considerable efforts from both experimental 
\cite{San80}-\cite{Kin97} and theoretical \cite{Sto78}-\cite{Fuc97} groups
were made to study  various aspects of the 
in-medium pion dispersion relation and pion dynamics, 
such as pion spectrum, pion and anti-pion flow in hot and dense nuclear matter.
Due to the high interaction cross section of 
the pion with the nuclear environment 
they are continuously absorbed  by forming $\Delta$-resonances which then decay
again into pions. Therefore, pions have a chance to be emitted during the whole
course of the reaction. While the high-energy tail of the pion spectrum  
 provides information about compressed
and excited nuclear matter in the early reaction stage, 
the low-energy part of the pion
spectrum and pion flow contain information of the in-medium pion potential and
nuclear equation of state (EOS) \cite{Bas95,Fuc97}. The low- and 
high-energy  pions originate from  different stages of the collision.
 A detailed study of the pion dynamics allows to extract the
time evolution of heavy-ion collisions. 

 \end{sloppypar}
 \begin{sloppypar}
On the other hand, 
dileptons produced from $\pi^{+}-\pi^{-}$ annihilation 
\cite{Ko89,Gal87,Li96,Ern97,Bra97} provide information on the high density
 phase at time scales
of 1 fm/c. Since dileptons can leave the 
reaction volume essentially undistorted by final-state interactions, 
as was first pointed out by Gale and Kapusta \cite{Gal87}, they are
expected to be a good tool for an investigation of the violent phases
of high-energy heavy-ion collisions. Recent data by the CERES collaboration
\cite{Aga95} show a substantial modification of the dileptons yield which
might be explained either by many-body effects \cite{Rap97} or 
by an enhanced $\rho$-meson production (via $\pi^{+}-\pi
^{-}$ annihilation) and a dropping $\rho$-mass in the medium \cite{Li95,Cas95}.
Indeed, the property of the $\rho$-meson as well as
the $\Delta$ resonance are strongly influenced by the change of pion property
 in the medium due to the large  $\rho
\rightarrow \pi^{+} \pi^{-}$ and $\Delta \rightarrow N \pi$ decay widths. 
A detailed knowledge of pion dynamics in heavy-ion collisions is a prerequisite
for a quantitative description of dilepton production at SIS and SPS energies.
 
 \end{sloppypar}
 \begin{sloppypar}
 It was recently proposed that the difference between
$\pi^{-}$- and $\pi^{+}$-spectra can be 
  attributed to the influence of isospin and
Coulomb fields \cite{Wag96}. This should allow to extract the effective Coulomb 
field at the instant of the average pion emission. The comparison of spectra of
positively and negatively charged pions can also be used to learn about the
freeze-out of the pions during the expansion phase \cite{Pel97,Tei97}.
  It then provides a method
to determine the size of fireball during the nuclear expansion process.

\end{sloppypar}
 \begin{sloppypar}
 Since the importance of pions in heavy-ion collisions has been recognized for
more than two decades one may believe that the elementary pion properties 
in the hot and dense nuclear matter are already well understood. Unfortunately,
the situation is quite different to this expectation:  understanding 
the pion dynamics in high-energy nucleus-nucleus collisions is still a major
challenge to modern nuclear physics. This paradoxical circumstance seems to be
mainly due to theoretical rather than experimental inadequacies. Experiments 
have been able to record pion spectrum \cite{Sch94,Ber91,Ody88} and pion flow 
and antiflow \cite{Kin97,Gos89} with rather high accuracy. A reasonably strict
treatment of pions in the transport theories is, however, 
still not available. Most
theoretical approaches included the interaction of the pions with the 
surrounding nuclear medium only by collision processes. A free-particle
assumption was usually assigned to pions, while it is well known that the pion 
dispersion relation will be changed substantially in the medium due to
the strong p-wave interaction. Some authors \cite{Ehe93,Xio93,Fuc97} 
implemented the real part of the pion optical potential from 
the non-relativistic 
$\Delta$-hole model \cite{Eri88} to study 
the pion spectrum. In Ref. \cite{Fuc97}
a phenomenological parameterization suggested by Gale and Kapusta \cite{Gal87}
was also tried. However, different model treatments gave rather different 
results. None of the currently available models are able to reproduce the
experimental spectrum over the entire range of energy. The source of the 
problem seems to be that the observed quantities are sensitive to several of the
unknown pion properties in the hot and dense matter. The most important ones
are the in-medium pion dispersion relation (the real part of 
the pion self-energy)
 and  the in-medium pion cross sections (the imaginary part of the pion
self-energy). A self-consistent treatment of both the real part and
imaginary part of the pion self-energy  is necessary 
to obtain  useful information from experimental observable. However, in
transport theories one usually uses the experimentally determined free cross
sections and incorporates a dispersion relation from the non-relativistic 
$\Delta$-hole model or simply employs the free-particle assumption.  
 A self-consistent
description of pions in transport models has not even been reached in the 
 non-relativistic case. In the present work, we still don't have a fully 
 self-consistent treatment. But we go a step further than the present 
available relativistic transport theories, i.e., we derive the in-medium pion
dispersion relation and the in-medium pion cross sections from the same 
Lagrangian and treat them in a relativistic description.

 \end{sloppypar}
 \begin{sloppypar}
Another daunting obstacle to a quantitative description of pions in heavy-ion
collisions is the width and shape of the $\Delta$. As has been pointed out 
before,
when a pion in the hot and dense matter collides with a nucleon, it will be
absorbed to create a $\Delta$. Then the $\Delta$ decays again into a 
pion-nucleon "pair".
Thus the amplitude for creating and absorbing  pions will be sensitive
to the in-medium $\Delta$-decay width, which must be modified by the presence
of  matter due to the potential energies of $N$, $\Delta$ and $\pi$. 
However, in the presently available transport models, the
  free $\Delta$-decay width 
is commonly employed. From the theoretical point of view, 
realistic models for describing pions in dynamical processes should in principle
at least also treat deltas and nucleons simultaneously in an unified framework.

 \end{sloppypar}
 \begin{sloppypar}
It is the purpose of this paper to develop the relativistic transport theory for
pions within the framework of the relativistic 
Boltzmann-Nordheim-Vlasov-Uehling-Uhlenbeck (RVUU/RBUU)
equation. The RVUU model has been applied successfully in studying of 
high-energy
heavy-ion collisions \cite{Elz87,Ko87,Ber88,Bla88,Sch89,Zho94}. 
 In Ref. \cite{Elz87} we
briefly discussed the possible extension to include the pion degree of freedom.
By means of the density matrix method Wang et al. \cite{Wan91} developed 
a transport theory for the $N$, $\Delta$ and $\pi$ system. In their work 
pions are treated  as a free particle,  the detailed expressions of 
the collision term are not given explicitly. On the other hand, in Refs. 
\cite{ZPA94}-\cite{PRCSUB} we developed a set of self-consistent equations
for $N$, $\Delta$ and $N^{*}$(1440) distribution functions in which both mean
field and collision term are derived from the same effective Lagrangian and 
expressed analytically. However, mesons ($\sigma$, $\omega$, $\pi$) were treated
as virtual particles. 
In a physically reasonable scenario, the 
creation and destruction of real as well as virtual mesons 
ought to be described simultaneously and on the same self-consistent footing
\cite{Dav91}. 
Hence, one is forced to solve coupled Boltzmann
equations not only for baryons but also for all relevant mesons. This will cause
significant numerical difficulties and might be beyond the ability of modern
computers. As a first practical step, let us here  treat the pions 
explicitly. The pion is the most frequently observed meson. The other mesons
 still remain treated as virtual mesons. This is the main
strategy of our present work. Here we should note that the transport equations
for $\sigma$ and $\omega$ mesons were discussed in Ref. \cite{Elz87,Mro94} based on 
the Walecka model \cite{Ser86}. In Ref. \cite{Mro94} the mesons
turn out to be treated as free particles due to the approximations 
  used in that work. 
Also, no concrete expressions for the collision term were given there. 
We will come back to this point in Sect. III.

 \end{sloppypar}
 \begin{sloppypar}
Starting from an effective Lagrangian of the QHD-II model \cite{Ser86} we here 
derive a  RVUU equation for the pion distribution function in which 
both the mean field and the collision term are derived simultaneously and
 expressed analytically. In our framework a fully relativistic treatment
is realized and  medium effects are included.                   
Furthermore, we treat $N$, $\Delta$, and $\pi$ in an unified framework based
on the same effective Lagrangian and
finally obtain a set of coupled equations for hadronic matter.
The paper is organized as follows: in Sect. II we briefly review the closed
time-path Green's function technique which plays a central role in our
derivation. An effective Lagrangian for the $N$, $\Delta$ and $\pi$ system
interacting through the exchange of virtual mesons is also presented there.
In Sect. III we derive the RVUU-type transport equation for the pion 
distribution function. The main ingredients of the equation are the 
relativistic mean field
and collision terms, which are calculated from the same effective Lagrangian
 and presented analytically in Sect. IV
and V, respectively. In Sect. VI we present the numerical results for the
in-medium pion dispersion relation and $\Delta$-formation cross section. 
Finally,
 the summary and outlook is given in Sect. VII.
 
 \end{sloppypar}
\begin{center}
{\bf II. PRELIMINARIES}
 \end{center}
 \begin{sloppypar}
In  the present work we  
employ the closed time-path 
Green's function
technique.  
 For a detailed description  of this Green's function technique
for non-equilibrium system, we refer to Refs. \cite{Dan84,Cho85}.
Here we give a brief review for the reader's convenience.
In the Heisenberg picture the Green's function $G_{F}(1,2)$ of fermions and
 $\Delta_{B}(1,2)$ of bosons can be defined on the time contour depicted
 in Fig.~1 as
  \begin{eqnarray}
 &&i G_{F}(1,2) \equiv \langle T \lbrack \Psi_{H}(1)\bar{\Psi}_{H}(2) \rbrack
 \rangle , \\
 &&i \Delta_{B}(1,2) \equiv \langle T \lbrack \Phi_{H}(1) \Phi_{H}(2) \rbrack
 \rangle - \langle \Phi_{H}(1) \rangle \langle \Phi_{H}(2) \rangle,
  \end{eqnarray}
where 1, 2 denote $x_{1}$, $x_{2}$;
$\Psi_{H}(1)$ and $\bar{\Psi}_{H}(2)$ represent the field operators of
the nucleon and delta in the Heisenberg picture and $\Phi_{H}(1)$ and
 $\Phi_{H}(2)$ are those of the $\sigma$, $\omega$, $\pi$ and $\rho$. 
Here we have specified the initial state by  assuming that  its
density operator  commutes
with the particle-number operator \cite{Dan84}.
Furthermore, we assume that the initial state admits the Wick decomposition
(is noncorrelated). 
Thus, in Eq. (1)
 the expectation value of a single fermionic field vanishes. 
In the case of bosonic Green's functions, the contributions from classical
expectation values have been subtracted in order to concentrate on the
field fluctuations around the classical values.
 On the other hand, the second term on the right-hand side 
 of Eq. (2) explicitly indicates the presence
 of the {\em mean field}. 
 According to the position of field operators on
the time contour, we have four different Green's functions for fermions
 \begin{eqnarray}
 && iG_{F}^{--}(1,2)=\langle T^{c}\Psi_{H}(1)\bar{\Psi}_{H}(2)\rangle , 
  \nonumber \\
 &&iG_{F}^{++}(1,2)=\langle T^{a}\Psi_{H}(1)\bar{\Psi}_{H}(2)\rangle ,
  \nonumber \\
 &&iG_{F}^{+-}(1,2)=\langle \Psi_{H}(1)\bar{\Psi}_{H}(2)\rangle , 
  \nonumber \\
 &&iG_{F}^{-+}(1,2)=-\langle \bar{\Psi}_{H}(2)\Psi_{H}(1)\rangle , 
 \end{eqnarray}
 and four for bosons
 \begin{eqnarray}
 && i\Delta_{B}^{--}(1,2)=\langle T^{c}\Phi_{H}(1)\Phi_{H}(2) \rangle  
    - \langle \Phi_{H}(1) \rangle \langle \Phi_{H}(2) \rangle, \nonumber \\
 &&i\Delta_{B}^{++}(1,2)=\langle T^{a}\Phi_{H}(1)\Phi_{H}(2) \rangle  
    - \langle \Phi_{H}(1) \rangle \langle \Phi_{H}(2) \rangle , \nonumber \\
 &&i\Delta_{B}^{+-}(1,2)=\langle \Phi_{H}(1)\Phi_{H}(2) \rangle  
    - \langle \Phi_{H}(1) \rangle \langle \Phi_{H}(2) \rangle ,\nonumber \\
 &&i\Delta_{B}^{-+}(1,2)=\langle \Phi_{H}(2)\Phi_{H}(1) \rangle  
    - \langle \Phi_{H}(1) \rangle \langle \Phi_{H}(2) \rangle . 
 \end{eqnarray}
Here $T^{c}$ is the chronological ordering operator, $T^{a}$ is
the anti-chronological ordering operator. The designations $-$ and $+$ are
attributed to the respective time path shown in Fig.~1. 
   \begin{center}
 \fbox{Fig.~1}
   \end{center}
\noindent  We further on 
 express the $G_{F}(1,2)$ and $\Delta_{B}(1,2)$ in a compact matrix form
 \begin{eqnarray}
  iG_{F}(1,2)=\left( \begin{array}{cc}
   iG^{--}_{F}(1,2) & iG^{-+}_{F}(1,2)  \\
   iG^{+-}_{F}(1,2) & iG^{++}_{F}(1,2) \end{array} \right)  
  \end{eqnarray}
 and
 \begin{eqnarray}
  i\Delta_{B}(1,2)=\left( \begin{array}{cc}
   i\Delta^{--}_{B}(1,2) & i\Delta^{-+}_{B}(1,2) \\
   i\Delta^{+-}_{B}(1,2) & i\Delta^{++}_{B}(1,2) \end{array} \right)  .
   \end{eqnarray}
It should be
pointed out that the four Green's functions in Eq. (5) are not independent.
They satisfy the following relations
 \begin{eqnarray}
&& iG_{F}^{--}(1,2)=\theta(t_{1}-t_{2}) iG_{F}^{+-}(1,2) + \theta(t_{2} -t_{1}) 
 iG_{F}^{-+}(1,2), \\
&& iG_{F}^{++}(1,2)=\theta(t_{1}-t_{2}) iG_{F}^{-+}(1,2) + \theta(t_{2} -t_{1})
 iG_{F}^{+-}(1,2).
 \end{eqnarray}
 Here $\theta(t_{1}-t_{2})$ is defined as
   \begin{equation}
 \theta(t_{1}-t_{2}) = \left\{ \begin{array}{cl}
   1 & \qquad  t_{1} \mbox{ \rm is later on a contour than } t_{2} \\
     0 & \qquad t_{1} \mbox{ \rm is earlier on a contour than } t_{2}
    \end{array} \right. .   
  \end{equation}
 The same relations hold for the boson Green's functions in Eq. (6). 

 \end{sloppypar}
 \begin{sloppypar}
 In order to use the powerful perturbation expansion method of field theory,
 we choose the interaction picture. The time-ordered
 products in Eqs. (1) and (2) can then be rewritten as 
 \begin{eqnarray}
 && \langle T \lbrack \Psi_{H}(1)\bar{\Psi}_{H}(2) \rbrack \rangle 
  =\langle T \lbrack exp(-i \not \!\!\int\,dx H_{I}(x))
  \Psi_{I}(1)\bar{\Psi}_{I}(2)  \rbrack \rangle , \\
 && \langle T \lbrack \Phi_{H}(1)\Phi_{H}(2) \rbrack \rangle 
  =\langle T \lbrack exp(-i \not \!\!\int\,dx H_{I}(x))
  \Phi_{I}(1)\Phi_{I}(2)  \rbrack \rangle , \\
 && \langle \Phi_{H}(1) \rangle 
  =\langle T \lbrack exp(-i \not \!\!\int\,dx H_{I}(x))
  \Phi_{I}(1) \rbrack \rangle .
 \end{eqnarray}
Here $\psi_{I}(1)$, $\bar{\psi}_{I}(2)$ and $\Phi_{I}(1)$, $\Phi_{I}(2)$
represent the field operators in the interaction picture;
$\not\! \int\,dx\equiv \not\! \int\, dt d {\bf x}$, $\not\!\int$ stands
 for an integral along the time axis given in Fig.~1.
The definition of Eqs. (3) and (4) and the relations of Eqs. (7), (8) are
still valid in the interaction picture for both the full Green's functions
 $G_{F}(1,2)$, $\Delta_{B}(1,2)$ and zeroth-order Green's functions (i.e.,
 non-interacting Green's functions) $G_{F}^{0}(1,2)$, $\Delta_{B}^{0}(1,2)$.
 The detailed expressions of the zeroth-order Green's functions as well as
 $H_{I}$ in Eqs. (10) - (12) are determined by the specific effective 
 Lagrangian used in the model. As a
 preliminary step towards a complete description of hadronic matter, we first
consider a system consisting of real nucleons, deltas and pions  
interacting through the exchange of virtual $\sigma$, $\omega$, $\pi$ and
$\rho$ mesons. In order to avoid extensive cancelations of large terms to 
correctly describe the small S-wave $\pi N$ scattering length, we choose the
phenomenological pseudovector form for the $\pi NN$ and $\pi\Delta\Delta$ 
coupling.  With this choice of coupling, the value of the S-wave
$\pi N$ scattering length turns out to be $-0.010$ \cite{Ser86} while the
empirical value is $-0.015 \pm 0.015$ \cite{Pil73}. The inclusion
of the $\rho$-meson degree of freedom is important for the $\pi\pi$ scattering
due to vector meson dominance \cite{Zou94}. We furthermore  include two  
non-linear meson coupling terms $\sigma\pi\pi$ and $\rho\pi\pi$ which are 
applied only to the $\pi\pi$ scattering.
The total effective
Lagrangian can then be written as
\begin{equation}
{\cal L}={\cal L}_{\rm F}+{\cal L}_{\rm I}.
\end{equation}
Here ${\cal L}_{F}$ is the Lagrangian density for free nucleon, delta,
 and meson fields
\begin{eqnarray}
{\cal L}_{\rm F}&=&\bar{\psi}[i\gamma_{\mu}\partial^{\mu}-M_{N}]\psi
  + \bar{\psi}_{\Delta \nu}[i\gamma_{\mu}\partial^{\mu}-M_{\Delta}]
\psi^{\nu}_{\Delta} \nonumber \\ && + \frac{1}{2}
\partial_{\mu}\sigma\partial^{\mu}\sigma-U(\sigma) 
 -\frac{1}{4}\omega_{\mu\nu}\omega^{\mu\nu}+U(\omega) \nonumber \\
 &&+ \frac{1}{2}(\partial_{\mu} \mbox{\boldmath $\pi$} \partial^{\mu}
\mbox{\boldmath $\pi$}
-m_{\pi}^{2}\mbox{\boldmath $\pi$}^{2})
 - \frac{1}{4} \mbox{\boldmath $\rho$}_{\mu\nu}\mbox{\boldmath $\rho$}^{\mu\nu}
 +\frac{1}{2}m_{\rho}^{2}\mbox{\boldmath $\rho$}_{\mu}  \cdot
 \mbox{\boldmath $\rho$}^{\mu}
    \end{eqnarray}
and U($\sigma$), U($\omega$) are the self-interaction part of the scalar field
\cite{Bog83} and vector field \cite{Bod91,Sug94}
\begin{eqnarray}
  && U(\sigma)=
   \frac{1}{2}m_{\sigma}^{2}\sigma^{2}+\frac{1}{3}b({\rm g}_{NN}^{\sigma}
\sigma)^{3}+\frac{1}{4}c({\rm g}_{NN}^{\sigma}\sigma)^{4}, \\
  && U(\omega)=\frac{1}{2}m_{\omega}^{2}\omega_{\mu}\omega^{\mu}
     (1+\frac{({\rm g}_{NN}^{\omega})^{2}}{2}\frac{\omega_{\mu}\omega^{\mu}}
     {Z^{2}}),
   \end{eqnarray}
respectively.  Here the field tensor for the rho and omega are given in terms 
of their potential fields by
  \begin{equation}
 \mbox{\boldmath $\rho$}_{\mu\nu}=\partial_{\mu} \mbox{\boldmath $\rho$}_{\nu}
 -\partial_{\nu} \mbox{\boldmath $\rho$}_{\mu}
  \end{equation}
and
  \begin{equation}
  \omega_{\mu\nu}=\partial_{\mu}\omega_{\nu} -\partial_{\nu}\omega_{\mu}.
  \end{equation}
The interaction Lagrangian ${\cal L}_{I}$ consists of baryon-baryon, 
baryon-meson and
meson-meson terms, which are given by
\begin{eqnarray}
 {\cal L}_{I}&=&{\cal L}_{NN}
 +{\cal L}_{\Delta \Delta}+{\cal L}_{\Delta N}
 + {\cal L}_{\sigma\pi} + {\cal L}_{\rho\pi} \nonumber \\
     &=&{\rm g}^{\sigma}_{NN}\bar{\psi}(x)\psi(x)\sigma(x)
      - {\rm g}^{\omega}_{NN}\bar{\psi}(x)\gamma
  _{\mu}\psi(x)\omega^{\mu}(x) \nonumber \\
 &&+{\rm g}_{NN}^{\pi}\bar{\psi}(x)\gamma_{\mu}\gamma_{5}\mbox{\boldmath $\tau$}
\cdot \psi(x)\partial^{\mu}\mbox{\boldmath $\pi$}(x) 
- \frac{1}{2}{\rm g}_{NN}^{\rho}\bar{\psi}(x)\gamma_{\mu}\mbox{\boldmath $\tau$}
\cdot \psi(x)\mbox{\boldmath $\rho$}^{\mu}(x) \nonumber \\ 
 &&+{\rm g}^{\sigma}_{\Delta \Delta}
\bar{\psi}_{\Delta \nu}(x)\psi^{\nu}_{\Delta}(x)\sigma(x)
     -{\rm g}^{\omega}_{\Delta \Delta}\bar{\psi}_{\Delta \nu}(x)\gamma
  _{\mu}\psi^{\nu}_{\Delta}(x)\omega^{\mu}(x) \nonumber \\
&& +{\rm g}_{\Delta \Delta}^{\pi}\bar{\psi}_{\Delta\nu}(x)\gamma_{\mu}\gamma_{5} {\bf T }
 \cdot \psi^{\nu}_{\Delta}(x) \partial^{\mu}\mbox{\boldmath $\pi$}(x) 
-\frac{1}{2} {\rm g}_{\Delta \Delta}^{\rho}\bar{\psi}_{\Delta\nu}(x)\gamma_{\mu}{\bf T }
 \cdot \psi^{\nu}_{\Delta}(x) \mbox{\boldmath $\rho$}^{\mu}(x) \nonumber \\ 
 &&- {\rm g}_{\Delta N}^{\pi}\bar{\psi}_{\Delta \mu}(x)
 \partial^{\mu}\mbox{\boldmath $\pi$}(x) \cdot {\bf S}^{+}\psi(x)
 - {\rm g}_{\Delta N}^{\pi}\bar{\psi}(x){\bf S}\psi_{\Delta \mu}(x) \cdot
 \partial^{\mu}\mbox{\boldmath $\pi$}(x) \nonumber \\
&&+ \frac{1}{2}{\rm g}_{\sigma \pi}m_{\sigma} \sigma(x)\mbox{\boldmath $\pi$}(x)
 \cdot \mbox{\boldmath $\pi$}(x) + {\rm g}_{\rho \pi} \lbrack \partial^{\mu}
 \mbox{\boldmath $\pi$}(x) \times \mbox{\boldmath $\pi$} (x) \rbrack 
 \cdot \mbox{\boldmath $\rho$}_{\mu}(x) \nonumber \\
 &=& {\rm g}_{NN}^{A}\bar{\psi}(x)\Gamma_{A}^{N} \psi(x) \Phi_{A}(x)
 + {\rm g}_{\Delta \Delta}^{A} \bar{\psi}_{\Delta \nu}(x)\Gamma_{A}^{\Delta}\psi
 ^{\nu}_{\Delta}(x) \Phi_{A}(x) \nonumber \\
 &&- {\rm g}_{\Delta N}^{\pi}\bar{\psi}_{\Delta \mu}(x)
 \partial^{\mu}\mbox{\boldmath $\pi$}(x) \cdot {\bf S}^{+}\psi(x)
 - {\rm g}_{\Delta N}^{\pi}\bar{\psi}(x){\bf S}\psi_{\Delta \mu}(x) \cdot
 \partial^{\mu}\mbox{\boldmath $\pi$}(x) \nonumber \\
 && + {\rm g}_{\pi\pi}^{A} \pi_{i}(x) \Gamma_{A}^{\pi} \pi_{j}(x) \Phi_{A}(x)
   \end{eqnarray}
In the above expressions $\psi(x)$ is the Dirac spinor of the nucleon
 and $\psi_{\Delta \mu}(x)$ is the Rarita-Schwinger spinor of the 
 $\Delta$-baryon. $\mbox{\boldmath $\tau$}$ is the isospin operator of the 
 nucleon and
${\bf T}$ is the isospin operator of the $\Delta$. ${\bf S}$ and 
${\bf S}^{+}$ are the
isospin transition operator between the isospin 1/2 and 3/2 fields.
${\rm g}_{NN}^{\pi}=f_{\pi}/m_{\pi}$,
 ${\rm g}_{\Delta N}^{\pi}=f^{*}/m_{\pi}$;
 $\Gamma_{A}^{N}=
 \gamma_{A}\tau_{A}$, $\Gamma_{A}^{\Delta}=\gamma_{A}T_{A}$,
 $\Gamma_{A}^{\pi}=\gamma^{\pi}_{A}\tau^{\pi}_{A}$, 
  A=$\sigma$, $\omega$, $\pi$, $\rho$,
the symbols and notation are given   in Table I and II for the baryon-baryon-meson
 vertex and meson interaction vertex, respectively.
  \begin{center}
  \fbox{Table I}   \hspace{2cm}  \fbox{Table II}
  \end{center}
 The zeroth-order Green's functions of nucleons, deltas as well as mesons
 corresponding to the free Lagrangian density of Eq. (14) 
 are summarized in Appendix A, where the distribution functions  
 of negative-energy
states are neglected for fermions.  
 They are kept for bosons. 
 Considering that we will derive a transport equation for the pion in which we
 only treat the real pion with positive-energy states, we rewrite the 
 zeroth-order Green's functions of the pion as
   \begin{eqnarray}
  && \Delta^{0\mp\mp}_{\pi}(x,k)=\frac{\pm 1}{k^{2}-m_{\pi}^{2}\pm i\epsilon}
 - \frac{\pi i}{\omega(k)}\,\delta\!\left[ k_{0}-\omega(k)\right] f_{\pi}(x,k), \\
  && \Delta^{0+-}_{\pi}(x,k)=
 - \frac{\pi i}{\omega(k)}\,\delta\!\left[ k_{0}-\omega(k)\right] 
  \left[ 1 + f_{\pi}(x,k)\right] , \\
  && \Delta^{0-+}_{\pi}(x,k)=
 - \frac{\pi i}{\omega(k)}\,\delta\!\left[ k_{0}-\omega(k)\right] 
  f_{\pi}(x,k) , 
  \end{eqnarray}
 here $\omega(k)$ is the energy of the pion.        

 \end{sloppypar}
\begin{center}
{\bf III. DERIVATION OF THE QUANTUM TRANSPORT  EQUATION FOR PIONS}
 \end{center}
\begin{center}
{\bf A. Dyson Equation and Pion Self-energy}
 \end{center}
 \begin{sloppypar}
With the discussions of Sect. II we can write down the pion Green's function
in the interaction picture as
 \begin{eqnarray}
  i \Delta_{\pi}(1,2) \delta_{ij} & = &
  \langle T \lbrack exp(-i \not \!\!\int\,dx H_{I}(x))
  \mbox{\boldmath $\pi$}_{I}(1)\mbox{\boldmath $\pi$}_{I}(2)  \rbrack \rangle 
 \nonumber \\
  &-&\langle T \lbrack exp(-i \not \!\!\int\,dx H_{I}(x))
  \mbox{\boldmath $\pi$}_{I}(1) \rbrack \rangle  
  \langle T \lbrack exp(-i \not \!\!\int\,dx H_{I}(x))
  \mbox{\boldmath $\pi$}_{I}(2) \rbrack \rangle ,
 \end{eqnarray}
 here $i$, $j$=1, 2, 3 represent the isospin indices of pion. In the following
 we suppress this subscript because at the end we will obtain a RVUU-type
 transport equation which is averaged on the isospin. Furthermore, the second
term on the right-hand side of Eq. (23) vanishes in the spin- and isospin-saturated
system. By expanding Eq. (23) perturbatively one can  obtain the Dyson
equation for the pion Green's function, which reads as 
   \begin{equation}
 i\Delta_{\pi}(1,2)=i\Delta_{\pi}^{0}(1,2)+\not\!\! \int\,dx_{3}
\not\!\! \int\,dx_{4}\Delta_{\pi}^{0}(1,4) \Pi(4,3)i\Delta_{\pi}(3,2),
     \end{equation}
here $\Pi(4,3)$ is the self-energy of the pion, which is also a matrix on the
time contour
 \begin{eqnarray}
  \Pi(4,3)=\left( \begin{array}{cc}
   \Pi^{--}(4,3) & \Pi^{-+}(4,3) \\
   \Pi^{+-}(4,3) & \Pi^{++}(4,3) \end{array} \right) .
  \end{eqnarray}
Eq. (24) is coupled to the Dyson equation of the nucleon \cite{PRC94,PRC97}
   \begin{equation}
 iG(1,2)=iG^{0}(1,2)+\not\!\! \int\,dx_{3}
\not\!\! \int\,dx_{4}G^{0}(1,4) \Sigma(4,3)iG(3,2)
     \end{equation}
and delta \cite{PRC96}
 \begin{equation}
 iG_{\alpha \beta}(1,2)=iG_{\alpha \beta}^{0}(1,2)+\not\!\! \int\,dx_{3}
\not\!\! \int\,dx_{4}G_{\alpha \nu}^{0}(1,4) \Sigma^{\nu \mu}(4,3)iG_{\mu \beta}
(3,2)
     \end{equation}
through the self-energy terms of $\Pi(4,3)$, $\Sigma(4,3)$ and  $\Sigma_{\nu\mu}
 (4,3)$. Here $G(1,2)$, $G_{\alpha\beta}(1,2)$ are Green's functions of the nucleon
and delta, and $\Sigma(4,3)$, $\Sigma_{\nu\mu}(4,3)$ are the respective 
self-energies. Eqs. (24), (26) and (27) are a set of dynamical equations for
the hadronic matter. From Eqs. (26) and (27) we have derived the 
 RVUU-type transport equations for the nucleon 
\cite{Elz87,ZPA94,PRC94,Zhuxia,PRC97} and delta \cite{PRC96,PLB96} distribution
 functions. In this work we will develop a transport equation for the pion 
distribution function from Eq. (24), in which both the mean field and collision term will be
expressed analytically. Since the lowest-order Feynman diagrams contributing
 to the two-body scattering cross sections are the Born diagrams, we consider
 the pion self-energy $\Pi(4,3)$ up to the {\em Born approximation}.
In principle, one should calculate the in-medium cross sections and meson
fields for all the particles within a relativistic G-matrix theory.
However, since we have to deal with many reaction channels and many degrees
of freedom, such calculations seem to be out of the present practical 
possibilities. For a qualitative insight in the cross sections and potentials
we think that the Born approximation will be sufficient. A comparison between
the cross sections for $\sigma^{*}_{NN \rightarrow NN}$ and 
$\sigma^{*}_{NN \rightarrow N\Delta}$ calculated in G-matrix theory
\cite{Haa87,Li94} and in 
Born approximation \cite{ZPA94,PRC94}
 shows  differences only in the order of 10-20\%.

The pion self-energy up to the Born term can be written as
  \begin{equation}
 \Pi(4,3)=\Pi_{HF}(4,3) + \Pi_{Born}(4,3),
  \end{equation}
here $\Pi_{HF}(4,3)$ is the Hartree-Fock self-energy of the pion and $\Pi_{Born}
 (4,3)$ is the Born self-energy. The corresponding Feynman diagrams are given
 in Fig.~2, 3 and 4. 
  \begin{center}
  \fbox{Fig.~2}   \hspace{2cm}  \fbox{Fig.~3}   \hspace{2cm} \fbox{Fig.~4}
  \end{center}
 \noindent In Fig.~3 we only take the baryon-loops into account
since the contributions of meson-loops ($\sigma$-$\pi$ and $\rho$-$\pi$ mixed
loop) are negligible \cite{Gao95} at zero temperature (finite temperatures 
are not taken into account explicitly in the present framework of
 microscopic transport theory). Furthermore, since the pseudovector form is
chosen for the $\pi NN$ and $\pi\Delta\Delta$ 
 coupling, 
as discussed in Ref. \cite{Ser86} (Sec. 8.3)
the contribution of the sigma-pion coupling term to the $\pi N$
S-wave scattering lengths is small of order $m_{\pi}^2/(M_{N}m_{\sigma})$
and can be neglected, we drop 
the contribution of Fig.~2(a)  
 and Fig.~2(b) to the pion self-energy.   
Therefore, only Fig.~2(c) contributes to the Hartree-term of the pion self-energy,
 which plays a role in the case that a large amount of pions are produced
in relativistic heavy-ion collisions at very high energy. For the Born term
we consider the Feynman diagrams contributing to the $\pi + N \rightarrow \pi
 + N$, $\pi + \Delta \rightarrow \pi + \Delta$ and $\pi + \pi \rightarrow \pi
 + \pi$ elastic scattering processes as depicted in Fig.~4. 
For the same reason we neglect the contribution of the $\sigma$-exchange in 
Fig.~4(a) and 4(b).
The Hartree-Fock
self-energy $\Pi_{HF}(4,3)$ and Born self-energy $\Pi_{Born}(4,3)$ can then 
 be expressed as
  \begin{eqnarray}
 && \Pi_{HF}(4,3)=\Pi_{H}(4,3) + \Pi_{loop}(4,3), \\
 && \Pi_{loop}(4,3) = \Pi_{N N^{-1}}(4,3) + \Pi_{\Delta \Delta^{-1}}(4,3)
 + \Pi_{\Delta N^{-1}}(4,3) + \Pi_{N \Delta^{-1}}(4,3), \\
 && \Pi_{Born}(4,3) =\Pi_{a}(4,3) + \Pi_{b}(4,3) + \Pi_{c}(4,3) + \Pi_{d}(4,3),
  \end{eqnarray} 
where the lower subscripts $N^{-1}$ and $\Delta^{-1}$ in Eq. (30) denote the 
particles described by the nucleon and delta distribution functions 
rather than the anti-particles which are not taken into account in
this work because the gap between the effective masses of particle and 
anti-particle is much larger than the pion mass even at three times normal
density (see Fig.~5). For even higher densities and temperatures the production
of particle and anti-particle pairs becomes more important and the 
anti-particle degree of freedom should be taken into account. 
The lower subscripts a, b, c, d in Eq. (31) denote the terms 
contributed from Fig.~4(a)-(d), respectively.
$\Pi_{\Delta\Delta^{-1}}(4,3)$ (corresponding to Fig.~3(c) and 3(d))
 and $\Pi_{N \Delta^{-1}}(4,3)$ (corresponding to Fig.~3(g) and 3(h)) are 
usually neglected in the investigation of the influence of the in-medium 
 pion dispersion relation on  the pion dynamics in relativistic heavy-ion
collisions 
\cite{Xia88,Ko89,Ehe93,Xio93,Fuc97}. However, it was recently reported that
more than 30\% of nucleons are excited to the resonance-states, especially
 $\Delta$-resonance, in the $Au$ + $Au$ collisions at an incident energy of
2 GeV/nucleon \cite{Met93}. That means that the contributions of Fig.~3(c), 3(d) and 
3(g), 3(h) should be taken into account once the problem of the in-medium pion  
dispersion relation is concerned in {\em relativistic heavy-ion collisions}.
To our knowledge, up to now no investigation of this have been made
in transport theories. Here we note that this effect has been addressed in
some non-relativistic calculations of the pion self-energy
in static nuclear matter at finite temperature \cite{Hen94}.

 \end{sloppypar}
 \begin{sloppypar}
The concrete expressions of self-energies in Eqs. (29) - (31) can be
written down according to the standard Feynman rules 
 \begin{eqnarray}
 && \Pi_{H}(4,3)=\frac{3}{4}({\rm g}_{\sigma\pi}m_{\sigma})^{2}\delta(3,4)
 \not\!\!\int \,dx_{3}^{\prime} \Delta_{\pi}^{0}(3^{\prime},3^{\prime})
 i \Delta_{\sigma}^{0}(3^{\prime},4), \\  
 && \Pi_{NN^{-1}}(4,3)=-2 i ({\rm g}_{NN}^{\pi})^{2} tr \left[ \not\! P
 \gamma_{5} G^{0}(3,4) \not\! P \gamma_{5} G^{0}(4,3) \right], \\
 && \Pi_{\Delta\Delta^{-1}}(4,3)=-5 i ({\rm g}_{\Delta\Delta}^{\pi})^{2}
  tr \left[ \not\! P
 \gamma_{5} G^{0}_{\mu\nu}(3,4) \not\! P \gamma_{5} G^{0,\nu\mu}(4,3) \right],\\
 && \Pi_{\Delta N^{-1}}(4,3)=\frac{4}{3} i ({\rm g}_{\Delta N}^{\pi})^{2}
  tr \left[ 
  G^{0}(3,4) P^{\mu}P^{\nu}  G^{0}_{\nu\mu}(4,3) \right], \\
 && \Pi_{N\Delta^{-1}}(4,3)=\frac{4}{3} i ({\rm g}_{\Delta N}^{\pi})^{2} 
 tr \left[ P^{\mu} P^{\nu}
  G^{0}_{\mu\nu}(3,4) G^{0}(4,3) \right], 
 \end{eqnarray}
 \begin{eqnarray}
 \Pi_{a}(4,3) & = & \sum_{r_{4}t_{5}t_{6}} \not\!\!\int\, dx_{5} \not\!\!\int\,
 dx_{6} \langle r \mid {\rm g}_{\pi\pi}^{\rho} \Gamma_{\rho}^{\pi} \mid r_{4} 
 \rangle \Delta_{\pi}^{0}(4,3) \langle r_{4} \mid {\rm g}_{\pi\pi}^{\rho} 
 \Gamma_{\rho}^{\pi} \mid r \rangle \nonumber \\
 && tr \{ \langle t_{6} \mid {\rm g}_{NN}^{\rho} \Gamma_{\rho}^{N} \mid t_{5}
 \rangle G^{0}(5,6) \langle t_{5} \mid {\rm g}_{NN}^{\rho} \Gamma_{\rho}^{N}\mid
 t_{6} \rangle G^{0}(6,5) \} \nonumber \\
 && \Delta_{\rho}^{0}(4,6) \Delta_{\rho}^{0}(5,3) D_{\rho}D_{\rho},\\
 \Pi_{b}(4,3) & = & \sum_{r_{4}T_{5}T_{6}} \not\!\!\int\, dx_{5} \not\!\!\int\,
 dx_{6} \langle r \mid {\rm g}_{\pi\pi}^{\rho} \Gamma_{\rho}^{\pi} \mid r_{4} 
 \rangle \Delta_{\pi}^{0}(4,3) \langle r_{4} \mid {\rm g}_{\pi\pi}^{\rho} 
 \Gamma_{\rho}^{\pi} \mid r \rangle \nonumber \\
&& tr \{ \langle T_{6} \mid {\rm g}_{\Delta\Delta}^{\rho} \Gamma_{\rho}^{\Delta}
  \mid T_{5} \rangle 
 G^{0,\sigma\rho}(5,6) \langle T_{5} \mid {\rm g}_{\Delta\Delta}^{\rho}
  \Gamma_{\rho}^{\Delta} \mid
 T_{6} \rangle G^{0}_{\rho\sigma}(6,5) \} \nonumber \\
 && \Delta_{\rho}^{0}(4,6) \Delta_{\rho}^{0}(5,3) D_{\rho}D_{\rho},\\
 \Pi_{c}(4,3) & = & -\sum_{r_{4}r_{5}r_{6}} \not\!\!\int\, dx_{5} \not\!\!\int\,
 dx_{6} \langle r \mid {\rm g}_{\pi\pi}^{A} \Gamma_{A}^{\pi} \mid r_{4} 
 \rangle \Delta_{\pi}^{0}(4,3) \langle r_{4} \mid {\rm g}_{\pi\pi}^{A} 
 \Gamma_{A}^{\pi} \mid r \rangle \nonumber \\
 &&  \langle r_{6} \mid {\rm g}_{\pi\pi}^{A} \Gamma_{A}^{\pi} \mid r_{5}
 \rangle \Delta_{\pi}^{0}(5,6) \langle r_{5} \mid {\rm g}_{\pi\pi}^{A} 
 \Gamma_{A}^{\pi} \mid
 r_{6} \rangle \Delta_{\pi}^{0}(6,5)  \nonumber \\
 && \Delta_{A}^{0}(4,6) \Delta_{A}^{0}(5,3) D_{A}D_{A},\\
 \Pi_{d}(4,3) & = & -\sum_{r_{4}r_{5}r_{6}} \not\!\!\int\, dx_{5} \not\!\!\int\,
 dx_{6} \langle r \mid {\rm g}_{\pi\pi}^{A} \Gamma_{A}^{\pi} \mid r_{4} 
 \rangle \Delta_{\pi}^{0}(4,5) \langle r_{4} \mid {\rm g}_{\pi\pi}^{B} 
 \Gamma_{B}^{\pi} \mid r_{5} \rangle \nonumber \\
 &&  \Delta_{\pi}^{0}(5,6) \langle r_{5} \mid {\rm g}_{\pi\pi}^{A} 
 \Gamma_{A}^{\pi} \mid r_{6}
 \rangle \Delta_{\pi}^{0}(6,3) \langle r_{6} \mid {\rm g}_{\pi\pi}^{B} 
 \Gamma_{B}^{\pi} \mid
 r \rangle  \nonumber \\
 && \Delta_{A}^{0}(4,6) \Delta_{B}^{0}(5,3) D_{A}D_{B},
 \end{eqnarray}
In Eqs. (37) - (40) $A$, $B=\sigma$, $\rho$; $r$, $r_{4}$, $r_{5}$, $r_{6}$ 
represent the
 isospin of pions, $t_{5}$, $t_{6}$ denote the isospin of nucleons and $T_{5}$,
 $T_{6}$ of deltas. The definition of the symbols is given in Table I and II.
The transformed four-momentum $P_{\mu}$ in Eqs. (33) - (36) stems from the 
derivative coupling of baryon-baryon-pion vertex used in our calculations.

 \end{sloppypar}
\begin{center}
{\bf B. Kadanoff-Baym Equation}
 \end{center}
 \begin{sloppypar}
Introducing the  differential operator of the Klein-Gordon field
 \begin{equation}
 \stackrel{\wedge}
 {\Delta}^{-1}_{01} = \partial^{1}_{\mu} \partial ^{\mu}_{1} + m_{\pi}^{2}
 \end{equation}
and applying it to the both sides of Eq. (24), with the help of relation 
\cite{Lur68}
 \begin{equation}
 \stackrel{\wedge}
 {\Delta}^{-1}_{01} \Delta_{\pi}^{0}(1,2) = -\delta(1,2),
 \end{equation}
we obtain
 \begin{equation}
 \stackrel{\wedge}
 {\Delta}^{-1}_{01} i \Delta_{\pi}(1,2) = -i \delta(1,2)
 - \not\!\!\int \, dx_{3} \Pi(1,3) i \Delta_{\pi}(3,2).
 \end{equation}
It has been shown in Sect. II that only two components of $\Delta_{\pi}(1,2)$
 are independent, from which the dynamical equations for the distribution
 function and the spectral function can be constituted \cite{Bot88}.
 Since we will use the {\em quasi-particle approximation} in the derivation,
 the spectral function turns out to be a $\delta$ function on the mass shell.
Thus, in the present work it will be sufficient to consider only one
component of $\Delta_{\pi}(1,2)$, i.e., $\Delta_{\pi}^{-+}(1,2)$, which is 
directly related to the single-particle density matrix in the case of 
$t_{1}=t_{2}$ \cite{Dan84}. The equation of motion for $\Delta_{\pi}^{-+}(1,2)$
 can be extracted from Eq. (43). Before doing it, let us firstly look at the
Feynman diagrams in Figs. 2, 3 and 4 which will be taken into account under
Born approximation.

As is well known, the RVUU-type transport equation contains two important
ingredients, i.e., the transport part related to the real part of the pion
self-energy and the collision term corresponding to  the imaginary part.
The Hartree-term of Fig.~2(c) only contributes to the real part.
However, the loop diagrams in Fig. 3 and the Born diagrams in Fig. 4 include 
both  real and imaginary part.
 It should be pointed out that 
the baryon-lines in Fig.~3 denoted by the symbols $N$ or $\Delta$ represent
 virtual baryons (nucleon or delta) when one calculates the real part of
the self-energies. They are not on-shell particles. The corresponding terms for
the $\sigma$ and $\omega$ self-energies are neglected in the derivation
of Ref. \cite{Mro94} for the $\sigma$ and $\omega$ transport equations
because they used the restriction that all Green's functions in the Feynman
diagrams  should be on the
mass-shell. Consequently, mesons ($\sigma$ and $\omega$) became free particles
in their framework. To our opinion, in computing the real part of self-energies 
which mainly relates to the virtual processes, it is not necessary  to keep
all particles on the mass-shell which will essentially give the imaginary part. 
It is well known that the particle-hole excitation is very important for the 
in-medium pion dispersion relation which will certainly have influence on the
pion spectra  and pion flow in relativistic heavy-ion collisions \cite{Fuc97} 
and should be taken into account in any realistic transport models for pions.
For the imaginary part of the self-energies from one-loop diagrams in Fig.~3,  
we include only the contributions of Fig.~3(e) and (h),
which contribute to the important $\Delta$-formation process
of $N + \pi \rightarrow \Delta$ and $\Delta$-decay process of $\Delta
\rightarrow N + \pi$, respectively. 
 The reason is as follows: the contributions
of the imaginary part of Fig.~3(a)-(d), 
in which both the baryon-lines are on the 
mass-shell, correspond to the process that a nucleon (delta) decays into a 
nucleon (delta) and a pion, which is forbidden due to energy-momentum 
conservation (here we don't consider the Cherenkov radiation discussed in
Ref. \cite{Bro89}, this process might be possible at high densities where  
the pion has a space-like four momentum due to its large potential); 
Fig.~3(f) and 3(g) don't correspond to  realistic physical processes when the 
pion has a positive energy. Since the matrix elements are the same for both
$\Delta$-formation and $\Delta$-decay process, we only need to calculate
the imaginary part of Fig.~3(e) explicitly. 
In view of the Born diagrams we take only the
imaginary parts into account and drop all the real parts which in principle
 are the corrections to the real part of the Hartree-Fock self-energies. 
The imaginary part of the self-energies can be expressed by 
 $\Pi_{coll}^{\pm\mp}(1,3)$, which is defined as 
 \begin{equation}
 \Pi_{coll}^{\pm\mp}(1,3)=\Pi_{Born}^{\pm\mp}(1,3) + \Pi_{3(e)}^{\pm\mp}(1,3),
 \end{equation}
here $\Pi_{3(e)}^{\pm\mp}(1,3)$ represents the imaginary part of Fig.~3(e).
The equation of motion for $\Delta_{\pi}^{-+}(1,2)$ can then be written as
 \begin{eqnarray}
 \left[ \partial^{1}_{\mu}\partial^{\mu}_{1} \right.  
 & + & \left.  m_{\pi}^{2} +\Pi_{H}(1)
 \right] i \Delta_{\pi}^{-+}(1,2) \nonumber \\
 &=& - \int_{t_{0}}^{t}\, dx_{3}(Re\Pi_{loop}^{--}(1,3))i\Delta_{\pi}^{-+}(3,2)
 \nonumber \\
 && - \int_{t}^{t_{0}}\, dx_{3} \Pi_{coll}^{-+}(1,3)i(Re\Delta_{\pi}^{++}(3,2))
 \nonumber \\
 && - \int_{t_{0}}^{t_{1}}\, dx_{3} \left[ \Pi_{coll}^{+-}(1,3) - \Pi_{coll}
 ^{-+}(1,3) \right] i \Delta_{\pi}^{-+}(3,2) \nonumber \\
 && + \int_{t_{0}}^{t_{2}} \, dx_{3} \Pi_{coll}^{-+}(1,3) \left[ 
 i \Delta_{\pi}^{+-}(3,2) -i\Delta_{\pi}^{-+}(3,2) \right] .
 \end{eqnarray}
Eq. (45) is the 
so-called Kadanoff-Baym equation \cite{Kad62}.
Here the symbol $Re$ denotes the real part of the corresponding self-energies.
The second term on the right-hand
side of Eq. (45) corresponds to the spreading width in the spectral function.
It should be dropped under the quasi-particle approximation \cite{Bot88}
which will be introduced later.
The structure of the third and fourth term on the right-hand side of Eq. (45) 
 implies that they contribute to the
collision term of the transport equation.
The concrete expressions of 
the self-energies  read as
  \begin{eqnarray}
 \Pi_{H}(1) &=& \frac{3}{4}({\rm g}_{\sigma\pi}m_{\sigma})^{2}
 \{ \int_{t_{0}}^{t}\, dx_{3}^{\prime} \Delta_{\pi}^{0--}(3^{\prime},3^{\prime})
 i \Delta_{\sigma}^{0--}(3^{\prime},1) \nonumber \\
 && + \int_{t}^{t_{0}}\, dx_{3}^{\prime} \Delta_{\pi}^{0++}(3^{\prime},
 3^{\prime}) i\Delta_{\sigma}^{0+-}(3^{\prime},1) \}, \\
  \Pi_{NN^{-1}}^{--}(1,3)&=&-2 i ({\rm g}_{NN}^{\pi})^{2} tr \left[ \not\! P
 \gamma_{5} G^{0--}(3,1) \not\! P \gamma_{5} G^{0--}(1,3) \right], \\
  \Pi_{\Delta\Delta^{-1}}^{--}(1,3)&=&-5 i ({\rm g}_{\Delta\Delta}^{\pi})^{2}
  tr \left[ \not\! P
 \gamma_{5} G^{0--}_{\mu\nu}(3,1) \not\! P \gamma_{5} G^{0--,\nu\mu}(1,3) 
\right],\\
  \Pi_{\Delta N^{-1}}^{--}(1,3)&=&\frac{4}{3} i ({\rm g}_{\Delta N}^{\pi})^{2}
  tr \left[ 
  G^{0--}(3,1) P^{\mu}P^{\nu}  G^{0--}_{\nu\mu}(1,3) \right], \\
  \Pi_{N\Delta^{-1}}^{--}(1,3)&=&\frac{4}{3} i ({\rm g}_{\Delta N}^{\pi})^{2} 
 tr \left[ P^{\mu} P^{\nu}
  G^{0--}_{\mu\nu}(3,1) G^{0--}(1,3) \right], \\
  \Pi_{3(e)}^{\pm\mp}(1,3)&=&\frac{4}{3} i ({\rm g}_{\Delta N}^{\pi})^{2}
  tr \left[ 
  G^{0\mp\pm}(3,1) P^{\mu}P^{\nu} G^{0\pm\mp}_{\nu\mu}(1,3) \right]. 
  \end{eqnarray}
The expressions of the Born terms are rather complicated. If one does not write
out the isospin factor explicitly, $\Pi_{a}^{\pm\mp}(1,3)$ can be expressed as
  \begin{eqnarray}
 \Pi_{a}^{\pm\mp}(1,3) & \sim & \int_{t_{0}}^{t}\, dx_{5}\int_{t_{0}}^{t}\,
  dx_{6} \Delta_{\pi}^{0\pm\mp}(1,3) \nonumber \\
 && \{ tr \left[ G^{0\mp\mp}(5,6)G^{0\mp\mp}(6,5) \right] \Delta_{A}^{0\pm\mp}
 (1,6)\Delta_{A}^{0\mp\mp}(5,3) \nonumber \\
 && + tr \left[ G^{0\pm\pm}(5,6)G^{0\pm\pm}(6,5) \right] \Delta_{A}^{0\pm\pm}
 (1,6)\Delta_{A}^{0\pm\mp}(5,3) \nonumber \\
 && -tr \left[ G^{0\mp\pm}(5,6)G^{0\pm\mp}(6,5) \right] \Delta_{A}^{0\pm\pm}
 (1,6)\Delta_{A}^{0\mp\mp}(5,3) \nonumber \\
 && - tr \left[ G^{0\pm\mp}(5,6)G^{0\mp\pm}(6,5) \right] \Delta_{A}^{0\pm\mp}
 (1,6)\Delta_{A}^{0\pm\mp}(5,3) \}.      
  \end{eqnarray}
Other Born terms can be written down in the same way. 

 \end{sloppypar}
\begin{center}
{\bf C. RVUU Equation of the Pion}
 \end{center}
 \begin{sloppypar}
 Defining $X=\frac{1}{2}(x_{1}+x_{2})$, $y=x_{1}-x_{2}$, $x^{\prime}=x_{3}
 -x_{2}$ and taking the Wigner transformation on the both side of Eq. (45),
 we arrive at
 \begin{eqnarray}
 \left[ \frac{1}{4} \partial_{\mu}^{X}\partial_{X}^{\mu} \right.
 &- &iP^{\mu}\partial_{\mu}
 ^{X} -P^{2} +m_{\pi}^{2} + \Pi_{H}(X) +Re\Pi_{loop}^{--}(X,P)
 -\frac{i}{2}\partial_{X}^{\mu}\Pi_{H}(X)\partial_{\mu}^{P} \nonumber \\
 &+& \left. \frac{i}{2}\partial_{P}^{\mu}Re\Pi_{loop}^{--}(X,P)\partial_{\mu}^{X}
 -\frac{i}{2}\partial_{X}^{\mu}Re\Pi_{loop}^{--}(X,P)\partial_{\mu}^{P}
 \right] i\Delta_{\pi}^{-+}(X,P) \nonumber \\
&=& - \int\, dy e^{iPy}\int_{- \infty}^{\tau}\, dx^{\prime}\left[\Pi_{coll}^{+-}
 (y-x^{\prime},X) i\Delta_{\pi}^{-+}(x^{\prime},X) \right. \nonumber \\
  && \left. - \Pi_{coll}^{-+}(y -
 x^{\prime}, X) i \Delta_{\pi}^{+-}(x^{\prime},X) \right].
 \end{eqnarray}
Here  we have adopted the {\em semi-classical approximation}, in which the
Green's functions and self-energies are assumed to be peaked around the
relative coordinate and smoothly changing with the center-of-mass coordinate.
The details of the Wigner transformation are given in Appendix B. The Hermitian
conjugate equation of Eq. (53) reads as
 \begin{eqnarray}
 \left[ \frac{1}{4} \partial_{\mu}^{X}\partial_{X}^{\mu} \right.
 &+ &iP^{\mu}\partial_{\mu}
 ^{X} -P^{2} +m_{\pi}^{2} + \Pi_{H}(X) +Re\Pi_{loop}^{--}(X,P)
 +\frac{i}{2}\partial_{X}^{\mu}\Pi_{H}(X)\partial_{\mu}^{P} \nonumber \\
 &-& \left. \frac{i}{2}\partial_{P}^{\mu}Re\Pi_{loop}^{--}(X,P)\partial_{\mu}^{X}
 +\frac{i}{2}\partial_{X}^{\mu}Re\Pi_{loop}^{--}(X,P)\partial_{\mu}^{P}
 \right] i\Delta_{\pi}^{-+}(X,P) \nonumber \\
&=& - \int\, dy e^{iPy}\int^{ \infty}_{\tau}\, dx^{\prime}\left[\Pi_{coll}^{-+}
 (y-x^{\prime},X) i\Delta_{\pi}^{+-}(x^{\prime},X) \right. \nonumber \\
  && \left. - \Pi_{coll}^{+-}(y -
 x^{\prime}, X) i \Delta_{\pi}^{-+}(x^{\prime},X) \right].
 \end{eqnarray}
We drop the term of $\partial_{X}^{\mu}\partial_{\mu}^{X}$ in Eqs. (53) and (54)
since it may be viewed as of higher order than the other terms within the
{\em gradient expansion} used in the Wigner transformation. 
If one would keep this term, the pion Green's function could
  be non-zero for off-shell
four-momenta \cite{Mro90}. In this paper we only consider real on-shell pions.
The summation of
Eqs. (53) and (54) gives
  \begin{equation}
 \left[ P^{2} -m_{\pi}^{2} -\Pi_{H}(X) -Re\Pi_{loop}^{--}(X,P) \right]
 i \Delta_{\pi}^{-+}(X,P)=0,
 \end{equation}
and the subtraction of them yields
 \begin{eqnarray}
\{ P^{\mu}\partial_{\mu}^{X} &+&\frac{1}{2}\partial_{X}^{\mu}\Pi_{H}(X)\partial
 _{\mu}^{P} + \frac{1}{2}\partial_{X}^{\mu}Re\Pi_{loop}^{--}(X,P)\partial_{\mu}
 ^{P} - \frac{1}{2}\partial_{P}^{\mu}Re\Pi_{loop}^{--}(X,P)\partial_{\mu}^{X}
 \} i \Delta_{\pi}^{-+}(X,P) \nonumber \\
 &=& \frac{1}{2} \left[ \Pi_{coll}^{+-}(X,P)\Delta_{\pi}^{-+}(X,P)
 - \Pi_{coll}^{-+}(X,P)\Delta_{\pi}^{+-}(X,P) \right].
 \end{eqnarray}
Now we  introduce the {\em quasi-particle approximation} 
and dress the masses and
momenta in the zeroth-order Green's functions appearing in the self-energies
 with the effective masses and momenta. The canonical variables $X$, $P$ are 
then transformed to the kinetic variables $x$, $p$ which will be used in
 the RVUU code for the simulation of 
  relativistic heavy-ion collisions. Since the pion is
a pseudoscalar particle, we have $P_{\mu}=p_{\mu}$. Medium effects are 
included through the effective mass which is defined  as
 \begin{equation}
 m^{*2}_{\pi}(x,p)=m_{\pi}^{2} + \Pi_{H}(x) + Re\Pi_{loop}^{--}(x,p).
 \end{equation}
The on-shell condition
is guaranteed by Eq. (55)
 \begin{equation}
 p_{0}^{2}- \omega^{*2}(p)=0,
 \end{equation}
here
 \begin{equation}
 \omega^{*}(p) = \left[ {\bf p}^{2} + m^{*2}_{\pi}(x,p) \right] ^{\frac{1}{2}}.
 \end{equation}
We further define a distribution function 
 \begin{equation}
 i\Delta_{\pi}^{-+}(x,p) =\frac{\pi}{\omega^{*}(p)} Z_{B}
 \,\delta\!\left[p^{0}-\omega^{*}
 (p) \right] f_{\pi}({\bf x},{\bf p},\tau),
 \end{equation}
where
 \begin{equation}
 Z_{B}^{-1}= 1- \frac{1}{2\omega^{*}(p)} \frac{\partial Re\Pi_{loop}^{--}(x,p)}
 {\partial p_{0}} \mid _{p_{0}=\omega^{*}(p)}.
 \end{equation}
In the following we drop the derivative term in Eq. (61) 
since it will cause significant difficulty in deriving the collision
term.  
In the nuclear medium the quantum numbers of the pion can be either
transported as a physical pion or as a delta-hole bound state. In several
studies of the non-relativistic delta-hole model one considers the mixing
between these two branches of the pion dispersion relation 
\cite{Ehe93,Xio93,Fuc97}.
In this case strength is redistributed between the two branches as
a function of momentum. Therefore, the wave function renormalization factor
is essential and in principle it can be calculated from the energy dependence
of the pion self-energy. However, in practical application in transport
theories only phenomenological simulations of this mixing have been
investigated, Since we neglect the delta-hole branch in our relativistic
dynamical treatment, we put $Z_{B}^{-1}=1$. This does not cause any difficulty
with the conservation laws as can be seen from Eq. (F3). We simply obtain a
different but still conserved current. An improvement over this not very
satisfactory situation might be achieved if one studies relativistic transport
theories beyond the quasi-particle approximation \cite{Bot88}. Especially,
the inclusion of bound states in transport theory is, however, studied only in
a very few non-relativistic case near equilibrium up to now, e.g., for the
formation of deutron in nuclear matter \cite{Bey97}. It is clear that the
relativistic bound-state problem is much more involved than the
non-relativistic one. Therefore, we neglect this problem here.
Through 
inserting Eq. (60) into Eq. (56) one obtains the self-consistent RVUU equation
for the pion distribution function
 \begin{eqnarray}
\{ p^{\mu}\partial_{\mu}^{x} &+&\frac{1}{2}\partial_{x}^{\mu}\Pi_{H}(x)\partial
 _{\mu}^{p} + \frac{1}{2}\partial_{x}^{\mu}Re\Pi_{loop}^{--}(x,p)\partial_{\mu}
 ^{p} \nonumber \\ 
 &-& \frac{1}{2}\partial_{p}^{\mu}Re\Pi_{loop}^{--}(x,p)\partial_{\mu}^{x}
 \} \frac{f_{\pi}({\bf x},{\bf p},\tau)}{\omega^{*}(p)}  \nonumber \\
 &=& F_{c}(x,p).
 \end{eqnarray}
In Appendix F we show that this equation satisfies the conservation laws
of current and energy-momentum tensor.
The left-hand side of Eq. (62) is the transport part and the right-hand side
 is the collision term, which includes two parts,  
 \begin{equation}
 F_{c}(x,p)= F_{N\pi \rightarrow \Delta}(x,p) + F_{el}(x,p)
 \end{equation}
stemming from the $N \pi \rightarrow \Delta$ process and $\pi$-hadron elastic
scattering processes, respectively. Other reactions are not 
 included in the present work. 
The collision term  can be further expressed via in-medium
differential cross sections (Sect. V). 

 \end{sloppypar}
\begin{center}
{\bf IV. CALCULATION OF THE MEAN FIELD}
 \end{center}
 \begin{sloppypar}
In Sect. III we derived the RVUU-type transport equation for the pion 
distribution function. The left-hand side of the equation is the transport part
and the right-hand side is the collision term. The heart of the equation is the
 mean field, which relates to the in-medium pion dispersion relation, and the
$\pi$-relevant in-medium differential cross sections. In this section and the
next section we will evaluate the concrete expressions of them. Before coming
to it, we would like to emphasize again that in the present work we  
consider only the $\pi$ meson as a real meson. $\sigma$, $\omega$ and $\rho$ mesons
are still viewed as virtual ones. In other words, the terms relating to the 
distribution functions of $\sigma$, $\omega$ and $\rho$ mesons vanish. After
Wigner transformation, Eqs. (46) - (50) turn out to be  
  \begin{eqnarray}
 && \Pi_{H}(x)=\frac{3}{4}({\rm g}_{\sigma\pi}m_{\sigma})^{2}
 \int\, \frac{d^{4}q}{(2\pi)^{4}} \Delta_{\pi}^{0--}(x,q) i \Delta_{\sigma}
 ^{0--}(x,0), \\
 && \Pi_{NN^{-1}}^{--}(x,p)=-2i ({\rm g}_{NN}^{\pi})^{2} \int\, \frac{d^{4}q}
 {(2\pi)^{4}} tr \left[ \not\! p \gamma_{5} G^{0--}(x,q) \not\! p \gamma_{5}
 G^{0--}(x,p+q) \right], \\
 && \Pi_{\Delta\Delta^{-1}}^{--}(x,p)=-5i ({\rm g}_{\Delta\Delta}^{\pi})^{2} 
 \int\, \frac{d^{4}q}
 {(2\pi)^{4}} tr \left[ \not\! p \gamma_{5} G^{0--}_{\mu\nu}(x,q) 
  \not\! p \gamma_{5}
 G^{0--,\nu\mu}(x,p+q) \right], \\
 && \Pi_{\Delta N^{-1}}^{--}(x,p)=\frac{4}{3}i ({\rm g}_{\Delta N}^{\pi})^{2} 
  \int\, \frac{d^{4}q}
 {(2\pi)^{4}} tr \left[ G^{0--}(x,q) p^{\mu} p^{\nu}    
 G^{0--}_{\nu\mu}(x,p+q) \right], \\
 && \Pi_{N\Delta^{-1}}^{--}(x,p)=\frac{4}{3}i ({\rm g}_{\Delta N}^{\pi})^{2} 
  \int\, \frac{d^{4}q}
 {(2\pi)^{4}} tr \left[ p^{\mu}p^{\nu}  G^{0--}_{\mu\nu}(x,q)
 G^{0--}(x,p+q) \right]. 
  \end{eqnarray}
In the next step we insert the zeroth-order Green's functions  for
baryons (Appendix A)
and pion (Eqs. (20)- (22)) into Eqs. (64) - (68) to obtain concrete expressions of
the real part of the pion self-energies. Several approximations are made here.
Firstly, we take the quasi-particle approximation in which the
  free masses and momenta
in the zeroth-order Green's functions are addressed by the effective masses 
and momenta. Secondly, the first term on the right-hand side of Eq. (20), 
which will appear in Eq. (64) in the calculation of the Hartree term, is dropped
as usually done according to the {\em physical} arguments (otherwise, it will
cause divergence) \cite{Hor83}. Thirdly, in computing Eqs. (65) - (68), we drop
the contributions of anti-particles contained in the baryon Green's functions 
of $G^{0--}(x,q)$ and $G^{0--}_{\mu\nu}(x,q)$. 
The zeroth-order Green's functions used in this section then read as
  \begin{eqnarray}
 && \Delta_{\sigma}^{0--}(x,q)= \frac{1}{q^{2} -m_{\sigma}^{2} + i\epsilon},\\
 && \Delta_{\pi}^{0--}(x,q)= - \frac{\pi i}{\omega^{*}(q)} \,\delta \!\left[
 q_{0} - \omega^{*}(q) \right] f_{\pi}({\bf x},{\bf q},\tau), \\
 && G^{0--}(x,q)=\frac{\not\! q + m^{*}}{2 E^{*}(q)} \left[ \frac{1}
 {q_{0} -E^{*}(q) + i\epsilon } + 2\pi i\,\delta\!\left[ q_{0} -E^{*}(q)\right] 
 f({\bf x},{\bf q},\tau) \right], \\
 && G^{0--}_{\mu\nu}(x,q)=\frac{\not\! q + m_{\Delta}^{*}}{2 E_{\Delta}^{*}(q)}
 D_{\mu\nu}(q) \left[ \frac{1}
 {q_{0} -E_{\Delta}^{*}(q) + i\epsilon } + 2\pi i\, \delta\! 
 \left[ q_{0} -E_{\Delta}^{*}(q) \right] 
 f_{\Delta}({\bf x},{\bf q},\tau) \right], 
  \end{eqnarray}
where $E^{*}(q)=\left[ {\bf q}^{2} + m^{*2} \right] ^{1/2} $, 
$E^{*}_{\Delta}(q)=\left[ {\bf q}^{2} + m_{\Delta}^{*2} \right] ^{1/2} $, 
$D_{\mu\nu}(q)$ is given in Appendix A. The definition of $m^{*}$ and $m^{*}
 _{\Delta}$ will be given later.
It is interesting to notice that only the Green's functions on the upper branch
 of the time contour, which are similar to the ones 
 used in the standard effective field theory, 
  enter in the calculations.

 \end{sloppypar}
 \begin{sloppypar}
 The Hartree term
can be directly worked out 
 \begin{equation}
 \Pi_{H}(x)= - \frac{1}{4} ({\rm g}_{\sigma\pi})^{2} \rho_{S}(\pi), 
 \end{equation}
here $\rho_{S}(\pi)$ is the scalar density of pion
 \begin{equation}
 \rho_{S}(\pi)=\frac{3}{2(2\pi)^{3}} \int\, d{\bf q} \frac{1}{\sqrt{{\bf q}^{2}
 + m^{*2}_{\pi}}} f_{\pi}({\bf x},{\bf q},\tau).
 \end{equation}
For the one-loop diagrams we have to distinguish the real and virtual baryons.
The first terms on the right-hand side of Eqs. (71) and (72) 
describe the virtual
nucleon and delta, which are denoted by the $N$ and $\Delta$ on the Feynman
diagrams in Fig.~3. The second terms with distribution functions represent
the real nucleon and delta, and denoted by the $N^{-1}$ and $\Delta^{-1}$
on the Feynman diagrams. Through inserting Eqs. (71) and (72) into Eqs. (65) - 
(68),
 after some straightforward algebra we obtain the real part of the self-energies
  \begin{eqnarray}
 Re\Pi^{--}_{NN^{-1}}(x,p)=-2({\rm g}_{NN}^{\pi})^{2} \int \frac{d^{3}q}
 {(2\pi)^{3}} \left[ \frac{A(p,q)}{4E^{*}(q)E^{*}(p+q)} \frac{f({\bf x},{\bf q},
 \tau)}{E^{*}(p+q)-E^{*}(q)-p_{0}} + p_{0} \rightarrow - p_{0} \right] , \\
 Re\Pi^{--}_{\Delta\Delta^{-1}}(x,p)=-5({\rm g}_{\Delta\Delta}^{\pi})^{2}
   \int \frac{d^{3}q}
 {(2\pi)^{3}} \left[ \frac{B(p,q)}{4E^{*}_{\Delta}(q)E^{*}_{\Delta}(p+q)} 
  \frac{f_{\Delta}({\bf x},{\bf q},
 \tau)}{E^{*}_{\Delta}(p+q)-E^{*}_{\Delta}(q)-p_{0}} 
  + p_{0} \rightarrow - p_{0} \right] , \\
 Re\Pi^{--}_{\Delta N^{-1}}(x,p)=\frac{4}{3}({\rm g}_{\Delta N}^{\pi})^{2} 
 \int \frac{d^{3}q}
 {(2\pi)^{3}} \left[ \frac{C(p,q)}{4E^{*}(q)E^{*}_{\Delta}(p+q)} 
  \frac{f({\bf x},{\bf q},
 \tau)}{E^{*}_{\Delta}(p+q)-E^{*}(q)-p_{0}} + p_{0} \rightarrow - p_{0} \right] , \\
 Re\Pi^{--}_{N\Delta^{-1}}(x,p)=\frac{4}{3}({\rm g}_{\Delta N}^{\pi})^{2} 
 \int \frac{d^{3}q}
 {(2\pi)^{3}} \left[ \frac{D(p,q)}{4E^{*}_{\Delta}(q)E^{*}(p+q)} 
 \frac{f_{\Delta}({\bf x},{\bf q},
 \tau)}{E^{*}(p+q)-E^{*}_{\Delta}(q)-p_{0}} + p_{0} \rightarrow - p_{0} \right],
  \end{eqnarray}
here
  \begin{eqnarray}
 A(p,q) &=& 4\left[ p_{\mu}^{2} (p.q)_{A} -2m^{*2}p_{\mu}^{2} +2(p.q)^{2}_{A}
 \right] , \\
 (p.q)_{A} &=& \frac{1}{2} \left[ (E^{*}(q) + p_{0})^{2} -E^{*2}(p+q) -p_{\mu}
 ^{2} \right] , \\
 B(p,q) &=& - \frac{8}{9m^{*4}_{\Delta}} \{ 2p_{\mu}^{4}m^{*2}_{\Delta} \left[
 (p.q)_{B} -m^{*2}_{\Delta} \right] -2p_{\mu}^{2}(p.q)_{B}^{2} \left[
 (p.q)_{B} -3m^{*2}_{\Delta} \right] \nonumber \\
 && -5p_{\mu}^{2} m^{*4}_{\Delta} \left[ (p.q)_{B} -2m^{*2}_{\Delta} \right]
 -2(p.q)^{2}_{B} \left[ 2(p.q)_{B}^{2} +2(p.q)_{B}m^{*2}_{\Delta} +5m^{*4}
 _{\Delta} \right] \}, \\
 (p.q)_{B} &=& \frac{1}{2} \left[ (E^{*}_{\Delta}(q) + p_{0})^{2}
 -E^{*2}_{\Delta}(p+q) - p_{\mu}^{2} \right], \\
 C(p,q) &=& - \frac{8}{3m^{*2}_{\Delta}} \left[ p_{\mu}^{4} +2p_{\mu}^{2}(p.q)
 _{C} -m^{*2}_{\Delta}p_{\mu}^{2} +(p.q)^{2}_{C} \right] 
 \left[ (p.q)_{C} +m^{*2} +m^{*}m^{*}_{\Delta} \right] , \\
 (p.q)_{C} &=& \frac{1}{2} \left[ (E^{*}(q) + p_{0})^{2} -E^{*2}_{\Delta}(p+q)
 +m^{*2}_{\Delta}-m^{*2}-p_{\mu}^{2} \right] , \\
 D(p,q) &=& \frac{8}{3m^{*2}_{\Delta}} \left[ m^{*2}_{\Delta}p_{\mu}^{2} 
 -(p.q)_{D}^{2} \right] \left[ (p.q)_{D} + m^{*}m^{*}_{\Delta} +m^{*2}_{\Delta}
 \right] , \\
 (p.q)_{D} &=& \frac{1}{2} \left[ (E^{*}_{\Delta}(q) +p_{0})^{2} -E^{*2}(p+q)
 +m^{*2}-m^{*2}_{\Delta} -p_{\mu}^{2} \right] .
  \end{eqnarray}
Here we already dropped the contributions from virtual particle-particle
excitations (which are divergent)
 in consistently with the 
 mean field approximation. Otherwise, one has to renormalize it which
may be difficult in many situations.
The effective mass of pion is defined in Eq. (57). The effective masses and 
momenta of nucleon and delta are defined as \cite{PRC94,PRC96,PRC97,PRCSUB}
  \begin{eqnarray}
&& m^{*}(x) = M_{N} -{\rm g}_{NN}^{\sigma} \sigma(x) , \\
&& m^{*}_{\Delta}(x) = M_{\Delta} -{\rm g}_{\Delta\Delta}^{\sigma} \sigma(x) ,\\
&& p_{N}^{\mu}(x) = P^{\mu}_{N} -{\rm g}_{NN}^{\omega} \omega^{\mu}(x) , \\
&& p_{\Delta}^{\mu}(x) = P^{\mu}_{\Delta} -{\rm g}_{\Delta\Delta}^{\omega} 
 \omega^{\mu}(x) , 
  \end{eqnarray}
The mean fields of $\sigma(x)$ and $\omega^{\mu}(x)$ are obtained through
the following field equations within the {\em local density approximation}
 \begin{eqnarray}
 m_{\sigma}^{2}\sigma(x)+b({\rm g}^{\sigma}_{N N})^{3}\sigma^{2}(x)
+c({\rm g}^{\sigma}_{N N})^{4}\sigma^{3}(x)={\rm g}^{\sigma}_{N N}\rho_{S}(N)
+{\rm g}^{\sigma}_{\Delta \Delta}
\rho_{S}(\Delta) + \frac{1}{2}{\rm g}_{\sigma\pi}m_{\sigma} \rho_{S}(\pi), \\
 m_{\omega}^{2}\omega^{\mu}(x)+\frac{({\rm g}_{N N}^{\omega})^{2}m_{\omega}^{2}}
{Z^{2}}(\omega^{\mu}(x))^{3}
={\rm g}^{\omega}_{N N}\rho_{V}^{\mu}(N)
 + {\rm g}^{\omega}_{\Delta \Delta}\rho_{V}^{\mu}(\Delta).
 \end{eqnarray}
The scalar and vector densities
of the nucleon, and delta are defined as
 \begin{eqnarray}
&& \rho_{S}(i)=\frac{\gamma(i)}{(2 \pi)^{3}}\int d{\bf q} \frac{m^{*}_{i}}
 {\sqrt{{\bf q}^{2}+m^{*2}_{i}}} f_{i}({\bf x},{\bf q},\tau), \\
&& \rho_{V}^{\mu}(i)=\frac{\gamma(i)}{(2 \pi)^{3}}\int d{\bf q} \frac{q^{\mu}}
 {\sqrt{{\bf q}^{2}+m^{*2}_{i}}} f_{i}({\bf x},{\bf q},\tau).
 \end{eqnarray}
The abbreviations i=N, $\Delta$, and $\gamma$(i)= 4, 16, correspond
 to the nucleon and delta, respectively. 

 \end{sloppypar}
 \begin{sloppypar}
In Eqs. (87) - (90) we have dropped the Fock term of 
the  nucleon and delta self-energies since it makes only a small contribution.
The  Feynman diagrams for the Hartree term of nucleon and delta self-energies
can be drawn in the same way  as in
  Fig.~2 by replacing the pion external line with the nucleon and delta line. 
One may notice that the main contributions to the mean field of the nucleon and
 delta come from the Hartree term 
 while to the pion from the Fock term (one-loop diagrams).
The situation is caused by the pseudovector coupling for the pion adopted in our
considerations. If one uses pseudoscalar coupling for $\pi NN$ and $\pi\Delta
\Delta$ vertex, the pion will have a scalar 
self-energy from the Hartree term similar to the nucleon's and delta's. But this
term turns out to be so large that the effective mass of the pion will become
almost five times more massive at the normal density of nuclear matter than
in the vacuum \cite{Gao95}! This may not be the realistic case.

 \end{sloppypar}
\begin{center}
{\bf V. CALCULATION OF THE COLLISION TERM}
 \end{center}
\begin{center}
{\bf A. In-medium $N \pi \rightarrow \Delta$ Cross Section and $\Delta$-decay
 Width}
 \end{center}
 \begin{sloppypar}
Now we come to calculate the right-hand side of Eq. (62), i.e., the collision
term. The part corresponding to the $\Delta$-formation cross section reads as
 \begin{eqnarray}
F_{N\pi \rightarrow \Delta}(x,p)&=& \frac{1}{2} \left[ \Pi_{3(e)}^{+-}(x,p) 
 \Delta_{\pi}^{-+}(x,p)
 - \Pi_{3(e)}^{-+}(x,p) \Delta_{\pi}^{+-}(x,p) \right] \nonumber \\
 &=& \int\, \frac{d^{3}q}{(2\pi)^{3}} \int\, \frac{d^{3}k}{(2\pi)^{3}}
  (2\pi)^{4}\delta^{(4)}(p+q-k)
  W(p,q,k) (F_{2}^{0}-F_{1}^{0}),
 \end{eqnarray}
here $F^{0}_{2}$, $F^{0}_{1}$ are the Nordheim-Uehling-Uhlenbeck factors of the gain
($F^{0}_{2}$) and loss ($F_{1}^{0}$) terms,
 \begin{eqnarray}
 && F_{2}^{0} = \left[ 1+f_{\pi}({\bf x},{\bf p},\tau) \right]\left[ 1-
 f({\bf x},{\bf q},\tau)\right] f_{\Delta}({\bf x},{\bf k},\tau), \\
 && F_{1}^{0} = f_{\pi}({\bf x},{\bf p},\tau) 
 f({\bf x},{\bf q},\tau)\left[ 1- f_{\Delta}({\bf x},{\bf k},\tau)\right].
 \end{eqnarray}
$W(p,q,k)$ is the transition probability of the $N \pi \rightarrow \Delta$
process 
 \begin{eqnarray}
 W(p,q,k) &=& - \frac{({\rm g}_{\Delta N}^{\pi})^{2}}{6 \omega^{*}(p) E^{*}(q)
 E^{*}_{\Delta}(k)} tr \left[ (\not\! q +m^{*}) p^{\mu}p^{\nu} (\not\! k +
 m^{*}_{\Delta} ) D_{\nu\mu}(k) \right] \nonumber \\
 &=& \frac{({\rm g}_{\Delta N}^{\pi})^{2}}{\omega^{*}(p)E^{*}(q)E^{*}_{\Delta}
 (k) }\frac{1}{18m^{*2}_{\Delta}} \left[ (m^{*}_{\Delta} +m^{*})^{2} -m^{*2}
 _{\pi} \right]^{2} \left[ (m^{*}-m^{*}_{\Delta})^{2} -m^{*2}_{\pi} \right].
 \end{eqnarray}
In working out the second equality of Eq. (98)  we have used the relations
 $k=p+q$ and $p.q= (m^{*2}_{\Delta}-m^{*2}-m^{*2}_{\pi})/2$ from the
energy-momentum conservation and on-shell conditions.

In the above derivation all baryons are treated as elementary particles as 
usually done in quantum field theory. However the delta is a resonance that
can  decay. 
A mass distribution  function of the Breit-Wigner form is commonly 
introduced to describe the resonances with broad-widths \cite{Dan91,Wol92}.
However, one mostly discusses the problem
  in the free space. Here we assume that the same
form of distribution function applies to the medium with the free quantities
replaced by the effective quantities. That is, we introduce a Breit-Wigner
function for the $\Delta$-resonance in the medium 
  \begin{equation}
 F(m^{*2}_{\Delta})=\frac{1}{\pi}\frac{m^{*}_{0} \Gamma^{*}(\mid {\bf p} \mid)}
 {(m^{*2}_{\Delta} -m^{*2}_{0})^{2} +m^{*2}_{0} \Gamma^{*2}(\mid {\bf p} \mid)},
  \end{equation}
where $\Gamma^{*}(\mid {\bf p} \mid)$ is the in-medium momentum-dependent 
 $\Delta$-decay width. ${\bf p}$ is the relative momentum between nucleon and
pion in the $\Delta$-rest system 
 \begin{equation}
 {\bf p}^{2}=\frac{\left[ m^{*2}_{\Delta} - (m^{*} + m^{*}_{\pi})^{2} \right]
 \left[ m^{*2}_{\Delta}-(m^{*}-m^{*}_{\pi})^{2} \right] }{4m^{*2}_{\Delta}}.
 \end{equation}
$m^{*}_{0}$ is defined by Eq. (88) with the free $\Delta$-mass $M_{\Delta}$
 replaced by its resonance-mass $M_{0}$. Inserting the mass-distribution 
function of Eq. (99) into Eq. (95), we have
 \begin{eqnarray}
F_{N\pi \rightarrow \Delta}(x,p) 
 &=& \int\, \frac{d^{3}q}{(2\pi)^{3}} \int\, \frac{d^{3}k}{(2\pi)^{3}}
 \int\, dm^{*2}_{\Delta}
  (2\pi)^{4}\delta^{(4)}(p+q-k)
  W(p,q,k) F(m^{*2}_{\Delta})(F_{2}^{0}-F_{1}^{0}) \nonumber \\
 &=& \int\, \frac{d^{3}q}{(2\pi)^{3}} \upsilon \sigma_{abs}(s) (F_{2}^{0}
 -F_{1}^{0} ).
 \end{eqnarray}
In the second line of the above equation  we already expressed the collision 
term with the
cross section \cite{Gro80}. Since we are now in the 
$\Delta$-rest system, the effective 
 total energy of the system $s$ equals to the effective 
$\Delta$-mass $m^{*}_{\Delta}$. $\sigma_{abs}(m^{*}_{\Delta})$ reads as
  \begin{equation}
 \sigma_{abs}(m^{*}_{\Delta}) = \frac{2\pi}{\mid {\bf p} \mid}\frac{({\rm g}
 _{\Delta N}^{\pi})^{2}}{9 m^{*3}_{\Delta}} \left[ (m^{*}_{\Delta} +m^{*})^{2}
 -m^{*2}_{\pi} \right] ^{2} \left[ (m^{*}-m^{*}_{\Delta})^{2}-m^{*2}_{\pi}
 \right] F(m^{*2}_{\Delta}).
  \end{equation}
Performing an average over the initial states and writing 
out $F(m^{*2}_{\Delta})$
explicitly, we arrive at the  cross section of the $N\pi \rightarrow
 \Delta$ process
  \begin{eqnarray}
\sigma^{*}_{N \pi \rightarrow \Delta}(m^{*}_{\Delta})& =&\frac{({\rm g}_{\Delta N} 
 ^{\pi})^{2} }{18 m^{*3}_{\Delta} \mid {\bf p}\mid} 
 \left[ (m^{*}_{\Delta} +m^{*})^{2}
 -m^{*2}_{\pi} \right] ^{2} \left[ (m^{*}-m^{*}_{\Delta})^{2}-m^{*2}_{\pi}
 \right] \nonumber \\
  && \times \frac{m^{*}_{0}\Gamma^{*}(\mid {\bf p} \mid) }{(m^{*2}_{\Delta}
 -m^{*2}_{0})^{2} +m^{*2}_{0} \Gamma^{*2}(\mid {\bf p} \mid )}.
  \end{eqnarray}
According to Ref. \cite{Pil79}, the resonant cross section can also be 
expressed by means of the decay width
  \begin{equation}
 \sigma^{*}_{N\pi \rightarrow \Delta}=\frac{4\pi}{\mid {\bf p} \mid ^{2}}
 \frac{(2J+1)}{(2s+1)}\frac{(2I+1)}{(2t+1)(2r+1)}\frac{m^{*2}_{0}\Gamma^{*2}
 (\mid {\bf p} \mid) }{(m^{*2}_{\Delta}
 -m^{*2}_{0})^{2} +m^{*2}_{0} \Gamma^{*2}(\mid {\bf p} \mid )},
 \end{equation}
here $I$, $J$ are the isospin and spin of the $\Delta$, $t$, $s$ are that of
the nucleon, and $r$ is the isospin of the pion. 
Comparing Eqs. (103) and (104) we 
obtain the in-medium $\Delta$-decay width, which reads as
  \begin{equation}
 \Gamma^{*}(\mid {\bf p} \mid) = \frac{\mid {\bf p} \mid}{16\pi m^{*}_{0}}
 \frac{({\rm g}
 _{\Delta N}^{\pi})^{2}}{6 m^{*3}_{\Delta}} \left[ (m^{*}_{\Delta} +m^{*})^{2}
 -m^{*2}_{\pi} \right] ^{2} \left[ (m^{*}-m^{*}_{\Delta})^{2}-m^{*2}_{\pi}
 \right] .
  \end{equation}
In this way, we give a clear-cut relation between
in-medium $\Delta$-decay width and $\Delta$-formation cross section.

 \end{sloppypar}
\begin{center}
{\bf B. Elastic Pion-hadron Scattering}
 \end{center}
 \begin{sloppypar}
In this subsection we derive the analytical expressions for calculating the 
in-medium $\pi + N \rightarrow \pi + N$, $\pi  + \Delta \rightarrow \pi +
 \Delta$ and $\pi + \pi \rightarrow \pi + \pi$ elastic scattering cross 
sections. The corresponding part of the collision term can be written as
 \begin{eqnarray}
F_{el}(x,p)&=& \frac{1}{2} \left[ \Pi_{Born}^{+-}(x,p) 
 \Delta_{\pi}^{-+}(x,p)
 - \Pi_{Born}^{-+}(x,p) \Delta_{\pi}^{+-}(x,p) \right] \nonumber \\
 &=& \int\, \frac{d^{3}p_{2}}{(2\pi)^{3}} \int\, \frac{d^{3}p_{3}}{(2\pi)^{3}}
 \int\, \frac{d^{3}p_{4}}{(2\pi)^{3}} (2\pi)^{4}\delta^{(4)}(p+p_{2}-p_{3}
 -p_{4})  \nonumber \\
 && \times W(p,p_{2},p_{3},p_{4}) (F_{2}-F_{1}),
 \end{eqnarray}
here $F_{2}$, $F_{1}$ are again the Nordheim-Uehling-Uhlenbeck factors
 \begin{eqnarray}
 &&F_{2}= \left[ 1+f_{\pi}({\bf x},{\bf p},\tau) \right] \left[ 1 \pm f_{H_{2}}
 ({\bf x},{\bf p}_{2},\tau)\right] f_{H_{3}}({\bf x},{\bf p}_{3},\tau)
 f_{H_{4}}({\bf x},{\bf p}_{4},\tau), \\
 &&F_{1}= f_{\pi}({\bf x},{\bf p},\tau)  f_{H_{2}}
 ({\bf x},{\bf p}_{2},\tau)\left[ 1 \pm  f_{H_{3}}({\bf x},{\bf p}_{3},\tau)
 \right] \left[ 1 \pm f_{H_{4}}({\bf x},{\bf p}_{4},\tau) \right] , 
 \end{eqnarray}
$H_{2}$, $H_{3}$, $H_{4}$ can be $\pi$, $N$ or $\Delta$; the symbol $+$ assigns
to bosons and $-$ to fermions. The transition probability $W(p,p_{2},p_{3},
 p_{4})$ for different channels reads as
  \begin{eqnarray}
  W_{\pi N\rightarrow \pi N}(p,p_{2},p_{3},p_{4})&=&
 \frac{({\rm g}_{\pi\pi}^{\rho}
 {\rm g}_{NN}^{\rho})^{2}}{16\omega^{*}(p)E^{*}(p_{2})\omega^{*}(p_{3})E^{*}(p_{4})
 } T_{a}\Phi_{a}, \\
  W_{\pi \Delta\rightarrow \pi \Delta}(p,p_{2},p_{3},p_{4})&=&
 \frac{({\rm g}_{\pi\pi}^{\rho}
 {\rm g}_{\Delta\Delta}^{\rho})^{2}}{16\omega^{*}(p)E^{*}_{\Delta}(p_{2})
 \omega^{*}(p_{3})E^{*}_{\Delta}(p_{4})
 } T_{b}\Phi_{b}, \\
  W_{\pi \pi\rightarrow \pi \pi}(p,p_{2},p_{3},p_{4})&=&\frac{1}
 {16\omega^{*}(p)\omega^{*}(p_{2})\omega^{*}(p_{3})\omega^{*}(p_{4})} \sum_{AB}
  \left[ ({\rm g}_{\pi \pi}^{A})^{4}T_{c}\Phi_{c} + ({\rm g}_{\pi\pi}^{A}
 {\rm g}_{\pi\pi}^{B})^{2}T_{d}\Phi_{d} \right] \nonumber \\
 & & +  p_{3} \leftrightarrow  p_{4}. 
   \end{eqnarray}
where $A$, $B=\sigma$, $\rho$. $T_{a-d}$ is the isospin matrix and $\Phi_{a-d}$
 is the spin matrix. The subscripts $a$,$b$,$c$,$d$ denote the terms 
contributed from Fig.~4(a)-(d), respectively. The concrete expressions for 
$T_{a-d}$ and $\Phi_{a-d}$ are
  \begin{eqnarray}
 && T_{a}=\sum_{t_{2}r_{3}t_{4}} \langle r \mid \tau_{\rho}^{\pi} \mid r_{3}
 \rangle \langle r_{3} \mid \tau_{\rho}^{\pi} \mid r \rangle \langle t_{4} \mid
 \tau_{\rho} \mid t_{2} \rangle \langle t_{2} \mid \tau_{\rho}\mid t_{4} \rangle
 D_{\rho}^{i}D_{\rho}^{j}, \\
 && T_{b}=\sum_{T_{2}r_{3}T_{4}} \langle r \mid \tau_{\rho}^{\pi} \mid r_{3}
 \rangle \langle r_{3} \mid \tau_{\rho}^{\pi} \mid r \rangle \langle T_{4} \mid
 T_{\rho} \mid T_{2} \rangle \langle T_{2} \mid T_{\rho} \mid T_{4} \rangle
 D_{\rho}^{i}D_{\rho}^{j}, \\
 && T_{c}=\sum_{r_{2}r_{3}r_{4}} \langle r \mid \tau_{A}^{\pi} \mid r_{3}
 \rangle \langle r_{3} \mid \tau_{A}^{\pi} \mid r \rangle \langle r_{4} \mid
 \tau_{A}^{\pi}\mid r_{2} \rangle \langle r_{2} \mid \tau_{A}^{\pi} \mid
  r_{4} \rangle
 D_{A}^{i}D_{A}^{j}, \\
 && T_{d}=\sum_{r_{2}r_{3}r_{4}} \langle r \mid \tau_{A}^{\pi} \mid r_{4}
 \rangle \langle r_{4} \mid \tau_{B}^{\pi} \mid r_{2} \rangle \langle r_{2} \mid
\tau_{A}^{\pi} \mid r_{3} \rangle \langle r_{3} \mid \tau_{B}^{\pi} \mid r
 \rangle D_{A}^{i}D_{B}^{j}, 
  \end{eqnarray}
 \begin{eqnarray}
\Phi_{a}&=&(\gamma_{\rho}^{\pi})^{2}D_{\rho}^{\mu}D_{\rho}^{\nu} tr \left[ \gamma_{\rho}
 (\not\! p_{2} +m^{*}) \gamma_{\rho} (\not\! p_{4} +m^{*}) \right] 
 \left[ \frac{1}{(p-p_{3})^{2}-m_{\rho}^{2}}\right] ^{2} , \\
\Phi_{b}&=&(\gamma_{\rho}^{\pi})^{2}D_{\rho}^{\mu}D_{\rho}^{\nu} tr \left[ \gamma_{\rho}
 (\not\! p_{2} +m^{*}_{\Delta}) D^{\sigma\rho}(p_{2}) \gamma_{\rho} (\not\! p_{4}
 +m^{*}_{\Delta} ) D_{\rho\sigma}(p_{4}) \right] \nonumber \\
 && \times   \left[ \frac{1}{(p-p_{3})^{2}-m_{\rho}^{2}}\right] ^{2}, \\
  \Phi_{c}&=& (\gamma_{A}^{\pi})^{4}D_{A}^{\mu}D_{A}^{\nu} \left[ \frac{1}
 {(p-p_{3})^{2}-m_{A}^{2}} \right] ^{2}, \\
  \Phi_{d}&=&(\gamma_{A}^{\pi}\gamma_{B}^{\pi})^{2}D_{A}^{\mu}D_{B}^{\nu}
 \frac{1}{(p-p_{3})^{2}-m_{B}^{2}} \frac{1}{(p-p_{4})^{2}-m_{A}^{2}}.
 \end{eqnarray}
We further express the right-hand side of Eq. (106) by the differential cross 
sections \cite{Gro80}
  \begin{equation}
 F_{el}(x,p)=\int\, \frac{d^{3}p_{2}}{(2\pi)^{3}} \upsilon_{\pi} \sigma_{\pi}
 (s,t) (F_{2}-F_{1}) d\Omega,
  \end{equation}
here $\sigma_{\pi}(s,t)$ represents the in-medium differential cross sections
of $\pi + N\rightarrow \pi + N$, $\pi + \Delta \rightarrow \pi + \Delta$
and $\pi + \pi \rightarrow \pi + \pi$ elastic scattering. Its concrete 
expressions can be obtained through computing Eqs. (112) - (119) and finally 
transforming it into the center of mass system. We give the explicit expressions
of $\sigma_{\pi N \rightarrow \pi N}(s,t)$, $\sigma_{\pi \Delta \rightarrow
 \pi \Delta}(s,t)$ and $\sigma_{\pi \pi \rightarrow \pi \pi}(s,t)$ in Appendix
C. After averaging over initial states, the in-medium total cross sections can be
calculated through the following equations
  \begin{eqnarray}
&& \sigma^{*}_{\pi N \rightarrow \pi N} =\frac{1}{4} \int\, \sigma_{\pi N 
 \rightarrow \pi N}(s,t)d \Omega, \\
&& \sigma^{*}_{\pi \Delta \rightarrow \pi \Delta} =\frac{1}{16} \int\,
  \sigma_{\pi  \Delta
 \rightarrow \pi \Delta}(s,t)d \Omega, \\
&& \sigma^{*}_{\pi \pi \rightarrow \pi \pi} =\frac{1}{6} \int\, \sigma_{\pi  \pi
 \rightarrow \pi \pi}(s,t)d \Omega .
  \end{eqnarray}
Of course, in calculating the $\sigma_{\pi\Delta \rightarrow \pi\Delta}^{*}$
one should also take the $\Delta$-decay width into account. However,
 the strict treatment of Breit-Wigner distribution function as in Sect. IV
might cause complexity in the derivation procedure 
 since we are now concerning two deltas
in a scattering process. Practically, we 
usually introduce a centroid $\Delta$-mass in numerical calculations  
which can include the influence of $\Delta$-decay effectively.
For the detailed description of the method 
 we refer to Refs. \cite{PRC94,PRC96,PRC97}.
At the end we can rewrite Eq. (62), the RVUU-type transport 
equation of the pion, 
in the following form
 \begin{eqnarray}
\{ p^{\mu}\partial_{\mu}^{x} &+&\frac{1}{2}\partial_{x}^{\mu}\Pi_{H}(x)\partial
 _{\mu}^{p} + \frac{1}{2}\partial_{x}^{\mu}Re\Pi_{loop}^{--}(x,p)\partial_{\mu}
 ^{p}  
 - \frac{1}{2}\partial_{p}^{\mu}Re\Pi_{loop}^{--}(x,p)\partial_{\mu}^{x}
 \} \frac{f_{\pi}({\bf x},{\bf p},\tau)}{\omega^{*}(p)}  \nonumber \\
 &=& \int\, \frac{d^{3}q}{(2\pi)^{3}} \upsilon \sigma_{abs}(s) (F_{2}^{0}
 -F_{1}^{0} )
  + \int\, \frac{d^{3}p_{2}}{(2\pi)^{3}} \upsilon_{\pi} \sigma_{\pi}
 (s,t) (F_{2}-F_{1}) d\Omega.
 \end{eqnarray}
The first term on the right-hand side of Eq. (124) stems from the $N \pi
 \rightarrow \Delta$ process (it is angle independent in the center of mass system)
and the second term represents the $\pi$-hadron elastic scattering. This is 
of course interesting to investigate the $\pi$-hadron inelastic scattering
processes, which may warrant further studies.

 \end{sloppypar}
\begin{center}
{\bf VI. NUMERICAL RESULTS AND DISCUSSIONS}
 \end{center}
 \begin{sloppypar}
 In this section we present our numerical results for the in-medium pion
dispersion relation, $\Delta$-formation cross section and momentum-dependent
$\Delta$-decay width. The calculations are  performed in symmetric nuclear 
matter at zero temperature. The baryon distribution functions in Eqs. (75) - 
(78) and (93), (94) are replaced by the corresponding $\theta$ functions. 
The coupling strengths of ${\rm g}_{NN}^{\sigma}$, ${\rm g}_{NN}^{\omega}$
and b, c, Z are determined by fitting the known ground-state properties
for infinite nuclear matter. In this work we take the parameter set 2 of
Ref. \cite{Zli97}, which gives ${\rm g}_{NN}^{\sigma}$=11.77,
${\rm g}_{NN}^{\omega}$=13.88, $b({\rm g}_{NN}^{\sigma})^{3}$=13.447,
$c({\rm g}_{NN}^{\sigma})^{4}$=10.395 and $Z$=3.655. 
The corresponding saturation
 properties are: binding energy $E_{bin}$=$-$15.75 MeV, saturated effective 
nucleon mass $m^{*}_{N}/M_{N}$=0.6, compressibility $K$=200 MeV and the 
saturation density is $\rho_{0}$=0.1484 $fm^{-3}$.

For the coupling strengths of ${\rm g}_{\Delta\Delta}^{\sigma}$ and
${\rm g}_{\Delta\Delta}^{\omega}$, no direct information from experiments 
is available. For simplicity, we employ the argument of universal coupling 
strengths, i.e., ${\rm g}_{\Delta\Delta}^{\sigma}={\rm g}_{NN}^{\sigma}$ and
${\rm g}_{\Delta\Delta}^{\omega}={\rm g}_{NN}^{\omega}$ \cite{Mos74}. 
Other choices of the delta coupling strengths now in the literatures
\cite{Bog83,Wal87}
are not considered here since they give unreasonable results for the pion
dispersion relation.  
In this case, we
don't have a nonzero delta distribution in relativistic mean field calculations
\cite{Zli97,Wal87}. The contributions of Eqs. (76) and (78) to the real part
of pion self-energy vanish in the present calculations of the pion dispersion
relation. It is, of course, not the realistic situation of the dynamical process
of energetic heavy-ion collisions, where a rather large amount of nucleons
are excited to the resonance states \cite{Met93}. The contributions of Eqs. (76)
 and (78) will certainly enter the pion dispersion relation and might play an
important role because of the large spin- and isospin-factor of the
$\Delta$-resonance. Therefore, for the use of the in-medium pion dispersion 
relation presented in this work in the study of high-energy heavy-ion 
collisions, it should be viewed as a preliminary step approaching to the 
realistic description.

For the coupling strength of ${\rm g}_{NN}^{\pi}$, we take the most commonly 
used value $f_{\pi}^{2}/4\pi$=0.08 \cite{Ser86}. 
The coupling strength of ${\rm g}_{\Delta
N}^{\pi}$ can be fixed through using the Eq. (105) in the free space. If one
takes $M_{N}$=939 MeV, $M_{\Delta}$=1232 MeV, $m_{\pi}$=138 MeV and the 
empirical value of $\Gamma_{0}$=115 MeV, it turns out $f^{*2}/4\pi$=0.362,
very close to the commonly used value 0.37 \cite{Eri88} (if one uses this
value in Eq. (105), it gives $\Gamma_{0}$=118 MeV, still in the error bar of
experimental data). In computing the  real part of the pion self-energy we use a
cut-off factor of $\Lambda^{2}=exp(-2 {\bf p}^{2}/b^{2})$ with b=7 $m_{\pi}$ as
usually done \cite{Xia88,Ko89}.
 
 \end{sloppypar}
\begin{center}
{\bf A. In-medium Pion Dispersion Relation}
 \end{center}
 \begin{sloppypar}
Fig.~5 displays the gap between the effective masses of particles and
anti-particles
at different densities. One can see that the mass gap is much larger than 
the pion mass even at three times normal density. That means that the 
anti-particles contribute only to the high branch of collective modes, which
can be neglected in the present consideration. 

Fig.~6 depicts the in-medium pion dispersion relation calculated with Eq. (59)
at normal density. Here we already drop the contribution of 
the Hartree term since
we do not expect a large pion distribution at the energy and density considered.
The areas of non-vanishing imaginary part (NIP) of the pion self-energy are also
indicated in the figure. The imaginary part of the pion self-energy in  nuclear
matter can be derived through inserting Eqs. (71) and (72) into Eqs. (65) - (68)
 and taking both the baryons on the mass-shell. The analytical expressions are 
given in Appendix D.  It should be pointed out that in calculating the pion 
dispersion relation of Eq. (59) the self-consistency is realized only in the
real part of the pion self-energy.
From the figure one can find that the $\Delta$-h branch
is above the area of NIP. The sound branch is buried in the region of NIP
contributed from the nucleon-hole excitation. For the pion branch, it firstly
increases with the increase of momentum and then disappears in the region of
NIP contributed from the $\Delta$-hole excitation when the momentum is larger
than three times pion mass. That means that at larger momenta the pion can be
bounded in the $\Delta$-resonance. 
This scenario is commonly used in transport models. In the 
figure one can also find another dotted curve buried in the region of NIP from
$\Delta$-hole excitation, which is related to the case that a formed-$\Delta$
decays immediately. 

  \begin{center}
  \fbox{Fig.~5}   \hspace{2cm}  \fbox{Fig.~6}   \hspace{2cm} \fbox{Fig.~7}
  \end{center}

Since we are mostly interested in the pion branch, in the following we discuss
it in detail. In Fig.~7(a) we show the pion dispersion relation at different 
densities. The corresponding pion self-energy is displayed in Fig.~7(b).
At $\rho = 2\rho_{0}$ one can see the numerical unstable since at that density
the pion self-energy is very large compared to the pion mass. Contrary to the
results of the non-relativistic model \cite{Xia88} where the pion branch always
starts from the point of $\omega^{*}=m_{\pi}$ since the pion self-energy is
an explicit function of ${\bf p}^{2}$ in that model, 
our results exhibit that the pion 
dispersion relation (pion branch) has a rather different behavior for different
momenta. At lower momenta the pion has a positive self-energy, which causes
the in-medium pion dispersion relation to be harder than the free one. The pion
self-energy decreases with the increase of momentum. When $p$ exceeds the point
around 100 MeV the self-energy becomes negative and correspondingly the 
dispersion relation becomes softer than the free one. 
One may argue that nucleons are not very relativistic in nuclear matter, little
difference is expected, for slow pions, between the non-relativistic and
relativistic results. In Appendix E we reduce our relativistic formulae to the
non-relativistic limit. It is shown that the relativistic effect stemming from
the Fermi motion of nucleons is negligible. But there does exist  an evident
difference to the non-relativistic model mainly coming from the relativistic
kinetics where $p_{\mu}^{2}$ instead of ${\bf p}^{2}$ is used.
The dispersion relation of Fig.~7(a)
 may provide a possible explanation to the pion spectrum over whole 
energy range. In Ref. \cite{Fuc97} it was shown that the pion yield is 
overestimated at low momentum whereas underestimated at high momentum when a 
free dispersion relation was used.  
Fig.~7(c) displays the contributions of different excitation modes 
to the real part
 of the pion self-energy. One can find the main contribution comes from the 
$\Delta$-hole excitation as expected. The self-energy from the nucleon-hole
excitation always has a positive value whereas the one from the $\Delta$-hole
excitation changes its sign from positive to negative at certain momentum
point, which controls the behavior of the pion dispersion relation.

  \begin{center}
  \fbox{Fig.~8}     
  \end{center}

The above calculations are performed through considering the lowest-order
Feynman diagrams. The short-range correlations have not been included. In
non-relativistic model, the short-range correlations are taken into account 
by means of the Migdal parameter ${\rm g}^{\prime}$. This method is frequently
employed in the relativistic model \cite{Her92} although it has never been  
checked carefully. We follow this way and take the pion self-energy under 
Random Phase Approximation as
   \begin{equation}
 \Pi^{\prime}(p_{\mu})=\frac{Re \Pi_{loop}^{--}(x,p)}
{1 + ({\rm g}^{\prime}/p_{\mu}^{2}) Re \Pi_{loop}^{--}(x,p)}.
   \end{equation}
With ${\rm g}^{\prime}$=0.6 we have recalculated the pion dispersion relation
which is plotted in Fig.~8. The behavior of the pion dispersion relation
becomes quite strange and difficult to be understood. Similar results were
obtained in Ref. \cite{Her92}. That might mean that, it is unsuitable to 
incorporate the short-range correlations in a relativistic model by means of
a non-relativistic approach. A fully self-consistent
inclusion of correlation effects might be necessary, which is, however, quite
complicated and needs to be discussed in a separate paper. Another possibility
is that the effective Lagrangian of Eq. (19) might be valid only at  
lower order. 

For a convenient use in the study of heavy-ion collisions, we parameterize the
results of Fig.~7(a) as
   \begin{equation}
 \omega^{*} = \left\{ \begin{array}{cl}
  1.10398+0.0790471p+0.232015p^{2}
   -0.049101p^{3}  & \qquad \rho=0.5\rho_{0} \\
  1.32175-0.13706p+0.165035p^{2}
    -0.0173563p^{3} & \qquad \rho=\rho_{0} \\
  1.56304-0.585238p+0.261195p^{2}
   -0.0370091p^{3}  & \qquad \rho=2\rho_{0} 
    \end{array} \right. .   
  \end{equation}
The unit of $\omega^{*}$ and $p$ is $m_{\pi}$.

 \end{sloppypar}
\begin{center}
{\bf B. $\Delta$-formation Cross Section and $\Delta$-decay Width}
 \end{center}
 \begin{sloppypar}
 Now let us turn to the collision term of the pion transport equation. In this 
subsection we study the $\pi N \rightarrow \Delta$ cross section and the
momentum-dependent $\Delta$-decay width, which are the most important channels
for pion absorption and production, both in the free space and nuclear medium.
As has been pointed out in Sect. V, the delta is a physically decay particle. 
A Breit-Wigner function is commonly introduced to describe the broad-width
of the $\Delta$-resonance when one considers the $\Delta$-relevant scattering
processes \cite{Dan91,PRC94,Wol92}. Consequently, the mass of the $\Delta$ has a
distribution with respect to the total energy of the 
system. But in the framework of an
effective field theory one only treats a point particle with fixed mass. If
one introduces an energy-dependent mass, correspondingly, one should introduce
certain corrections on the interaction vertex. Brueckner \cite{Bru52} suggested
a vertex function of
  \begin{equation}
 F = \left( \frac{1 + R^{2} (p_{\pi}^{2})_{0} }{ 1 +R^{2}p_{\pi}^{2} } \right)
 ^{1/2}
  \end{equation}
 to fit the phase shift of a broad resonance away from the resonance position.
Here $p_{\pi}$ is the relative momentum between nucleon and pion in the 
$\Delta$-rest system. $R$ is the radius of the boundary of the internal region.
The typical value for the $\Delta$ is 0.98 fm \cite{Pil79}. In the same spirit a
mixed version of the form factor for the $N\Delta\pi$ vertex was  used in our
previous works \cite{PRC94,PRC96,PRC97} which has essentially  similar effects.
More conveniently, here we phenomenologically introduce a mass-dependent
coupling strength (MDC) ${\rm g}_{\Delta N}^{\pi} (M_{\Delta}) = 
{\rm g}_{\Delta N}^{\pi} F$, which will be used in the following calculations. 
The original coupling strength ${\rm g}_{\Delta N}^{\pi}$ is afterwards referred
to mass-independent coupling strength (MIC).

Fig.~9 displays the $\pi^{+}p \rightarrow \Delta^{++}$ and $\pi^{-}p 
\rightarrow \Delta^{0}$ cross section in free space. The dots are the
experimental data from Ref. \cite{Car71}. The solid and dashed curves are
our numerical results calculated with the mass-dependent (solid) and
mass-independent (dashed) coupling strength, respectively. The results
with MDC can reproduce the experimental data nearly perfectly. Furthermore,
our calculations are almost parameter-free. Only ${\rm g}_{\Delta N}^{\pi}$
was fixed by fitting $\Gamma_{0}=115$ MeV. That implies that our theoretical
framework for describing the pion is quite reasonable although it should be further
checked in relativistic heavy-ion collisions.

  \begin{center}
  \fbox{Fig.~9}   \hspace{4cm}  \fbox{Fig.~10} \hspace{4cm} \fbox{Fig.~11}  
  \end{center}

In Fig.~10 we depict the momentum-dependent $\Delta$-decay width in  free
space, calculated with MDC and MIC, respectively. Some phenomenological 
parameterizations commonly used in the transport models are also presented in
the figure. Our results with MDC are comparable with these parameterizations. 
But the decay width calculated with MIC increases very rapidly with the increase
of the pion-momentum, which will open the possibility that a $\Delta$ may have a
mass much larger than its resonance mass \cite{PRC96}. This may not be the real
case. In the following calculations we will use the mass-dependent coupling
strength for the $N\Delta\pi$ vertex.

In Fig.~11 we show the in-medium $\Delta$-formation cross section and
$\Delta$-decay width. The effective masses of nucleon and $\Delta$ are 
determined by Eqs. (87) and (88). The free pion mass is used in  (a) and
(b) while the effective pion mass from the dispersion relation of Fig.~7(a) is 
used in (c) and (d). 
 From the figure one can find the strong medium corrections. In the case of
free pion mass, the
effective $\Delta$-decay width decreases rapidly with the increase of density.
The $\Delta$-formation cross section is enhanced near the $\Delta$-resonance
mass but suppressed at other region. The whole shape of curves becomes narrower
compared to the free one. The in-medium $\Delta$-decay width was ever studied 
by Kim et al. \cite{Kim97} where an effective pion mass stemming from 
the nucleon-hole excitation was used. This kind of effective mass 
is very close to the free mass as stated in their paper, and might be seen
from Fig.~7(c). They obtained a suppressed $\Delta$-decay width at 
normal density 
which is in qualitative agreement with our results. 
However, in our calculations the Pauli blocking of the final nucleon is not
taken into account, so a quantitative comparison is not possible. The Pauli
blocking is of course incorporated in the transport equation (124).
 The effects
of Pauli-blocking alone on the in-medium resonance decay width were ever 
investigated by Effenberger et al. in Ref. \cite{Eff97}. 

In Fig.~11(c) and (d) the medium effects on the pion are incorporated.
Compared to Fig.~11(a) and (b) one can find that the effective mass of the 
 pion can change 
the results completely, even the trend of the density dependence. 
It seems that if the smaller effective mass of the pion is taken into 
account the in-medium   
$\Delta$-decay width increases as compared to the free mass.  
The medium effects
on the $\Delta$-formation cross section are now exhibited to be important 
only in the region where the formed-$\Delta$ has the mass around the 
$\Delta$-resonance mass. It decreases with the increase of density. When the
formed-$\Delta$ is far away from its resonance-mass, the medium corrections
on the $\Delta$-formation cross section are negligible.

 \end{sloppypar}
 \begin{center}
{\bf VII. SUMMARY AND OUTLOOK}
 \end{center}
  \begin{sloppypar}
A large amount
of data with high accuracy for the pion spectrum and pion flow has been accumulated.
 A $\pi$-beam facility will be available at GSI which will provide new and
specific data to help to understand the pion dynamics in relativistic 
heavy-ion collisions. However, a self-consistent treatment of pions, together
with the nucleons and deltas, is still not realized in the current
 transport models.
In view of this fact,  we have developed a RVUU-type transport
equation for the pion distribution function based on an effective Lagrangian 
of the QHD-II model \cite{Ser86}. 
The closed time-path Green's function technique
is employed and the semi-classical, quasi-particle and Born approximation are
adopted in our derivation. 
We have presented an unified approach to the following problems:
First, both the mean field (the real part of the pion 
self-energy) and the collision term (the imaginary part of the pion self-energy)
of the transport equation are derived simultaneously from the same effective
Lagrangian and presented analytically.  
Second, we treat the real pion and virtual pion on the same footing.
 Third, the transport equation of pion is derived
within the same framework which we applied to the nucleon
\cite{Elz87,ZPA94,PRC94,Zhuxia,PRC97} and delta \cite{PRC96,PLB96} before.      
Therefore, we obtain a set of coupled equations for the $N$, $\Delta$ and
$\pi$ system which describes the hadronic matter in an unified way. 

Within this approach we have investigated the in-medium pion dispersion 
relation. In contrast to the results of the non-relativistic model where a softer
dispersion relation over the whole momentum range is exhibited, the predicted 
in-medium dispersion relation turns out to be harder at lower momenta and 
softer at higher momenta, compared to the free one. The main reason for the
difference relies on that in our relativistic model the pion self-energy 
has a relativistic kinetics $p_{\mu}$ while in the non-relativistic model it
explicitly depends on three-momentum ${\bf p}$ \cite{Eri88}, in which the
real-part of the pion self-energy goes to zero when ${\bf p} \rightarrow 0$.
However, a  pion in the nuclear medium should in principle suffer the 
interaction of surrounding particles whatever it moves or not. In our 
calculations the real part of the pion self-energy has a positive value at 
smaller momenta \cite{Fri98}. 
Consequently, the momentum dependence of the in-medium pion dispersion 
relation becomes very flat and quite different from the free one and that of
the non-relativistic model. This will certainly have effects on the pion spectrum,
 pion flow as well as on the 
 dilepton production since one of the important channels
$\pi^{+} \pi^{-} \rightarrow \rho \rightarrow  e^{+}e^{-}$ explicitly dependents
 on the slope of $d\omega / dp$ \cite{Gal87}. It would be very interesting to
check this kind of pion dispersion relation in the dynamical processes of 
relativistic heavy-ion collisions. Work on this aspect is in progress.
 
Considering that 
 in the nuclear medium the absorption and production of pions are mostly 
realized through the formation and decay of the $\Delta$-resonance, we have studied
 the $\Delta$-formation cross section and $\Delta$-decay width both in the free
space and in the medium. Our theoretical prediction for the free 
$\Delta$-formation cross section is nearly in perfect agreement with the 
experimental data. The computed free $\Delta$-decay width is comparable to 
several phenomenological parameterizations commonly used in the transport models.
 It is found that the effective pion mass has strong influence on the 
predicted in-medium $\Delta$-formation cross section and $\Delta$-decay width.
It can even change the trend of the density dependence. After taking into account
the medium corrections on the nucleon, delta and pion mass simultaneously, 
the $\Delta$-decay width turns out to be enhanced in the medium especially 
at higher momenta, while the $\Delta$-formation cross section is suppressed 
around the resonance mass. When the formed-$\Delta$ is far away from its 
resonance mass, the medium effects on the $\Delta$-formation cross section are
negligible.  

 In investigating the in-medium pion dispersion relation we have taken into 
account only the lowest-order diagrams stemming from the nucleon-hole and
$\Delta$-hole excitation. The effects of short-range correlations have 
been addressed through following the method of the non-relativistic model with the
Migdal parameter ${\rm g}^{\prime}$. But the results turn out to be quite 
strange. In fact, the short range correlations can be considered in a 
relativistic model self-consistently by implementing correlation terms like
$( \bar{\psi}\gamma_{\mu}\gamma_{5}\mbox{\boldmath $\tau$}\psi )^{2}$ and
$( \bar{\psi}_{\Delta\mu}{\bf S}^{+}\psi )^{2}$ in the effective Lagrangian 
of Eq. (19). In the mean field approximation one should calculate the 
expectation value of $\langle \bar{\psi}\gamma_{\mu}\gamma_{5}\mbox{\boldmath
$\tau$} \psi \rangle$ and $\langle \bar{\psi}_{\Delta \mu} {\bf S}^{+} \psi
\rangle $ which corresponds to the condensate of the pion field. Although the
detailed investigation of correlation effects may go beyond the scope of the RVUU
approach, it is nevertheless very interesting to check the effects of 
short-range correlations on the pion dispersion relation even in static nuclear 
matter within a relativistic model self-consistently. It 
would be extremely interesting if this collective instability could be studied
in a dynamical situation.

As usually done in a microscopic transport theory for non-equilibrium system,
the temperature degree of freedom is not taken into account in the present work.
 One might consider the temperature degree of freedom in  nuclear matter 
by simply replacing the
single-particle distribution functions with the Fermi-Dirac (for fermions) and
Bose-Einstein (for bosons) distribution functions. In this way one can study    
the effects of temperature on the predicted quantities. This is especially 
meaningful in the present case since the delta distribution will enter 
explicitly at certain temperatures \cite{Zli97}. Then the $\Delta \Delta^{-1}$
and $N\Delta^{-1}$ excitation of Fig.~3(c), (d) and (g), (h) might play an
important role on the in-medium pion dispersion relation. This would be
  more close
to the realistic situation of energetic heavy-ion collisions and 
requires further studies.

As frequently emphasized in the paper, up to now we only treat  the pion as a real
particle. Other mesons such as $\sigma$, $\omega$ and $\rho$ still remain to
be virtual ones. It is of course interesting to develop transport equations for
$\omega$ and $\rho$ mesons as well as for other experimental observable mesons
$K$, $K^{*}$, $\eta$, $\phi$ ... within the present framework. Among them,
the $\rho$ meson is especially important for dilepton production which can not be
explained by the transport models currently on the market. Medium corrections
on the properties of $\omega$, $\rho$ as well as $\eta$ mesons may provide
a possible explanation \cite{Rap97,Li95,Cas95,LiPre,BraPre,ErnPre}.

Almost in all practically used RVUU-type transport calculations, 
the local-density
approximation is employed in order to realize the numerical solution of the
equation. This drops the retardation effects of the mean field, although the
equation itself is constructed from the relativistic model. From Eq. (2) one
can find that it is possible to describe the propagation of real as well as
virtual $\sigma$ and $\omega$ mesons simultaneously. In this case one can 
obtain the time-dependent mean field and the transport equations for the 
$\sigma$ and $\omega$ distribution functions at the same time. It would be 
interesting to study this problem and explore it just 
from a theoretical point of view.

  \end{sloppypar}

 \begin{center}
{\bf ACKNOWLEDGMENTS}
 \end{center}
 \begin{sloppypar}
 The authors  thank C.~Ernst  and 
 I.N.~Mishustin for fruitful discussions.
 G.~Mao is  grateful to the Alexander von
Humboldt-Stiftung for financial support and to the people at the 
Institut f\"{u}r
Theoretische Physik der J.~W.~Goethe Universit\"{a}t for their hospitality.
This work was supported by DFG-Graduiertenkolleg Theoretische \& Experimentelle
Schwerionenphysik, GSI, BMBF, DFG and A.v.Humboldt-Stiftung.

 \end{sloppypar}

\setcounter{equation}{0}
\renewcommand{\theequation}{A\arabic{equation}}
 \begin{center}
{\bf APPENDIX A}
 \end{center}
 \begin{sloppypar}
 In this appendix we present the zeroth-order Green's functions of nucleon,
delta and mesons used in this work. \\
  (1) Nucleon
  \begin{equation}
 \langle T \lbrack \psi(1)\bar{\psi}(2) \rbrack \rangle =
  iG^{0}(1,2)\delta_{t_{1},t_{2}^{\prime}},
  \end{equation}
   \begin{eqnarray}
&& iG^{0}(1,2)=i\int\frac{d^{4}k}{(2\pi)^{4}}G^{0}(x,k)e^{-ik(x_{1}-x_{2})}, \\
&& G^{0 \mp \mp}(x,k)=(\not\!k+M_{N}) \left[ \frac{\pm 1}{k^{2}-M_{N}^{2}\pm i
\epsilon}
 + \frac{\pi i}{E(k)}\,\delta\!\left[k_{0}-E(k)\right]f(x,k) \right] , \\
&& G^{0+-}(x,k)=-\frac{\pi i}{E(k)}\,\delta\!\left[k_{0}-E(k)\right][1-f(x,k)]
 (\not\! k+M_{N}), \\ 
 && G^{0-+}(x,k)=\frac{\pi i}{E(k)}\,\delta\!\left[k_{0}-E(k)\right]
  f(x,k)(\not\! k+M_{N}).
   \end{eqnarray}
  (2) Delta 
  \begin{equation}
 \langle T\lbrack\psi_{\Delta\mu}(1)\bar{\psi}_{\Delta\nu}(2)\rbrack\rangle=
  -iG^{0}_{\mu\nu}(1,2)\delta_{T
  _{1},T_{2}^{\prime}},
  \end{equation}
  \begin{eqnarray}
  && iG^{0}_{\mu \nu}(1,2)=i\int\frac{d^{4}k}{(2\pi)^{4}}G^{0}_{\mu \nu}(x,k)
 e^{-ik(x_{1}-x_{2})}, \\
 && G^{0\mp\mp}_{\mu\nu}(x,k)=(\not\! k+M_{\Delta})D_{\mu \nu}\left[\frac{\pm 1}
 {k^{2}-M_{\Delta}^{2}\pm i\epsilon} +\frac{\pi i}{E(k)}\,\delta\!\left[k_{0}-E(k)\right]
 f_{\Delta}(x,k)\right], \\
 && G^{0+-}_{\mu \nu}(x,k)=-\frac{\pi i}{E(k)}\,\delta\!\left[ k_{0}-E(k)\right]
  [1-f_{\Delta}(x,k)]
 (\not\! k+M_{\Delta})D_{\mu \nu}, \\
 && G^{0-+}_{\mu \nu}(x,k)=\frac{\pi i}{E(k)}\,\delta\!\left[ k_{0}-E(k)\right]
  f_{\Delta}(x,k)
 (\not\! k+M_{\Delta})D_{\mu \nu}, \\
 && D_{\mu \nu}=g_{\mu \nu}-\frac{1}{3}\gamma_{\mu}\gamma_{\nu}-\frac{1}{3M_{
\Delta}}(\gamma_{\mu}k_{\nu}-\gamma_{\nu}k_{\mu})-\frac{2}{3M_{\Delta}^{2}}
 k_{\mu}k_{\nu}.
   \end{eqnarray}
(3)  Mesons
\begin{equation}
\langle T\lbrack\Phi_{A}(1)\Phi_{B}(2)\rbrack\rangle 
  =iD_{A}\Delta_{A}^{0}(1,2)\delta_{AB},
 \end{equation}
where $A$, $B=\sigma$, $\omega$, $\pi$, $\rho$, 
 \begin{equation}
 D_{A}=D_{A}^{\mu}D_{A}^{i},
 \end{equation}
D$_{A}^{\mu}$ and D$_{A}^{i}$ are defined and listed in Table I and
   \begin{eqnarray}
  && i\Delta_{A}^{0}(1,2)=i \int\frac{d^{4}k}{(2\pi)^{4}}\Delta^{0}_{A}(x,k)
 e^{-ik(x_{1}-x_{2})}, \\
  && \Delta^{0\mp\mp}_{A}(x,k)=\frac{\pm 1}{k^{2}-m_{A}^{2}\pm i\epsilon}
 -2\pi i \,\delta\!\left[ k^{2}-m_{A}^{2}\right] f_{A}(x,k), \\
  && \Delta^{0\pm\mp}_{A}(x,k)=-2\pi i \,\delta\!\left[ k^{2}-m_{A}^{2}\right]
 [\theta(\pm k_{0})+f_{A}(x,k)].
  \end{eqnarray}
Here the number 1, 2 represent $x_{1}$,
 $x_{2}$.
 t$_{1}$, t$_{2}^{\prime}$ denote the isospin of nucleons and T$_{1}$,
 T$_{2}^{\prime}$ denote those of deltas.
 f$(x,k)$, f$_{\Delta}(x,k)$, f$_{A}(x,k)$ are nucleon, delta and
meson distribution functions, respectively. The abbreviation for isospin
on the distribution function has been suppressed. 

 \end{sloppypar}
\setcounter{equation}{0}
\renewcommand{\theequation}{B\arabic{equation}}
 \begin{center}
{\bf APPENDIX B}
 \end{center}
 \begin{sloppypar}
 In this appendix we perform the Wigner transformation of Eq. (45), which 
can be easily realized by means of the following formulae \cite{Mro90}
 \begin{eqnarray}
 \partial_{x}^{\mu}f(x,y) & \longrightarrow & (-i P^{\mu} +\frac{1}{2}
 \partial^{\mu} ) f(X,P) ,  \\
 \partial_{y}^{\mu}f(x,y) & \longrightarrow & (i P^{\mu} +\frac{1}{2}
 \partial^{\mu} ) f(X,P) ,  \\
 h(x)g(x,y) & \longrightarrow &h(X)g(X,P) - \frac{i}{2} \frac{\partial h(X)}
 {\partial X^{\mu}} \frac{\partial g(X,P)}{\partial P_{\mu}}, \\
 h(y)g(x,y) & \longrightarrow &h(X)g(X,P) + \frac{i}{2} \frac{\partial h(X)}
 {\partial X^{\mu}} \frac{\partial g(X,P)}{\partial P_{\mu}}, \\
 \int d^{4}x^{\prime} f(x,x^{\prime})g(x^{\prime},y)
 & \longrightarrow & f(X,P)g(X,P) + \frac{i}{2} \left[ \frac{\partial f(X,P)}
 {\partial P_{\mu}} \frac{\partial g(X,P)}{\partial X^{\mu}} \right. \nonumber
 \\
  && \left. - \frac{\partial f(X,P)}{\partial X^{\mu}} \frac{\partial g(X,P)}
 {\partial P_{\mu}} \right],
 \end{eqnarray}
here 
 \begin{equation}
 X=\frac{1}{2}(x+y).
 \end{equation}
After Wigner transformation the different terms in Eq. (45) turn out to be
 \begin{eqnarray}
 \partial_{\mu}^{1}\partial_{1}^{\mu} \Delta_{\pi}^{-+}(1,2)
 & \longrightarrow & (\frac{1}{4}\partial_{\mu}^{X}\partial_{X}^{\mu}
 -i P^{\mu}\partial_{\mu}^{X} -P^{2} ) \Delta_{\pi}^{-+}(X,P), \\
 \Pi_{H}(1)\Delta_{\pi}^{-+}(1,2) & \longrightarrow & \Pi_{H}(X)\Delta_{\pi}
 ^{-+}(X,P) - \frac{i}{2} \partial_{X}^{\mu}\Pi_{H}(X)\partial_{\mu}^{P}
 \Delta_{\pi}^{-+}(X,P), \\ 
 Re\Pi_{loop}^{--}(1,3)\Delta_{\pi}^{-+}(3,2) & \longrightarrow &
 Re\Pi_{loop}^{--}(X,P)\Delta_{\pi}^{-+}(X,P) + \frac{i}{2} \left[ \partial_{P}
 ^{\mu} Re\Pi_{loop}^{--}(X,P)\partial_{\mu}^{X}\Delta_{\pi}^{-+}(X,P)
 \right. \nonumber \\
 && - \left. \partial_{X}^{\mu} Re\Pi_{loop}^{--}(X,P) \partial_{\mu}^{P}
 \Delta_{\pi}^{-+}(X,P) \right], \\
 \Pi_{coll}^{+-}(1,3) \Delta_{\pi}^{-+}(3,2) & \longrightarrow &
 \Pi_{coll}^{+-}(X,P) \Delta_{\pi}^{-+}(X,P).
 \end{eqnarray}
In Eq. (B10) we have dropped the contributions from derivative terms. That means
that collisions are performed at instantaneous time: {\em Boltzmann Ansatz}
 \cite{Dav91,Kad62}.

 \end{sloppypar}
\setcounter{equation}{0}
\renewcommand{\theequation}{C\arabic{equation}}
 \begin{center}
{\bf APPENDIX C}
 \end{center}
 \begin{sloppypar}
In this appendix we present analytical expressions of in-medium differential 
cross sections for $\pi$-hadron elastic scattering. \\
 (a) Differential cross section of in-medium $\pi N \rightarrow \pi N$
 scattering:
  \begin{eqnarray}
 \sigma_{\pi N \rightarrow \pi N}(s,t) & =& \frac{1}{(2\pi)^{2}s} 
  \frac{({\rm g}_{\rho\pi}{\rm g}_{NN}
 ^{\rho})^{2}}{2(t-m_{\rho}^{2})^{2}} \left[ (m^{*2}+m^{*2}_{\pi}-2s)
 \right. \nonumber \\ && \left. \times (m^{*2}+m^{*2}_{\pi} ) 
   +s^{2} +st -m^{*2}t \right] ,
  \end{eqnarray}
where
 \begin{eqnarray}
 && s=(p+p_{2})^{2}=\left[ \omega^{*}(p) +E^{*}(p_{2})\right] ^{2}
 - ({\bf p} + {\bf p}_{2})^{2}, \\
 && t=\frac{1}{2}(2m^{*2} +2m^{*2}_{\pi}-s) -\frac{1}{2s} (m^{*2}-m^{*2}_{\pi})
 ^{2} +2 \mid {\bf p} \mid \mid {\bf p}_{3} \mid \cos \theta, \\
 && \mid {\bf p} \mid = \mid {\bf p}_{3} \mid  =\frac{1}{2\sqrt{s}}
 \sqrt{(s-m^{*2}-m^{*2}_{\pi})^{2} -4m^{*2}_{\pi}m^{*2} },
 \end{eqnarray}
and $\theta$ is the scattering angle in c.m. system. \\
 (b) Differential cross section of in-medium $\pi \Delta \rightarrow \pi 
 \Delta $ scattering:
  \begin{eqnarray}
 \sigma_{\pi \Delta \rightarrow \pi \Delta}(s,t) & = & \frac{1}{(2\pi)^{2}s}
  \frac{5({\rm g}_{\rho\pi}{\rm g}_{\Delta\Delta}^{\rho})^{2}}{36m^{*4}
 _{\Delta}(t-m^{2}_{\rho})^{2}} \left[ 18 m^{*4}_{\Delta} (m^{*2}_{\Delta}
 +m^{*2}_{\pi})^{2} -2m^{*6}_{\Delta}(18s+11t) \right. \nonumber \\
 &&- m^{*4}_{\Delta}(36m^{*2}_{\pi}s +16 m^{*2}_{\pi}t -18s^{2} -26st -7t^{2})
   \nonumber \\
 && +m^{*2}_{\Delta}(8m^{*2}_{\pi}st -4m^{*4}_{\pi}t +2m^{*2}_{\pi}t^{2}
 -4s^{2}t -6st^{2} -t^{3} ) \nonumber \\
 && \left. + m^{*2}_{\pi}t^{2} (m^{*2}_{\pi} -2s) +st^{2}(s+t) \right] \},
  \end{eqnarray}
where
 \begin{eqnarray}
 && s=(p+p_{2})^{2}=\left[ \omega^{*}(p) +E^{*}_{\Delta}(p_{2})\right] ^{2}
 - ({\bf p} + {\bf p}_{2})^{2}, \\
 && t=\frac{1}{2}(2m^{*2}_{\Delta} +2m^{*2}_{\pi}-s) -\frac{1}{2s} 
 (m^{*2}_{\Delta}-m^{*2}_{\pi})
 ^{2} +2 \mid {\bf p} \mid \mid {\bf p}_{3} \mid \cos \theta, \\
 && \mid {\bf p} \mid = \mid {\bf p}_{3} \mid  =\frac{1}{2\sqrt{s}}
 \sqrt{(s-m^{*2}_{\Delta}-m^{*2}_{\pi})^{2} -4m^{*2}_{\pi}m^{*2}_{\Delta} },
 \end{eqnarray}
 (c) Differential cross section of in-medium $\pi \pi \rightarrow \pi \pi$
 scattering:
  \begin{equation}
 \sigma_{\pi\pi  \rightarrow \pi\pi}(s,t) = \frac{1}{(2\pi)^{2}s} \left[
 D(s,t) + E(s,t) + (s,t \leftrightarrow u) \right],
  \end{equation}
  \begin{eqnarray}
 D(s,t) &=& \frac{3({\rm g}_{\sigma\pi}m_{\sigma})^{4}}{256(t-m^{2}_{\sigma})
 ^{2}} + \frac{({\rm g}_{\rho\pi})^{4}}{4(t-m^{2}_{\rho})^{2}}(4m^{*2}_{\pi}
 -2s-t)^{2} , \\  
 E(s,t) &=& \frac{({\rm g}_{\sigma\pi}m_{\sigma})^{4}}{256 (t-m^{2}_{\sigma})
 (u-m^{2}_{\sigma})} + \frac{({\rm g}_{\rho\pi})^{4}}{8(t-m^{2}_{\rho})
 (u-m^{2}_{\rho})}(s-t)(2s+t-4m^{*2}_{\pi} ) \nonumber \\
 && + (\frac{1}{2}{\rm g}_{\sigma\pi}m_{\sigma}{\rm g}_{\rho\pi})^{2}
 \left[ \frac{4m^{*2}_{\pi} -2s-t}{8(t-m^{2}_{\rho})(u-m^{2}_{\sigma})}
 + \frac{t-s}{8(t-m^{2}_{\sigma})(u-m^{2}_{\rho}) } \right],
  \end{eqnarray}
where the function D represents the contribution of the direct term and E
is the exchange term and
 \begin{eqnarray}
 && s=(p+p_{2})^{2}=\left[ \omega^{*}(p) +\omega^{*}(p_{2})\right] ^{2}
 - ({\bf p} + {\bf p}_{2})^{2}, \\
 && t=\frac{1}{2}(4m^{*2}_{\pi}-s)  
  +2 \mid {\bf p} \mid \mid {\bf p}_{3} \mid \cos \theta, \\
 && u=4m^{*2}_{\pi} -s -t, \\
 && \mid {\bf p} \mid = \mid {\bf p}_{3} \mid  =\frac{1}{2}
 \sqrt{s-4m^{*2}_{\pi}}.
 \end{eqnarray}
 
 \end{sloppypar}
\setcounter{equation}{0}
\renewcommand{\theequation}{D\arabic{equation}}
 \begin{center}
{\bf APPENDIX D}
 \end{center}
 \begin{sloppypar}
In this appendix we present analytical expressions of the imaginary part of 
the pion self-energy in nuclear matter. \\
(a) For space-like $p_{\mu}$
  \begin{equation}
Im \Pi_{NN^{-1}}^{--}(x,p)=\frac{m^{*2}p_{\mu}^{2}}{\pi \mid {\bf p} \mid}
 ({\rm g}_{NN}^{\pi})^{2}(E_{F}-E^{*}),
  \end{equation}
where
  \begin{eqnarray}
 &&E_{F}=(m^{*2} +k_{F}^{2})^{1/2}, \\
 && E^{*}=min \;\;\; (E_{F},E_{max}), \\
 && E_{max}=max \;\;\; \left[ m^{*}, E_{F}-\mid p_{0} \mid, 
 -\frac{1}{2}\mid p_{0} \mid + \frac{1}{2}\mid {\bf p} \mid (1- \frac{4m^{*2}}
 {p_{\mu}^{2}})^{1/2} \right] .
  \end{eqnarray}
  \begin{equation}
Im \Pi_{\Delta\Delta^{-1}}^{--}(x,p)
  =\frac{5p_{\mu}^{2}}{18\pi \mid {\bf p} \mid m_{\Delta}^{*2}}
 (p_{\mu}^{4}-2p_{\mu}^{2}m^{*2}_{\Delta}+10m_{\Delta}^{*4})
 ({\rm g}_{\Delta\Delta}^{\pi})^{2}(E_{F}^{\Delta}-E^{*}_{\Delta}),
  \end{equation}
and
  \begin{eqnarray}
 &&E_{F}^{\Delta}=\left[ m^{*2}_{\Delta} +(k_{F}^{\Delta})^{2}\right]^{1/2}, \\
 && E^{*}_{\Delta}=min \;\;\; (E_{F}^{\Delta},E_{max}^{\Delta}), \\
 && E_{max}^{\Delta}=max \;\;\; \left[ m^{*}_{\Delta}, E_{F}^{\Delta}-\mid p_{0} \mid, 
 -\frac{1}{2}\mid p_{0} \mid + \frac{1}{2}\mid {\bf p} \mid 
 (1- \frac{4m^{*2}_{\Delta}}
 {p_{\mu}^{2}})^{1/2} \right] .
  \end{eqnarray}
  \begin{equation}
Im \Pi_{\Delta N^{-1}}^{--}(x,p)=\frac{({\rm g}_{\Delta N}^{\pi})^{2}}
 {12\pi \mid {\bf p} \mid}C(p,q)
 (E_{F}-E^{*}),
  \end{equation}
where
  \begin{eqnarray}
 && E^{*}=min \;\;\; (E_{F},E_{max}), \\
 && E_{max}=max \;\;\; \left[ m^{*}, E_{F}^{\Delta}-\mid p_{0} \mid, 
 -\frac{1}{2}\mid p_{0} \mid \epsilon+ \frac{1}{2}\mid {\bf p} \mid 
(\epsilon^{2}- \frac{4m^{*2}}
 {p_{\mu}^{2}})^{1/2} \right] ,
  \end{eqnarray}
and
  \begin{eqnarray}
&& \epsilon = \frac{p_{\mu}^{2} + m^{*2} -m_{\Delta}^{*2}}{p_{\mu}^{2}}, \\
&& C(p,q)=\frac{1}{3m_{\Delta}^{*2}} \left[ p_{\mu}^{2} -(m^{*}_{\Delta}
 -m^{*})^{2} \right] \left[p_{\mu}^{2} -(m^{*}_{\Delta}+m^{*})^{2} \right]^{2}.
  \end{eqnarray}
The definition of $E_{F}$ and $E^{\Delta}_{F}$ is the same as in Eqs. (D2) and
 (D6).
  \begin{equation}
Im \Pi_{N\Delta^{-1}}^{--}(x,p)=\frac{({\rm g}_{\Delta N}^{\pi})^{2}}
 {12\pi \mid {\bf p} \mid}C(p,q)
 (E_{F}^{\Delta}-E^{*}),
  \end{equation}
where
  \begin{eqnarray}
 && E^{*}=min \;\;\; (E_{F}^{\Delta},E_{max}), \\
 && E_{max}=max \;\;\; \left[ m^{*}_{\Delta}, E_{F}-\mid p_{0} \mid, 
 -\frac{1}{2}\mid p_{0} \mid \epsilon^{\prime}+ \frac{1}{2}\mid {\bf p} \mid 
(\epsilon^{\prime 2}- \frac{4m^{*2}_{\Delta}}
 {p_{\mu}^{2}})^{1/2} \right] ,
  \end{eqnarray}
and
  \begin{equation}
 \epsilon^{\prime} = \frac{p_{\mu}^{2} +m_{\Delta}^{*2}-m^{*2}}{p_{\mu}^{2}}. \\
  \end{equation}
(b) For time-like $p_{\mu}$
   \begin{equation}
  Im \Pi_{\Delta N^{-1}}^{--}(x,p) = \left\{ \begin{array}{cl}
\frac{({\rm g}_{\Delta N}^{\pi})^{2}}
 {12\pi \mid {\bf p} \mid}C(p,q) (E_{u} - E_{d})
    & \qquad  0 \leq p_{\mu}^{2} \leq (m_{\Delta}^{*}-m^{*})^{2}  \\
     0 & \qquad otherwise                                            
    \end{array} \right. ,   
  \end{equation}
where
  \begin{eqnarray}
 && E_{u}=min \;\;\; \left[  E_{F}, 
 -\frac{1}{2}\mid p_{0} \mid \epsilon+ \frac{1}{2}\mid {\bf p} \mid 
(\epsilon^{2}- \frac{4m^{*2}}
 {p_{\mu}^{2}})^{1/2} \right] , \\
 && E_{d} = min \;\;\; (E_{F}, E_{max}), \\
 && E_{max}=max \;\;\; \left[ m^{*}, E_{F}^{\Delta}-\mid p_{0} \mid, 
 -\frac{1}{2}\mid p_{0} \mid \epsilon- \frac{1}{2}\mid {\bf p} \mid 
(\epsilon^{2}- \frac{4m^{*2}}
 {p_{\mu}^{2}})^{1/2} \right] .
  \end{eqnarray}
For the time-like $p_{\mu}$, the contributions of 
$Im \Pi_{NN^{-1}}^{--}(x,p)$ and $Im \Pi_{\Delta\Delta^{-1}}^{--}$,
which describe the particle-antiparticle decay processes, 
 vanish since we neglect the
anti-particles in the present framework. In this work the numerical calculations
 are performed in  cold nuclear matter with the assumption of chemical
equilibrium, i.e., $E_{F}^{\Delta}=E_{F}$ when a delta is produced. In this case
 the $\Delta$-decay process is Pauli-blocked since a delta can only decay into
a pion and a nucleon with a momentum smaller than the nucleon Fermi-momentum, 
which leads to
$Im \Pi_{N \Delta^{-1}}^{--}(x,p)=0$ when $p_{0} > \mid {\bf p} \mid$.
 
 \end{sloppypar}
\setcounter{equation}{0}
\renewcommand{\theequation}{E\arabic{equation}}
 \begin{center}
{\bf APPENDIX E}
 \end{center}
 \begin{sloppypar}
In this appendix we introduce the non-relativistic approximation for Eqs. (75)
and (77). These two terms stemming from the particle-hole and $\Delta$-hole 
excitations are commonly considered in the 
non-relativistic approach
\cite{Eri88,Xia88,Ehe93}. The effective masses and energies in
the expressions are replaced by the corresponding free ones.                 
In order to make
a complete  non-relativistic reduction we have to start from the full Green's
functions including  the contribution of 
anti-particles.
Eqs. (75) and (77) then read as
  \begin{eqnarray}
 Re\Pi^{--}_{NN^{-1}}(x,p)=2({\rm g}_{NN}^{\pi})^{2} \int \frac{d^{3}q}
 {(2\pi)^{3}} \left[ \frac{A(p,q)}{2E(q)} \frac{f({\bf x},{\bf q},
 \tau)}{(p_{0}+E(q))^{2}-E^{2}(p+q)} + p_{\mu} \rightarrow - p_{\mu}\right] , \\
 Re\Pi^{--}_{\Delta N^{-1}}(x,p)=-\frac{4}{3}({\rm g}_{\Delta N}^{\pi})^{2} 
 \int \frac{d^{3}q}
 {(2\pi)^{3}} \left[ \frac{C(p,q)}{2E(q)} 
  \frac{f({\bf x},{\bf q},
 \tau)}{(p_{0}+E(q))^{2}-E^{2}_{\Delta}(p+q)} 
+ p_{\mu} \rightarrow - p_{\mu} \right] , 
  \end{eqnarray}
with the A(p,q) and C(p,q) defined by Eqs. (79) and (83). After some algebra 
we obtain
  \begin{eqnarray}
 Re\Pi^{--}_{NN^{-1}}(x,p)&=&-4({\rm g}_{NN}^{\pi})^{2} \int \frac{d^{3}q}
 {(2\pi)^{3}} \frac{M_{N}^{2}p_{\mu}^{2}}{E^{2}(q)} 
\left[ \frac{f({\bf x},{\bf q},
 \tau)}{p_{\mu}^{2}/2E(q) - {\bf p}\cdot{\bf q}/E(q) + p_{0}}
  + p_{\mu} \rightarrow - p_{\mu}\right] , \\
 Re\Pi^{--}_{\Delta N^{-1}}(x,p)&=&\frac{8}{9}({\rm g}_{\Delta N}^{\pi})^{2} 
 \int \frac{d^{3}q}
 {(2\pi)^{3}}  
  \frac{f({\bf x},{\bf q},
 \tau)}{E(q)}\left[ \frac{(p.q)^{2}-M_{N}^{2}p_{\mu}^{2}}{M_{\Delta}^{2}}
+\frac{2p_{\mu}^{2}M_{N}(M_{\Delta}+M_{N})}{M_{\Delta}^{2}} \right. \nonumber \\
 && + \frac{(p.q)^{2}-M_{N}^{2}p_{\mu}^{2}}{2M_{\Delta}^{2}E(q)}
 \frac{(M_{\Delta}+M_{N})^{2}-p_{\mu}^{2}}{p_{\mu}^{2}/2E(q) - {\bf p}\cdot
 {\bf q}/E(q) -(M_{\Delta}^{2}-M_{N}^{2})/2E(q) + p_{0} } \nonumber \\
&& \left. + p_{\mu} \rightarrow - p_{\mu} \right] , 
  \end{eqnarray}
which are the same as Eqs. (8) and (11) in Ref. \cite{Dmi85}. Taking the 
non-relativistic limit $E(q) \approx M_{N}$, Eq. (E3) becomes
 \begin{equation}
 Re\Pi^{--}_{NN^{-1}}(x,p)=({\rm g}_{NN}^{\pi})^{2} p_{\mu}^{2} 
 \frac{2\omega_{p}}{p_{0}^{2} - \omega_{p}^{2} }\rho_{N}
\end{equation}
with $\omega_{p}=p_{\mu}^{2}/2M_{N}$. If one further 
neglects the relativistic kinematics,
 i.e., $p_{\mu}^{2} \rightarrow - {\bf p}^{2}$, it returns to the standard 
non-relativistic formula stemming from the particle-hole excitation
\cite{Eri88,Xia88}.  Therefore, the relativistic effects stay in two  aspects:
one is the Fermi motion of nucleons in a nucleus which is small; another one is 
the relativistic kinetics which turns out to be substantial in our calculations
as can be seen later. In the non-relativistic limit, Eq. (E4) becomes 
 \begin{eqnarray}
 Re\Pi^{--}_{\Delta N^{-1}}(x,p)&=&\frac{4}{9}({\rm g}_{\Delta N}^{\pi})^{2}
 \frac{M_{N} {\bf p}^{2}}{M_{\Delta}^{2}} \rho_{N} + \frac{8}{9}
 ({\rm g}_{\Delta N}^{\pi})^{2} \frac{(M_{\Delta} + M_{N})}{M_{\Delta}^{2}}
 p_{\mu}^{2}\rho_{N} \nonumber \\
 && - \frac{1}{9} ({\rm g}_{\Delta N}^{\pi})^{2} \frac{(M_{\Delta} + M_{N})^{2}
 -p_{\mu}^{2} }{M_{\Delta}^{2}} \frac{2\omega_{R}}{p_{0}^{2} - \omega_{R}^{2}}
 {\bf p}^{2} \rho_{N},
 \end{eqnarray}
with
 \begin{equation}
 \omega_{R}=\frac{p_{\mu}^{2}}{2M_{N}} - \frac{M_{\Delta}^{2} - M_{N}^{2}}
 {2M_{N}}.
 \end{equation}
The first and second terms on the r.h.s. of Eq. (E6) are the non-resonant
terms, which have no analogy in the non-relativistic model. The third term can  
be reduced ($p_{\mu}^{2} \rightarrow -{\bf p}^{2}$) to a similar term in the
non-relativistic model stemming from the $\Delta$-hole excitation, but there
exist some differences mainly caused by the different masses of nucleons
and deltas.  The situation might be understood in view that the problem of 
describing
a spin 3/2 particle  in relativistic quantum field theory remains unsolved.
Fortunately, the difference between the non-relativistic limit of the 
relativistic
model and the standard non-relativistic model is quantitatively  insubstantial.
   \begin{center}
 \fbox{Fig.~12}
   \end{center}

Fig.~12(a) displays the pion dispersion relation (the pion branch) at normal 
density. The solid line denotes the free pion dispersion relation. The dotted
line is calculated with Eqs. (E5) and (E6), i.e., the non-relativistic limit
of the relativistic model but with the relativistic kinetics. The dashed line is
computed by taking $p_{\mu}^{2} \rightarrow -{\bf p}^{2}$ in Eqs. (E5) and
(E6). One can clearly see that the relativistic effect 
(mainly from the kinematic
origin) makes the pion dispersion relation harder at low momenta and softer
at high momenta. Furthermore, the relativistic effect at low momenta mainly
comes from the non-resonant terms, i.e., the first and second terms on the
r.h.s. of Eq. (E6). If one switches off these two terms, at low momenta the
results (the dash-dotted line) approach to the dashed line while at high momenta
to the dotted line. Fig.~12(b) depicts the pion dispersion relation (both the
pion and the $\Delta$-hole branches) at different densities calculated with
Eqs. (E5) and (E6) and $p_{\mu}^{2} \rightarrow -{\bf p}^{2}$. The short-range
correlation effect is included through Eq. (125) ($p_{\mu}^{2} \rightarrow
-{\bf p}^{2}$) with ${\rm g}^{\prime}=0.6$. As one can see from the figure,
the obtained pion dispersion relation is nearly the same as that of the
non-relativistic model \cite{Xia88,Ehe93} except the pion branch is a little
harder at high momenta. In this case one may conclude that the difference
between the results of Fig.~6 and that of the non-relativistic model mainly 
stems from the relativistic kinetics.
 
 \end{sloppypar}
\setcounter{equation}{0}
\renewcommand{\theequation}{F\arabic{equation}}
 \begin{center}
{\bf APPENDIX F}
 \end{center}
 \begin{sloppypar}
In this appendix we derive the conserved current and energy-momentum tensor.\\
(1) Current 

Make a four-dimension integration of momentum on the both side of Eq. (62), the
right-hand side (the collision term) goes to zero \cite{Gro80} and we have
 \begin{eqnarray}
\int d^{4}p \{ p^{\mu}\partial_{\mu}^{x} &+&\frac{1}{2}\partial_{x}^{\mu}\Pi_{H}(x)\partial
 _{\mu}^{p} + \frac{1}{2}\partial_{x}^{\mu}Re\Pi_{loop}^{--}(x,p)\partial_{\mu}
 ^{p} \nonumber \\ 
 &-& \frac{1}{2}\partial_{p}^{\mu}Re\Pi_{loop}^{--}(x,p)\partial_{\mu}^{x}
 \} \frac{f_{\pi}({\bf x},{\bf p},\tau)}{\omega^{*}(p)}=0 .            
 \end{eqnarray}
It is straightforward to find the current conservation
 \begin{equation}
 \partial ^{\mu}_{x} J_{\mu}(x) = 0
 \end{equation}
with
 \begin{equation}
 J_{\mu}(x) =\int d^{4}p \left[ p_{\mu} -\frac{1}{2} \left( \partial_{\mu}^{p}
 Re \Pi_{loop}^{--}(x,p) \right) \right] \frac{f_{\pi}({\bf x}, {\bf p}, \tau)}
 {\omega^{*}(p)} .
 \end{equation}
We note that each $f_{\pi}({\bf x}, {\bf p}, \tau)$ is in principle 
accompanied by a $\delta$ function $\delta(p_{0} - \omega^{*}(p))$ for on-shell
pion.

\noindent (2)Energy-momentum tensor

Multiply $p_{\mu}$ on both side of Eq. (62) and make a four-dimension
integration of momentum  we arrive at
 \begin{eqnarray}
\int d^{4}p p_{\mu} \{ p_{\nu}\partial^{\nu}_{x} &+&\frac{1}{2}\partial_{x}^{\nu}\Pi_{H}(x)\partial
 _{\nu}^{p} + \frac{1}{2}\partial_{x}^{\nu}Re\Pi_{loop}^{--}(x,p)\partial_{\nu}
 ^{p} \nonumber \\ 
 &-& \frac{1}{2}\partial^{p}_{\nu}Re\Pi_{loop}^{--}(x,p)\partial^{\nu}_{x}
 \} \frac{f_{\pi}({\bf x},{\bf p},\tau)}{\omega^{*}(p)}  =0 .            
 \end{eqnarray}
Our strategy is to extract the $\partial_{x}^{\nu}$ out of the whole equation.
For the first term it is straightforward. The second term can be rewritten as
 \begin{eqnarray}
 - {\rm g}_{\mu\nu} \int d^{4} p \frac{f_{\pi}({\bf x},{\bf p},\tau)}{2
 \omega^{*}(p)} \left(\partial_{x}^{\nu} \Pi_{H}(x)\right) .  \nonumber
 \end{eqnarray}
With the help of Eqs. (73) and (74), it becomes
 \begin{eqnarray}
\partial_{x}^{\nu} \left[ {\rm g}_{\mu\nu} \frac{1}{8} ({\rm g}_{\sigma\pi})^{2}
 \rho_{S}^{2}(\pi) \right]. \nonumber
 \end{eqnarray}
The third and fourth terms can be written as
 \begin{eqnarray}
&& \int d^{4}p p_{\mu}\partial _{\nu}^{p} \left[ \frac{1}{2}\left( \partial_{x}
 ^{\nu} Re\Pi_{loop}^{--}(x,p)\right) \frac{f_{\pi}({\bf x},{\bf p},\tau)}
{\omega^{*}(p)} \right] - \int d^{4}p p_{\mu}\partial_{x}^{\nu} \left[ 
 \frac{1}{2} \left( \partial_{\nu}^{p} Re\Pi_{loop}^{--}(x,p)\right)
 \frac{f_{\pi}({\bf x},{\bf p},\tau)}{\omega^{*}(p)} \right] \nonumber \\
&& = -{\rm g}_{\mu\nu}\int d^{4}p \frac{1}{2}\left( \partial_{x}^{\nu}Re\Pi
_{loop}^{--}(x,p)\right) \frac{f_{\pi}({\bf x},{\bf p},\tau)}{\omega^{*}(p)}
 - \partial_{x}^{\nu}\int d^{4}p \frac{1}{2}p_{\mu} 
 \left( \partial^{p}_{\nu}Re\Pi
_{loop}^{--}(x,p)\right) \frac{f_{\pi}({\bf x},{\bf p},\tau)}{\omega^{*}(p)}
 \nonumber \\
&& = -{\rm g}_{\mu\nu}\partial_{x}^{\nu} 
 \int d^{4}p \frac{1}{2}Re\Pi
_{loop}^{--}(x,p) \frac{f_{\pi}({\bf x},{\bf p},\tau)}{\omega^{*}(p)}
 +{\rm g}_{\mu\nu}\int d^{4}p \frac{1}{2}Re\Pi
_{loop}^{--}(x,p)\partial_{x}^{\nu}
  \frac{f_{\pi}({\bf x},{\bf p},\tau)}{\omega^{*}(p)} \nonumber \\
 && \;\;\;\;\;- \partial_{x}^{\nu}\int d^{4}p \frac{1}{2}p_{\mu} 
 \left( \partial^{p}_{\nu}Re\Pi
_{loop}^{--}(x,p)\right) 
\frac{f_{\pi}({\bf x},{\bf p},\tau)}{\omega^{*}(p)}.
 \end{eqnarray}
The second term of the above equation turns out to be 
 \begin{eqnarray}
&-& \frac{1}{4}{\rm g}_{\mu\nu} \int d^{4}p \left( \partial_{p}^{\lambda}
 p_{\lambda} \right) Re\Pi_{loop}^{--}(x,p)\partial_{x}^{\nu}
\frac{f_{\pi}({\bf x},{\bf p},\tau)}{\omega^{*}(p)} \nonumber \\
& =& \frac{1}{4}{\rm g}_{\mu\nu}\int d^{4}p p_{\lambda} \left(\partial_{p}
 ^{\lambda} Re \Pi_{loop}^{--}(x,p)\right) \partial_{x}^{\nu}
\frac{f_{\pi}({\bf x},{\bf p},\tau)}{\omega^{*}(p)}
 + \frac{1}{4}{\rm g}_{\mu\nu}\int d^{4}p p_{\lambda} 
  Re \Pi_{loop}^{--}(x,p) \partial_{p}^{\lambda}\partial_{x}^{\nu}
\frac{f_{\pi}({\bf x},{\bf p},\tau)}{\omega^{*}(p)} \nonumber \\
& =& \frac{1}{4}{\rm g}_{\mu\nu}\int d^{4}p p_{\lambda} \left(\partial_{p}
 ^{\lambda} Re \Pi_{loop}^{--}(x,p)\right) \partial_{x}^{\nu}
\frac{f_{\pi}({\bf x},{\bf p},\tau)}{\omega^{*}(p)}
 + \frac{1}{4}{\rm g}_{\mu\nu}\int d^{4}p p_{\lambda} \left(
 \partial_{x}^{\nu}\partial_{p}^{\lambda} Re \Pi_{loop}^{--}(x,p) \right)
\frac{f_{\pi}({\bf x},{\bf p},\tau)}{\omega^{*}(p)} \nonumber \\
& =& \partial_{x}^{\nu} \frac{1}{4}{\rm g}_{\mu\nu}\int d^{4}p p_{\lambda}
 \left( \partial_{p}^{\lambda} Re\Pi_{loop}^{--}(x,p)\right)
\frac{f_{\pi}({\bf x},{\bf p},\tau)}{\omega^{*}(p)}.             
\end{eqnarray}
In the first and second equality of Eq. (F6) we have used the fact
that the terms with the double derivative acting on the same quantity 
can be neglected in the 
gradient expansion. At the end we have the energy-momentum conservation
 \begin{equation}
\partial_{x}^{\nu}T_{\mu\nu}(x)=0
 \end{equation}
with
 \begin{eqnarray}
T_{\mu\nu}(x)&=&\int d^{4}p p_{\mu}p_{\nu} 
\frac{f_{\pi}({\bf x},{\bf p},\tau)}{\omega^{*}(p)} + {\rm g}_{\mu\nu}\frac{1}
{8}({\rm g}_{\sigma\pi})^{2}\rho_{S}^{2}(\pi) \nonumber \\
&-& {\rm g}_{\mu\nu}\frac{1}{2}\int d^{4}p  Re\Pi_{loop}^{--}(x,p)            
\frac{f_{\pi}({\bf x},{\bf p},\tau)}{\omega^{*}(p)} \nonumber \\            
&-& \frac{1}{2}\int d^{4}p  p_{\mu}\left( \partial_{\nu}^{p} Re\Pi_{loop}^{--}
 (x,p)\right)
\frac{f_{\pi}({\bf x},{\bf p},\tau)}{\omega^{*}(p)} \nonumber \\              
&+& {\rm g}_{\mu\nu}\frac{1}{4}\int d^{4}p  p_{\lambda}
\left( \partial^{\lambda}_{p} Re\Pi_{loop}^{--}
 (x,p)\right)
\frac{f_{\pi}({\bf x},{\bf p},\tau)}{\omega^{*}(p)}.              
 \end{eqnarray}
 \end{sloppypar}
 
 \newpage
  \vspace{3cm}
  \begin{center}
 {\bf TABLES}
 \end{center}
 \vspace{2cm}
{\bf TABLE I}: Symbols and notation used for the baryon-baryon-meson 
 vertex, $P_{\mu}$ is the 
 transformed four-momentum.
     \vspace{+0.5cm}
    \begin{center}
    \tabcolsep 0.15in
  \begin{tabular}{|c|c|c|c|c|c|c|c|c|c|}
      \hline
 {\rm A} & m$_{A}$ & g$^{A}_{NN}$ &  
 g$^{A}_{\Delta \Delta}$   & $\gamma_{A} $ & $\tau_{A}$ & $ T_{A} $ &
  $\Phi_{A} (x) $ & D$_{A}^{\mu}$ & D$_{A}^{i}$ \\
 \hline
 $\sigma$ & m$_{\sigma}$ & g$^{\sigma}_{NN}$ & 
 g$_{\Delta \Delta}^{\sigma}$ &  1 & 1 & 1 & $\sigma(x)$ &
 1 & 1   \\
 \hline
$\omega$ & m$_{\omega}$ & $-$ g$^{\omega}_{NN} $ &  
 $-$ g$_{\Delta \Delta}^{\omega}$ &   $\gamma_{\mu} $ &
 1 & 1 & $\omega^{\mu}(x)$ & $-$ g$^{\mu \nu}$ & 1 \\
 \hline
 $\pi$ & m$_{\pi}$ & g$_{NN}^{\pi}$&  
 g$_{\Delta \Delta}^{\pi}$ &  
 $\not\! P \gamma_{5} $
& \mbox{\boldmath $\tau$} &${\bf T}$
 & \mbox{\boldmath $\pi$}(x) & 1 & $\delta_{ij}$ \\
 \hline
 $\rho$ & m$_{\rho}$ & $-$ $\frac{1}{2}$g$_{NN}^{\rho}$&  
 $-$ $\frac{1}{2}$g$_{\Delta \Delta}^{\rho}$ &  
 $\gamma_{\mu}$ 
& \mbox{\boldmath $\tau$} &${\bf T}$
 & $\mbox{\boldmath $\rho$}^{\mu}(x)$ & $-$ g$^{\mu\nu}$ & $\delta_{ij}$ \\
       \hline
      \end{tabular}
          \end{center}
   \vspace{1.0cm}
{\bf TABLE II}: Symbols and notation used for the interaction
 vertex involving only mesons.
    \vspace{0.5cm}
    \begin{center}
    \tabcolsep 0.15in
  \begin{tabular}{|c|c|c|c|c|}
      \hline
 {\rm A} & g$^{A}_{\pi\pi}$ & $\gamma_{A}^{\pi}$ & $\tau_{A}^{\pi} $ &
 $\Phi_{A}(x) $ \\
 \hline
 $\sigma$  & $\frac{1}{2}$g$_{\sigma\pi}$m$_{\sigma}$  &  1  & 
 $\delta_{ij}$ & $\sigma(x)$ \\
 \hline
 $\rho$  & g$_{\rho\pi}$ &  $i(p+q)_{\mu}$  &  $\varepsilon_{ijk}$  & 
   $\rho_{k}^{\mu}(x)$  \\
       \hline
      \end{tabular}
          \end{center}

 \newpage
\begin{center}
{\bf CAPTIONS}
 \end{center}
 \begin{description}
 \item[\tt Fig.1 ] Contour along the axis for an evaluation of the operator
expectation value. In practice, $t_{0}$ is shifted to $- \infty$ and $t_{max}$
 to $+ \infty$.
 \item[\tt Fig.2 ] Feynman diagrams contribute to the Hartree term of
pion self-energy. A wavy line denotes the exchanged virtual meson,  
a solid  line, a doubled line and a dashed line represents the nucleon,
delta and pion, respectively.
 \item[\tt Fig.3 ] Feynman diagrams contribute to the Fock term (one 
 baryon-loop) of pion self-energy. Different lines denote the different 
 particles as described in the caption of Fig.~2.
 \item[\tt Fig.4 ] Feynman diagrams contribute to the Born term of pion    
 self-energy. Different lines denote the different particles as described
 in the caption of Fig.~2. The imaginary part of Fig.~4(a) contributes to the
 $\pi N \rightarrow \pi N$ elastic cross section, and Fig.~4(b) to the $\pi 
 \Delta \rightarrow \pi \Delta$, Fig.~4(c),(d) to the $\pi \pi
 \rightarrow \pi \pi$ elastic cross section, respectively.
 \item[\tt Fig.5 ] The gap between the effective masses of particles and 
 anti-particles as a function of density. The universal coupling strengths
for the nucleons and deltas are assumed.
 \item[\tt Fig.6 ] The pion dispersion relation in symmetric nuclear matter
at saturation density. The solid line shows the free dispersion relation.
The dotted curves represent the in-medium dispersion relation for different
branches as indicated in the figure. The upper and lower hatched areas 
indicate regions of non-vanishing imaginary parts of the $\Delta$-hole and
nucleon-hole polarizations, respectively.
 \item[\tt Fig.7 ] 
(a) The pion dispersion relation at different densities.    
(b) The real part of pion self-energy at different densities
 relating to the pion dispersion relation shown in (a).
 (c) The contributions of different excitation modes to the 
real part of pion self-energy. The calculations are performed at normal density.
 \item[\tt Fig.8 ] The pion dispersion relation (dotted curves) at different
densities. The short-range correlations are taken into account in a 
non-relativistic way with Migdal parameter ${\rm g}^{\prime}$=0.6. The solid
lines represent the free dispersion relation.
 \item[\tt Fig.9] Free cross sections for reactions $\pi^{+}p \rightarrow 
 \Delta ^{++}$ and $\pi^{-}p \rightarrow \Delta^{0}$. The dots are the 
experimental data from Ref. \cite{Car71}. The solid curve is our results
calculated with the mass-dependent coupling strength while the dashed curve
with the mass-independent coupling strength. 
 \item[\tt Fig.10] The momentum-dependent $\Delta$-decay width in the free 
 space. The solid line and long dash-dotted line are our results computed
 with MDC and MIC, respectively. Dotted \cite{Kit86}, dashed \cite{Wol90}
and dash-dotted \cite{BasThe} line represent several phenomenological 
parameterizations commonly used in the transport models.
 \item[\tt Fig.11] The in-medium $\Delta$-formation cross section and   
 $\Delta$-decay width. In (a) and (b) the free pion mass is used in the 
 calculations.
 In (c) and (d)  the effective pion mass is  
 taken into account via the dispersion relation of Fig.~7(a).
 \item[\tt Fig.12] (a)The pion dispersion relation (the pion branch) at normal 
density. Different lines correspond to the different situations as explained
in text. (b)The pion dispersion relation (the pion and $\Delta$-hole branch)
at different densities. The calculations are performed with Eqs. (E5) and (E6)
and $p_{\mu}^{2} \rightarrow -{\bf p}^{2}$. The short-range correlation effect
is taken into account by means of the Migdal parameter of ${\rm g}^{\prime}=
0.6$.
  \end{description}

 \newpage
 {\Large Fig. 1}
  \vspace{0cm}
 \begin{figure}[htbp]
 \hskip  -1cm \psfig{file=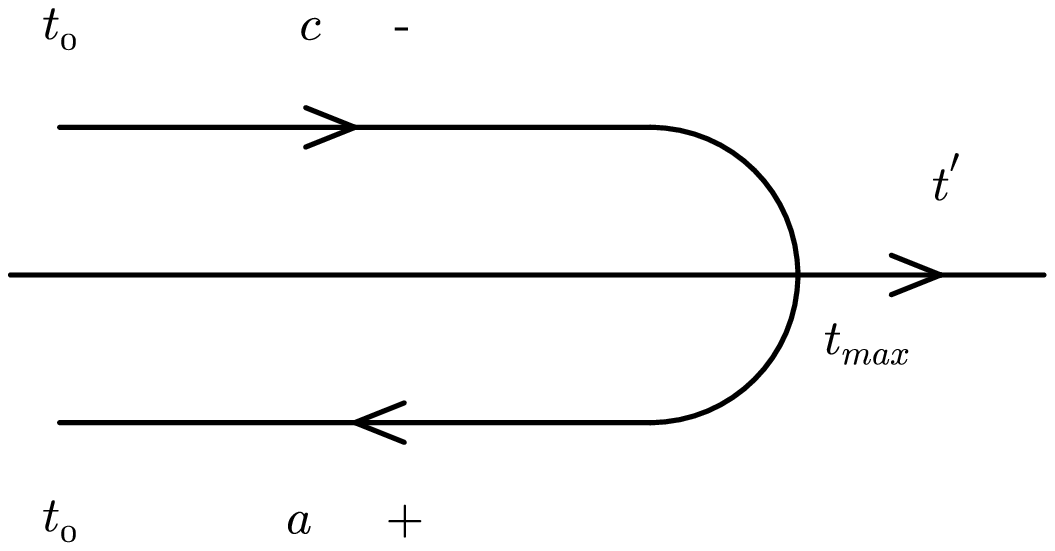,width=15cm,height=17cm,angle=-90}
\end{figure}
 \newpage
 {\Large Fig. 2}
 \begin{figure}[htbp]
  \vspace{-3.0cm}
 \hskip  -1cm \psfig{file=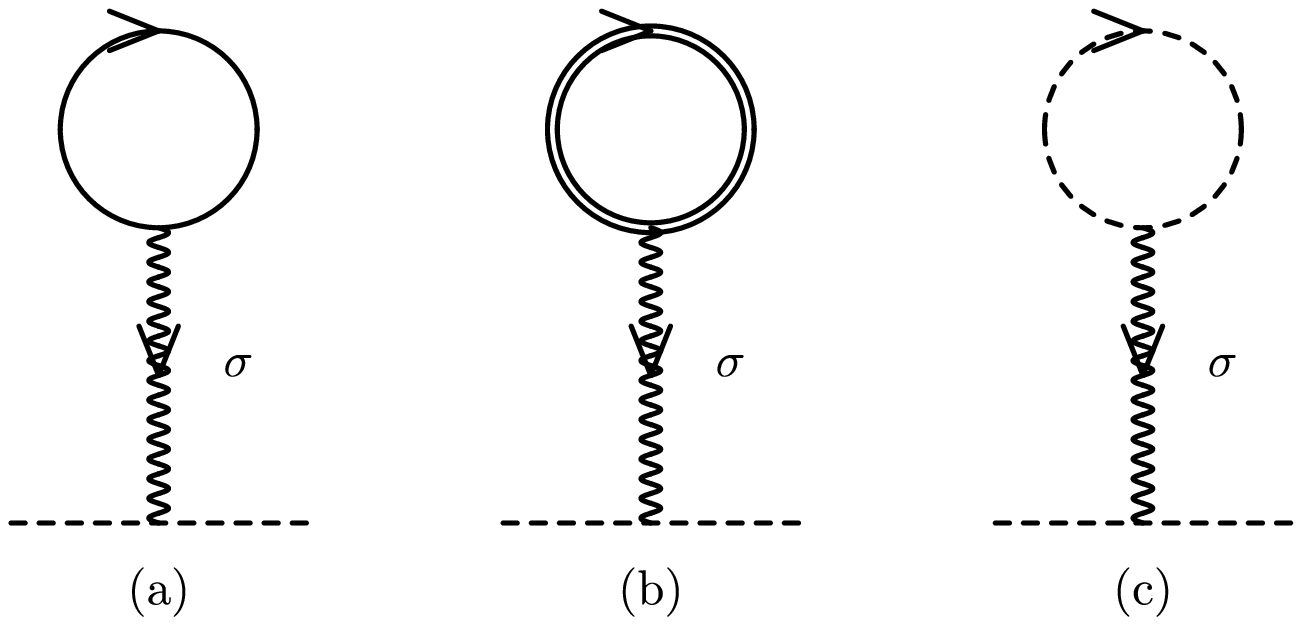,width=11.cm,height=10cm,angle=-90}
\end{figure}
 \vspace{-3.5cm}
 {\Large Fig. 3}
 \begin{figure}[htbp]
  \vspace{-2.5cm}
 \hskip  -0cm \psfig{file=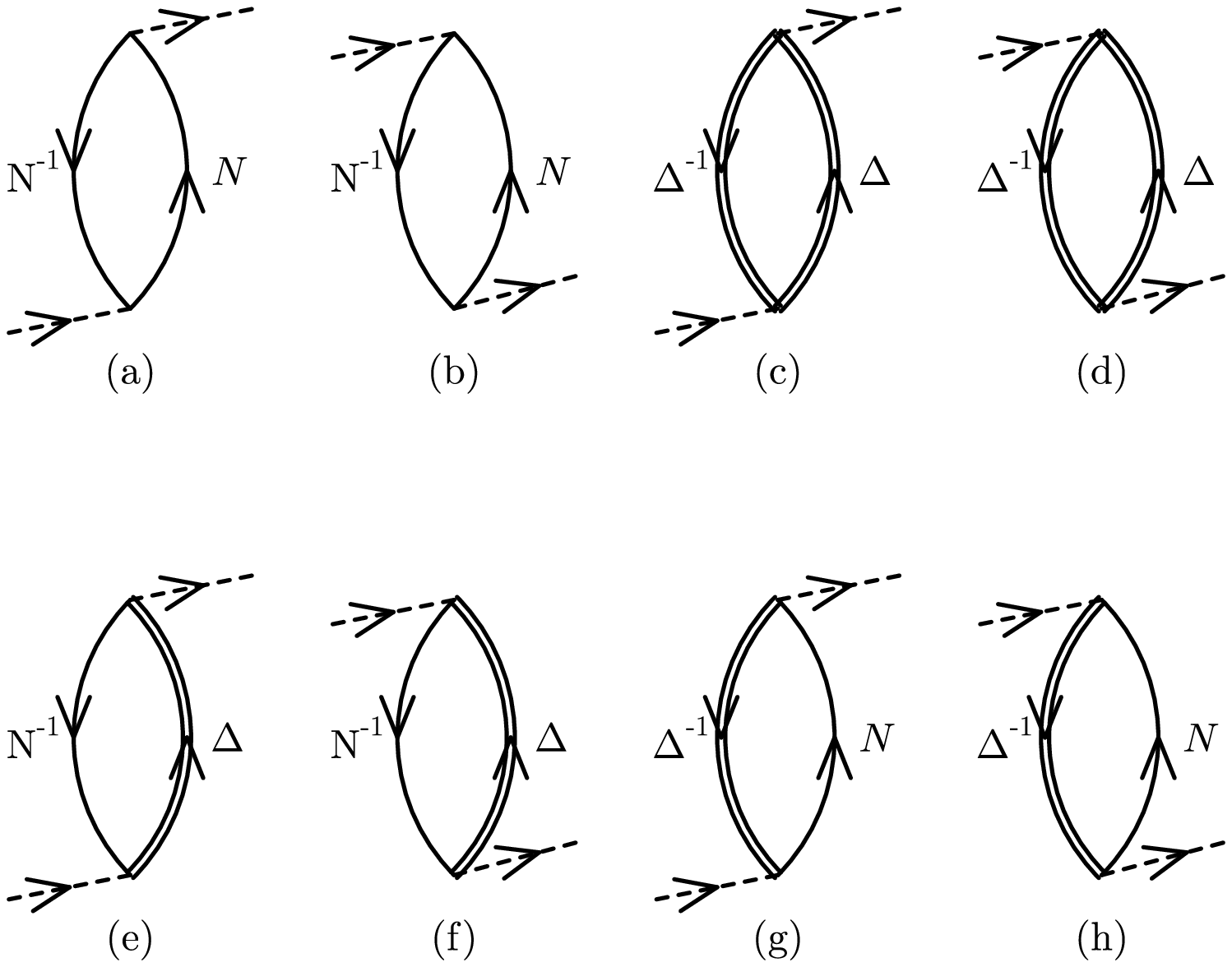,width=9.cm,height=13cm,angle=-90}
\end{figure}
 \vspace{-3.2cm}
 {\Large Fig. 4}
 \begin{figure}[htbp]
  \vspace{-4.5cm}
 \hskip  -0.2cm \psfig{file=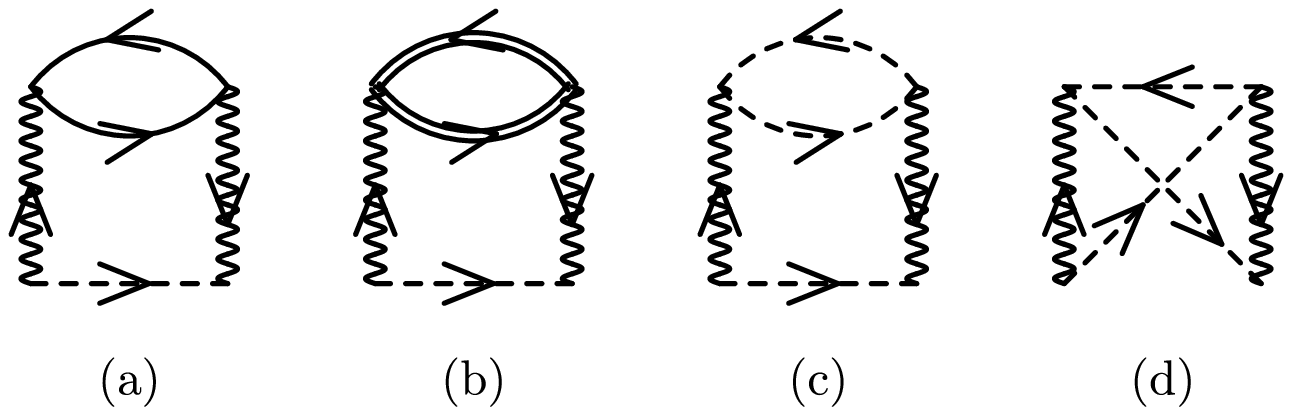,width=11.cm,height=11cm,angle=-90}
  \vspace{-22cm}
\end{figure}
 \newpage
 {\Large Fig. 5}
 \begin{figure}[htbp]
  \vspace{0cm}
 \hskip  -2cm \psfig{file=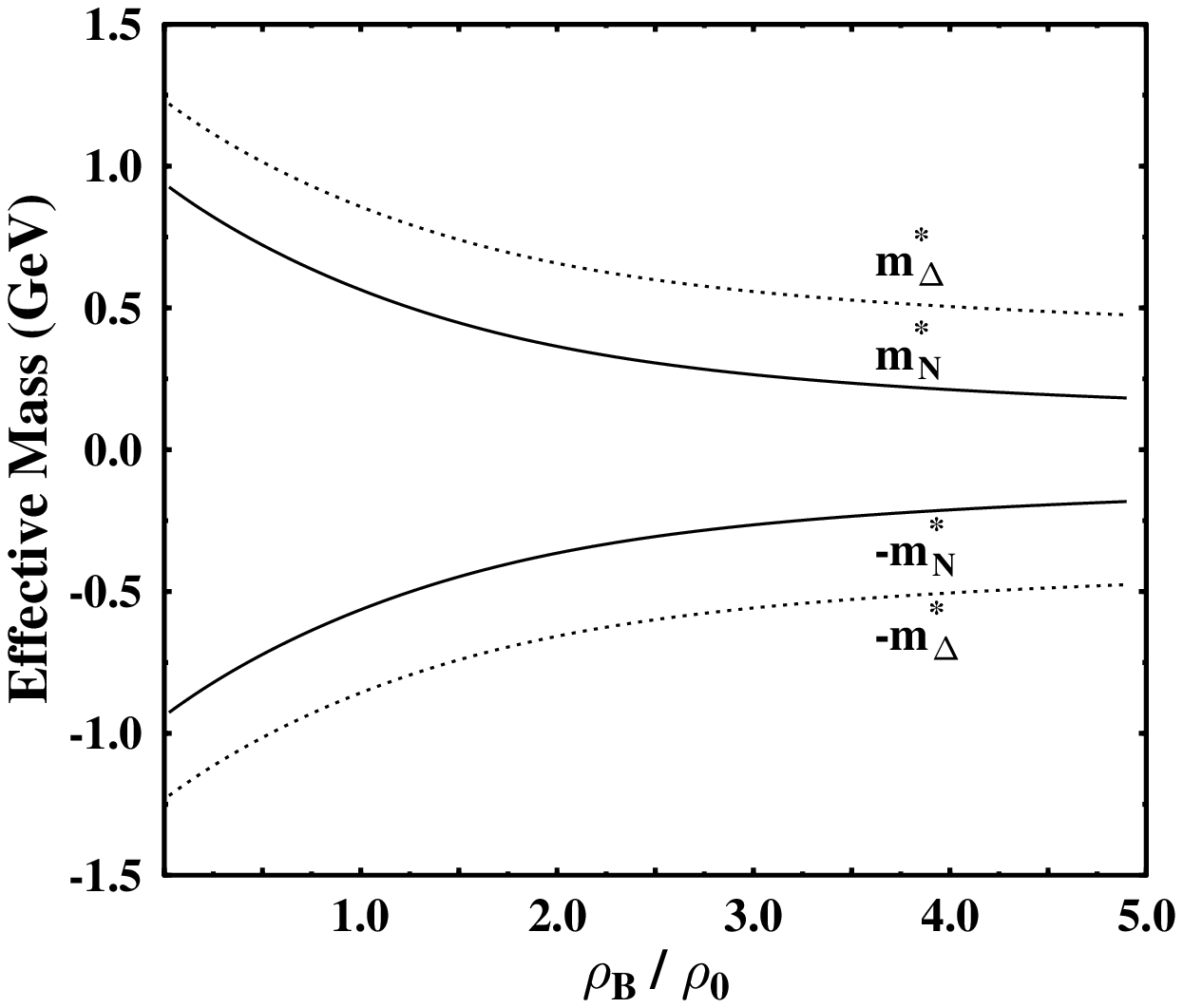,width=15.cm,height=17cm,angle=-90}
\end{figure}
 \newpage
 {\Large Fig. 6}
 \begin{figure}[htbp]
  \vspace{0cm}
 \hskip  -2cm \psfig{file=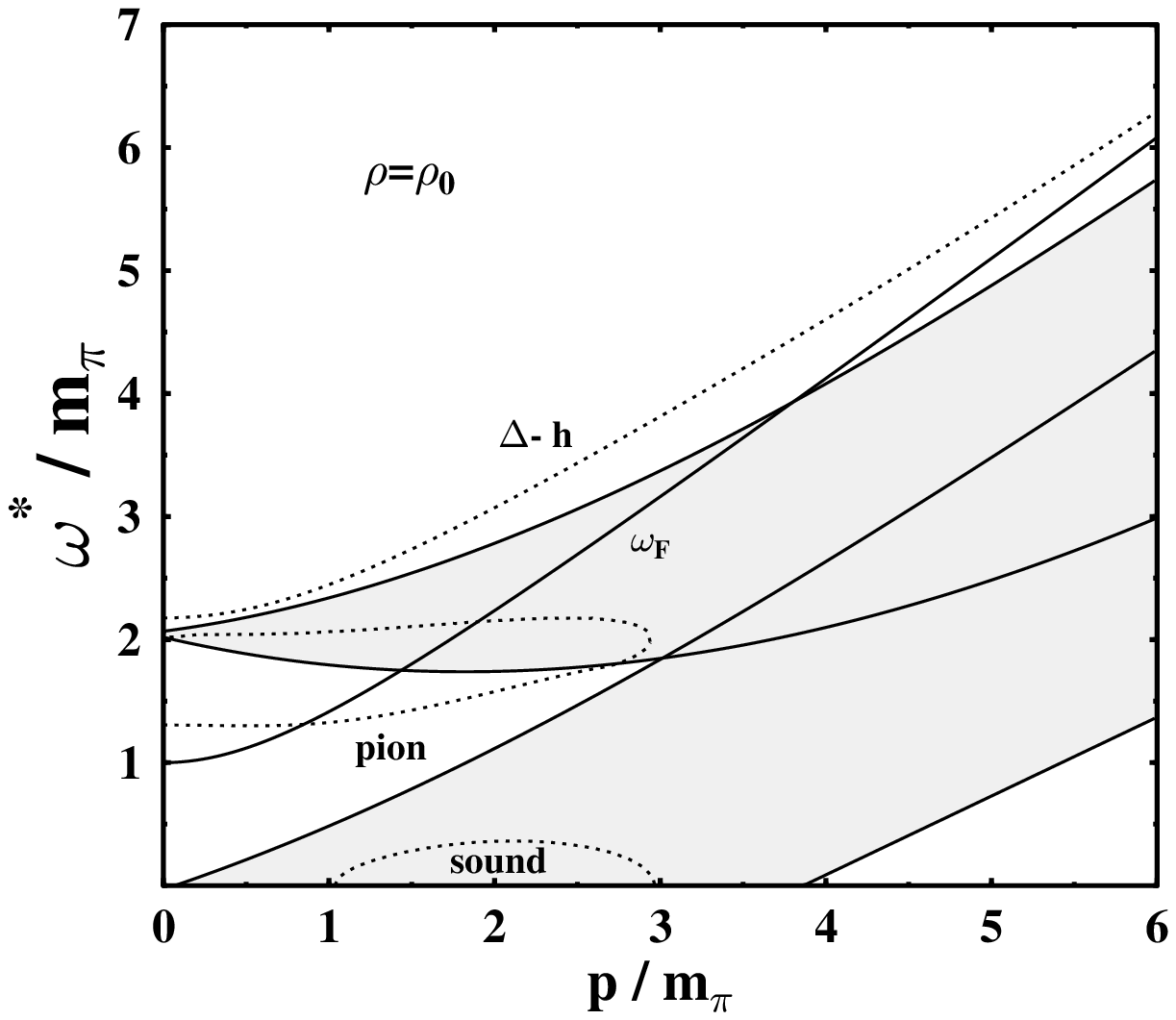,width=15.cm,height=17cm,angle=-90}
\end{figure}
 \newpage
 {\Large Fig. 7}
 \begin{figure}[htbp]
  \vspace{0cm}
 \hskip  0cm \psfig{file=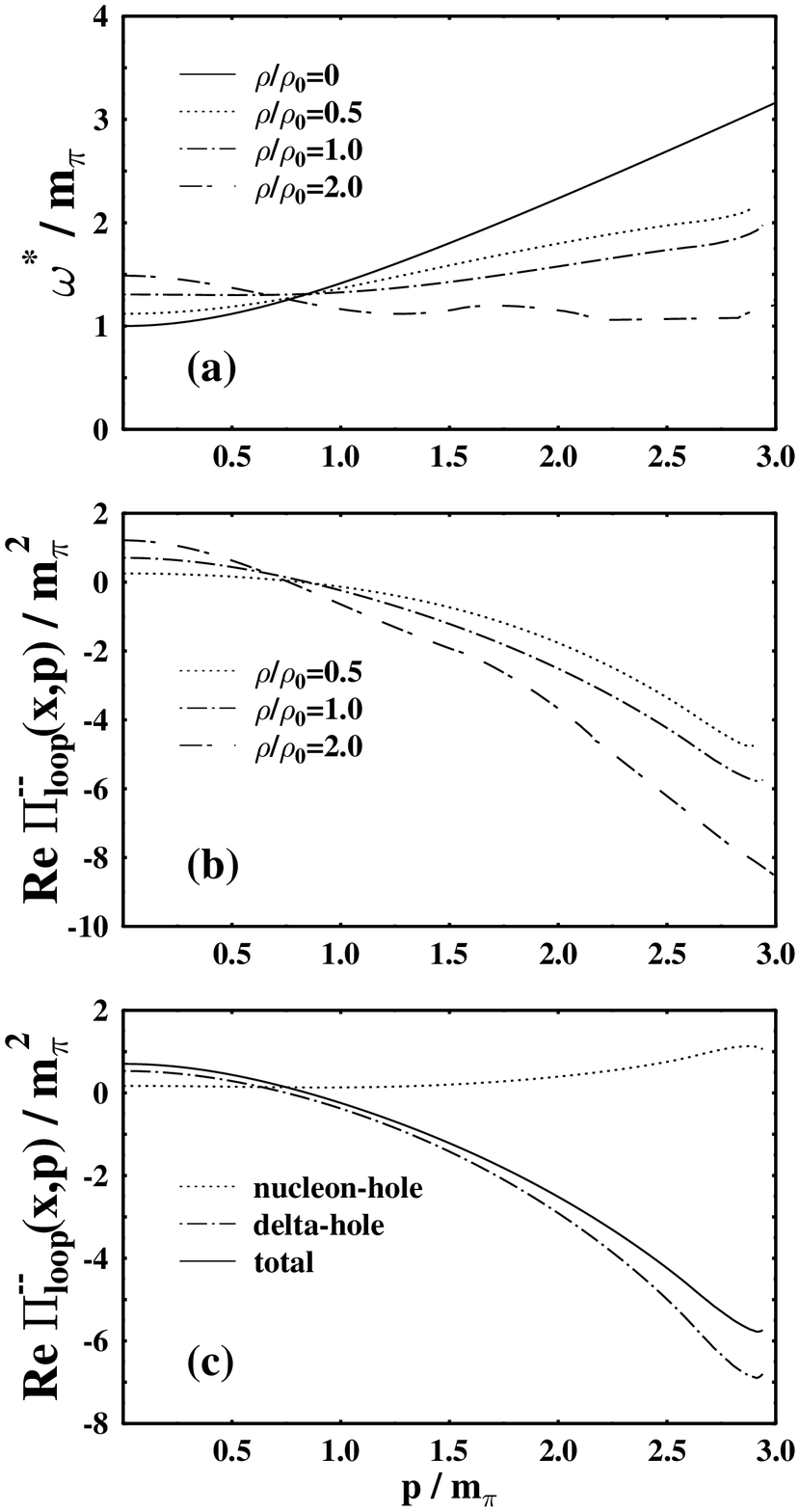,width=14.cm,height=21cm,angle=0}
\end{figure}
 \newpage
 {\Large Fig. 8}
 \begin{figure}[htbp]
  \vspace{0cm}
 \hskip  0cm \psfig{file=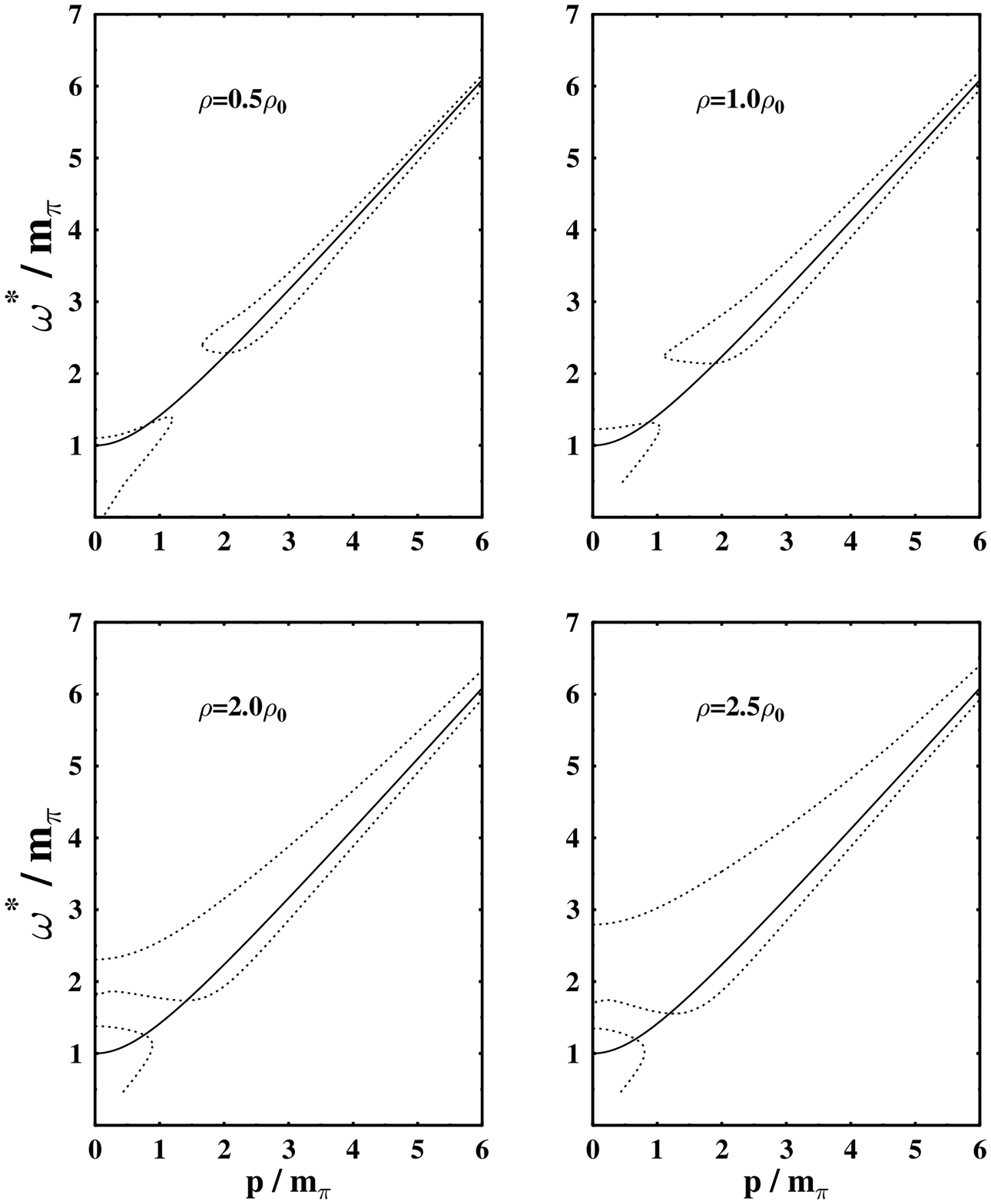,width=15.cm,height=21cm,angle=0}
\end{figure}
 \newpage
 {\Large Fig. 9}
 \begin{figure}[htbp]
  \vspace{0cm}
 \hskip  -1cm \psfig{file=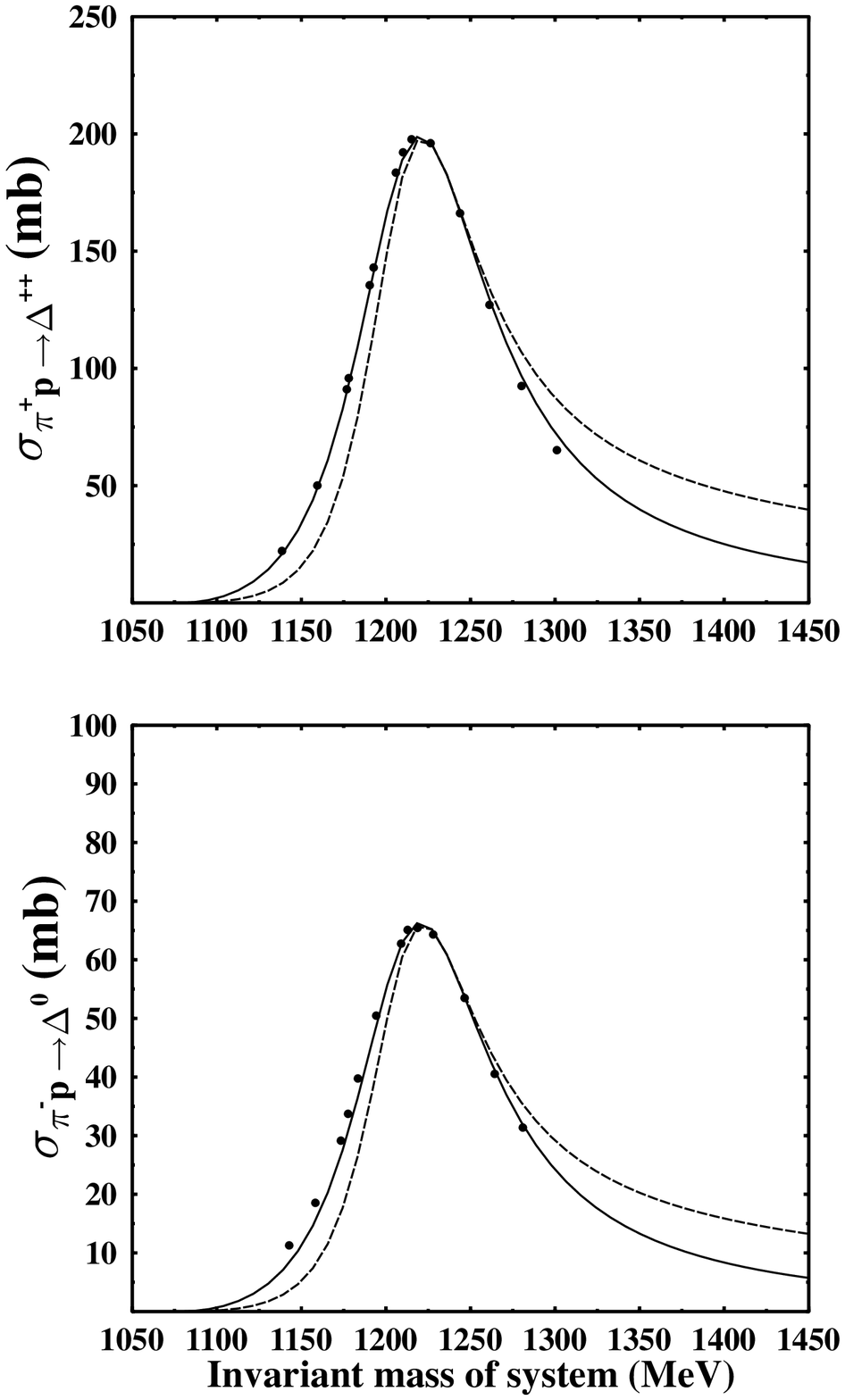,width=14.cm,height=21cm,angle=0}
\end{figure}
 \newpage
 {\Large Fig. 10}
 \begin{figure}[htbp]
  \vspace{0cm}
 \hskip  -2cm \psfig{file=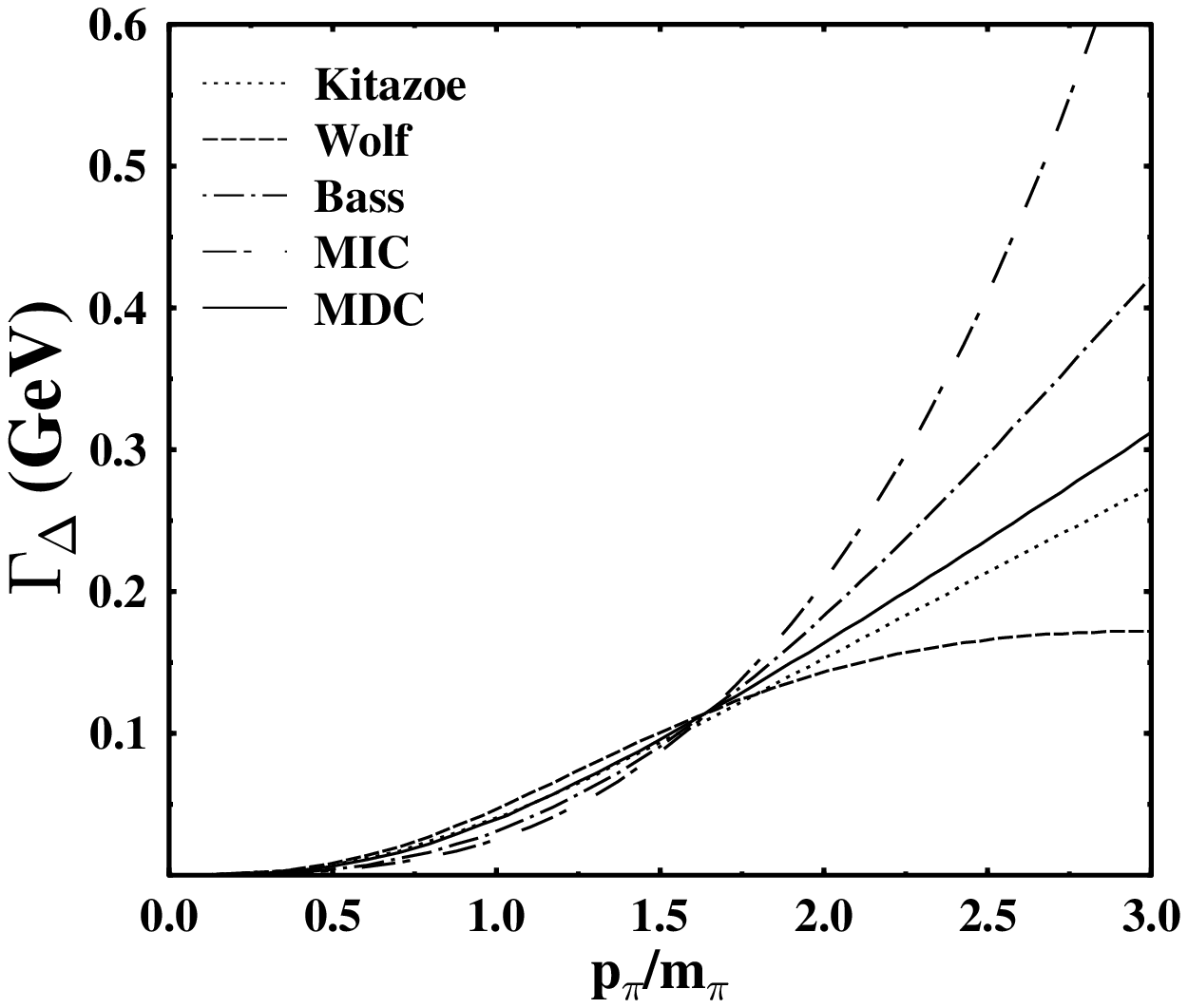,width=15.cm,height=17cm,angle=-90}
\end{figure}
 \newpage
 {\Large Fig. 11}
 \begin{figure}[htbp]
  \vspace{0cm}
 \hskip  -1cm \psfig{file=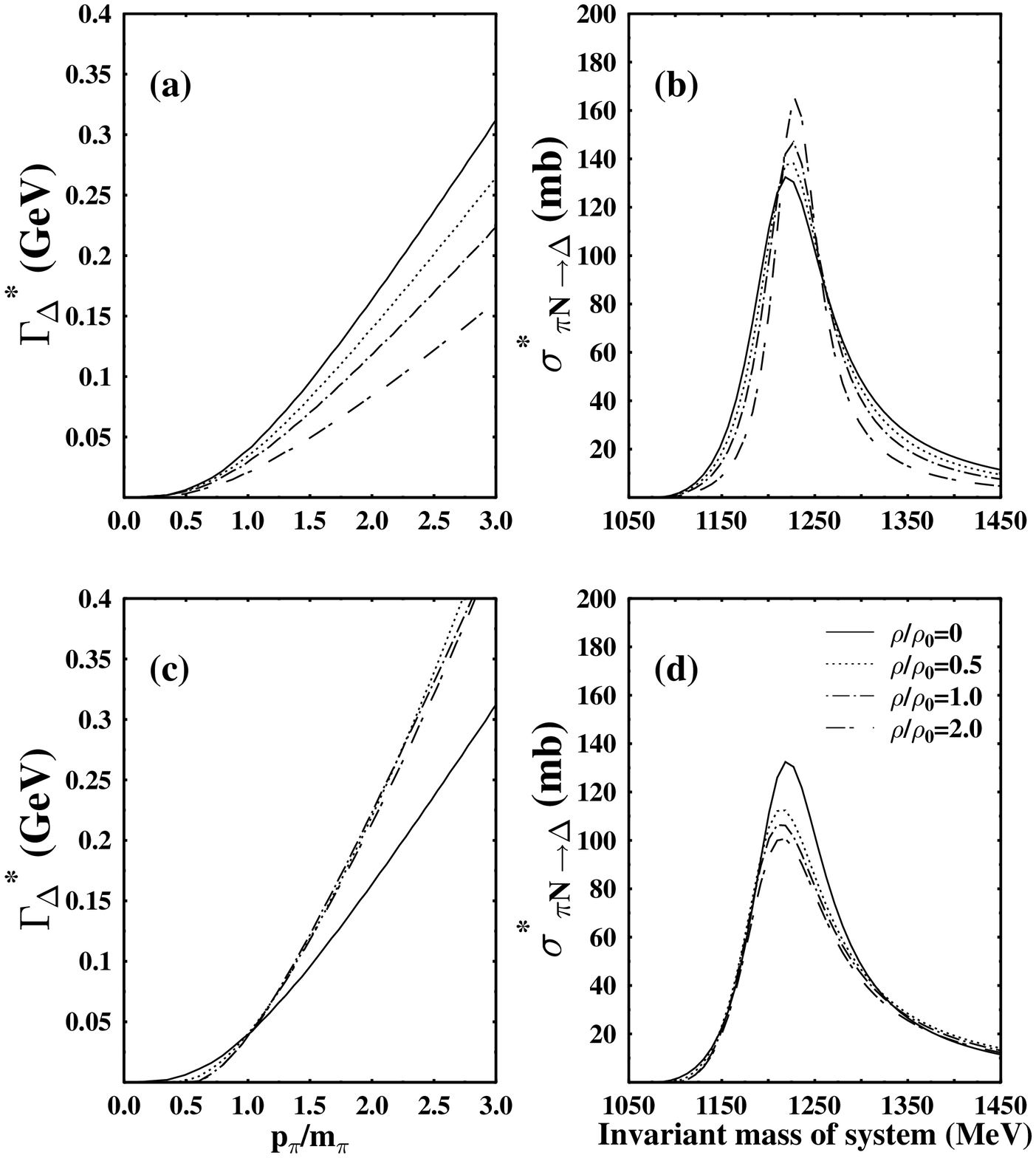,width=15.cm,height=21cm,angle=0}
\end{figure}
 \newpage
 {\Large Fig. 12}
 \begin{figure}[htbp]
  \vspace{0cm}
 \hskip  -1cm \psfig{file=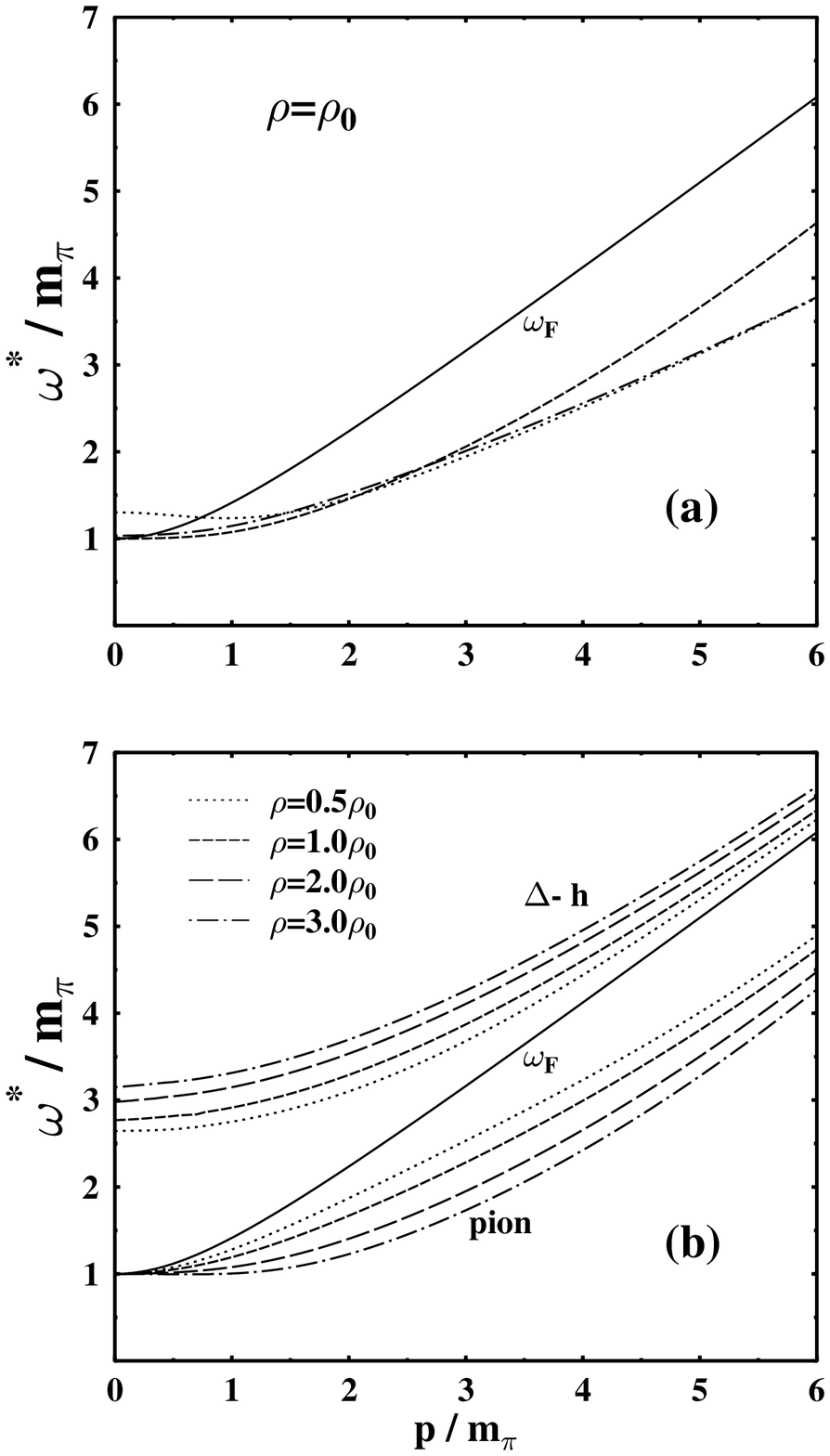,width=14.cm,height=21cm,angle=0}
\end{figure}

\newpage
\begin{thebibliography}{250}
\bibitem{Gyu77}
   M.~Gyulassy and W.~Greiner, 
   Ann. of Phys. (N.Y.) {\bf 109}, 485 (1977).
\bibitem{Mig78}
   A.~B.~Migdal,
   Rev. Mod. Phys. {\bf 50}, 107 (1978).
\bibitem{San80}
   A.~Sandoval, R.~Stock, H.~E.~Stelzer, R.~E.~Renfordt, J.~W.~Harris,
   J.~P.~Brannigan, J.~V.~Geaga, L.~J.~Rosenberg, L.~S.~Schroeder,
   and K.~L.~Wolf,
   Phys. Rev. Lett. {\bf 45}, 874 (1980).
\bibitem{Sto82}
   R.~Stock, R.~Bock, R.~Brockmann, J.~W.~Harris, A.~Sandoval, H.~Str\"{o}bele,
   K.~L.~Wolf, H.~G.~Pugh, L.~S.~Schroeder, M.~Maier, R.~E.~Renfordt,
   A.~Dacal, and M.~E.~Ortiz,
   Phys. Rev. Lett. {\bf 49}, 1236 (1982).
\bibitem{Har85}
   J.W.~Harris, R.~Bock, R.~Brockmann, A.~Sandoval, R.~Stock, H.~Stroebele, 
   G.~Odyniec, H.G.~Pugh, L.~Schroeder, R.~E.~Renfordt, D.~Schall, D.~Bangert, 
   W.~Rauch and K.~L.~Wolf,
   Phys. Lett. {\bf B153}, 377 (1985).
\bibitem{Sch94}
   O.~Schwalb and the TAPS collaboration, 
   Phys. Lett. {\bf B321}, 20 (1994).
\bibitem{Mun95}
   C.~M\"{u}ntz and the Kaos collaboration,
   Z. Phys. {\bf A352}, 175 (1995);
   Z. Phys. {\bf A357}, 399 (1997).
\bibitem{Ber91}
   F.~D.~Berg and the TAPS collaboration,
   Z. Phys. {\bf A340}, 297 (1991).
\bibitem{Kin97}
   J.~C.~Kintner, S.~Albergo, F.~Bieser et al., 
   Phys. Rev. Lett. {\bf 78}, 4165 (1997).
\bibitem{Sto78}
   H.~St\"{o}cker, W.~Greiner and W.~Scheid,
   Z. Phys. {\bf A286}, 121 (1978).
\bibitem{Dan79}
   P.~Danielewicz,
   Nucl. Phys. {\bf A314}, 465 (1979).
\bibitem{Sto81}
   H.~St\"{o}cker, A.~A.~Ogloblin, and W.~Greiner,
   Z. Phys. {\bf A303}, 259 (1981).
\bibitem{Wei85}
   W.~Weise,
   Nucl. Phys. {\bf A434}, 685c (1985).
\bibitem{Dmi85}
   V.F.~Dmitriev and T.~Suzuki,
   Nucl. Phys. {\bf A438}, 697 (1985).
\bibitem{Elz87}
   H.-Th.~Elze, M.~Gyulassy, D.~Vasak, H.~Heinz, H.~St\"{o}cker, and
   W.~Greiner,
   Mod. Phys. Lett. {\bf A2}, 451 (1987).
\bibitem{Xia88}
   L.~H.~Xia, C.~M.~Ko, L.~Xiong and J.~Q.~Wu,
   Nucl. Phys. {\bf A485}, 721 (1988).
\bibitem{Ko89}
   C.~M.~Ko, L.~H.~Xia and P.~J.~Siemens,
   Phys. Lett. {\bf B231}, 16 (1989).
\bibitem{Bro91}
   G.E.~Brown, V.~Koch and M.~Rho,
   Nucl. Phys. {\bf A535}, 701 (1991).
\bibitem{Dan91}
   P.~Danielewicz and G.~F.~Bertsch,
   Nucl. Phys. {\bf A533}, 712 (1991).
\bibitem{Li91}
   B.~A.~Li and W.~Bauer,
   Phys. Lett. {\bf B254}, 335 (1991);
   Phys. Rev. {\bf C44}, 450 (1991).
\bibitem{Ehe93}
   W.~Ehehalt, W.~Cassing, A.~Engel, U.~Mosel and Gy.~Wolf,
   Phys. Lett. {\bf B298}, 31 (1993).
\bibitem{Xio93}
   L.~Xiong, C.~M.~Ko, and V.~Koch,
   Phys. Rev. {\bf C47}, 788 (1993).
\bibitem{Her92}
   T.~Herbert, K.~Wehrberger, and F.~Beck,
   Nucl. Phys. {\bf A541}, 699 (1992).
\bibitem{Bas95}
   S.~A.~Bass, C.~Hartnack, H.~St\"{o}cker and W.~Greiner,
   Phys. Rev. {\bf C51}, 3343 (1995).
\bibitem{Liu95}
   Liang-gang~Liu, 
   Phys. Rev. {\bf C51}, 3421 (1995).
\bibitem{Fuc97}
   C.~Fuchs, L.~Sehn, E.~Lehmann, J.~Zipprich and A.~F\"{a}ssler,
   Phys. Rev. {\bf C55}, 411 (1997).
\bibitem{Gal87}
   C.~Gale and J.~Kapusta,
   Phys. Rev. {\bf C35}, 2107 (1987);
   Phys. Rev. {\bf C38}, 2659 (1988).
\bibitem{Li96}
   G.~Q.~Li, C.~M.~Ko, G.~E.~Brown, H.~Sorge,
   Nucl. Phys. {\bf A611}, 539 (1996).
\bibitem{Ern97}
   C.~Ernst, Diploma thesis, Frankfurt university, 1998.        
\bibitem{Bra97}
   E.~L.~Bratkovskaya, W.~Cassing,
   Nucl. Phys. {\bf A619}, 413 (1997).
\bibitem{Aga95}
   G.~Agakichiev and the CERES collaboration,
   Phys. Rev. Lett. {\bf 75}, 1272 (1995).
\bibitem{Rap97}
   R.~Rapp, G.~Chanfray, and J.~Wambach,
   Nucl. Phys. {\bf A617}, 472 (1997).
\bibitem{Li95}
   G.~Q.~Li, C.~M.~Ko, and G.~E.~Brown,
   Phys. Rev. Lett. {\bf 75}, 4007 (1995).
\bibitem{Cas95}
   W.~Cassing, W.~Ehehalt, and C.~M.~Ko,
   Phys. Lett. {\bf B363}, 35 (1995).
\bibitem{Wag96}
   A.~Wagner, et al., Gsi scientific report, 56 (1996).
\bibitem{Pel97}
   D. Pelte and the FOPI collaboration,
   Z. Phys. A, in press.
\bibitem{Tei97}
   S.~Teis, W.~Cassing, M.~Effenberger, A.~Hombach, U.~Mosel and Gy.~Wolf,
   Z. Phys. A, in press.
\bibitem{Ody88}
   C.~Odyniec et al., 
   LBL Report 24580, 215 (1988).
\bibitem{Gos89}
   J.~Gosset, O.~Valette, J.P.~Alard, et al., 
   Phys. Rev. Lett. {\bf 62}, 1251 (1989).
\bibitem{Eri88}
   T.~Ericson and W.~Weise,
   {\em Pions and Nuclei}, (Clarendon, Oxford, 1988).
\bibitem{Ko87}
   C.~M.~Ko, Q.~Li, and R.~Wang, 
   Phys. Rev. Lett. {\bf 59}, 1084 (1987);
   Q.~Li, J.~Q.~Wu, and C.~M.~Ko,
   Phys. Rev. {\bf C39}, 849 (1989).
\bibitem{Ber88}
   G.~F.~Bertsch and S.~das Gupta,
   Phys. Rep. {\bf 160}, 189 (1988)
\bibitem{Bla88}
   B.~Bl\"{a}ttel, V.~Koch, W.~Cassing, and U.~Mosel,
   Phys. Rev. {\bf C38}, 1767 (1988).
\bibitem{Sch89}
   M.~Sch\"{o}nhofen, M.~Cubero, M.~Gering, M.~Sambataro, H.~Feldmeier,
   and W.~N\"{o}renberg,
   Nucl. Phys. {\bf A504}, 875 (1989);
   M.~Sch\"{o}nhofen, M.~Cubero, B.~L.~Friman, W.~N\"{o}renberg, and Gy.~Wolf,
   Nucl. Phys. {\bf A572}, 112 (1994).
\bibitem{Zho94}
   Hongbo~Zhou, Zhuxia~Li, Yizhong~Zhuo and Guangjun~Mao,
   Nucl. Phys. {\bf A580}, 627 (1994).
\bibitem{Wan91}
   Shun-Jin~Wang, Bao-An~Li, Wolfgang~Bauer, and Jorgen~Randrup,
   Ann. Phys. (N.Y.) {\bf 209}, 251 (1991).
\bibitem{ZPA94}
   Mao Guangjun, Li Zhuxia, Zhuo Yizhong, Han Yinlu, Yu Ziqiang and M. Sano,
   Z. Phys. {\bf A347}, 173 (1994).
\bibitem{PRC94}
   Guangjun Mao, Zhuxia Li, Yizhong Zhuo, Yinlu Han, and Ziqiang Yu,
   Phys. Rev. {\bf C49}, 3137 (1994);
   Guangjun Mao, Zhuxia Li, Yizhong Zhuo, and Ziqiang Yu,
   Phys. Lett. {\bf B327}, 183 (1994).
\bibitem{Zhuxia}
   Zhuxia Li, Guangjun Mao, Yizhong Zhuo,
   in {\em Proceedings of the NATO ASI on Hot and Dense Nuclear Matter},
   Bodrum, Turkey, {\bf Vol. 335}, 659 (Plenum, New York, 1994).
\bibitem{PRC96}
   Guangjun Mao, Zhuxia Li, Yizhong Zhuo,
   Phys. Rev. {\bf C53}, 2933 (1996).
\bibitem{PLB96}
   Guangjun Mao, Zhuxia Li, Yizhong Zhuo, and Enguang Zhao,
   Phys. Lett. {\bf B378}, 5 (1996).
\bibitem{PRC97}
   Guangjun~Mao, Zhuxia~Li, Yizhong~Zhuo, and Enguang~Zhao,
   Phys. Rev. {\bf C55}, 792 (1997).
\bibitem{PRCSUB}
   Guangjun~Mao, L.~Neise, H.~St\"{o}cker, W.~Greiner, Zhuxia~Li,
   Phys. Rev. {\bf C57}, 1938 (1998).
\bibitem{Dav91}
   John~E.~Davis and Robert~J.~Perry,
   Phys. Rev. {\bf C43}, 1893 (1991).
\bibitem{Mro94}
   S.~Mr\'{o}wczy\'{n}ski and U.~Heinz,      
   Ann. Phys. (N.Y.) {\bf 229}, 1 (1994).
\bibitem{Ser86}
   B.~D.~Serot and J.~D.~Walecka,
   Adv. Nucl. Phys. {\bf 16}, 1 (1986).
\bibitem{Dan84}
   P.~Danielewicz,
   Ann. Phys. (N.Y.) {\bf 152}, 239 (1984).
\bibitem{Cho85}
   Kuangchao Chou, Zhaobin Su, Bailin Hao, and Lu Yu,
   Phys. Rep. {\bf 118}, 1 (1985).
\bibitem{Pil73}
   H.~Pilkuhn, W.~Schmidt, A.D.~Martin, C.~Michael, F.~Steiner,
   B.R.~Martin, M.M.~Nagels, and J.J.~de Swart,
   Nucl. Phys. {\bf B65}, 460 (1973).
\bibitem{Zou94}
   B.S.~Zou and D.V.~Bugg,
   Phys. Rev. {\bf D50}, 591 (1994).
\bibitem{Bog83}
   J.~Boguta and H.~St\"{o}cker,
   Phys, Lett. {\bf B120}, 289 (1983).
\bibitem{Bod91}
   A.~R.~Bodmer,
   Nucl. Phys. {\bf A526}, 703 (1991).
\bibitem{Sug94}
   Y.~Sugahara, H.~Toki,
   Nucl. Phys. {\bf A579}, 557 (1994).
\bibitem{Haa87}
   Bernard~ter~Haar and Rudi~Malfliet,
   Phys. Rev. {\bf C36}, 1611 (1987).
\bibitem{Li94}
   G.~Q.~Li and R.~Machleidt,
   Phys. Rev. {\bf C49}, 566 (1994).
\bibitem{Gao95}
   Song~Gao, Yi-Jun~Zhang and Ru-Keng~Su,
   Phys. Rev. {\bf C52}, 380 (1995).
\bibitem{Met93}
   V.~Metag,
   Nucl. Phys. {\bf A553}, 283c (1993);
   R.~Averbeck, R.~Holzmann, A.~Schubert et al.,
   GSI Science Report, 80 (1994).
\bibitem{Hen94}
   P.A.~Henning and H.~Umezawa,
   Nucl. Phys. {\bf A571}, 617 (1994);
   R.~Rapp and J.~Wambach,
   Nucl. Phys. {\bf A573}, 626 (1994);
   R.~Rapp, G.~Chanfray and J.~Wambach,
   Nucl. Phys. {\bf A617}, 472 (1997).
\bibitem{Lur68}
   David~Luri\'{e}, 
   {\em Particles and Fields} (Bristol, England, 1968).
\bibitem{Bot88}
   Wim~Botermans and Rudi~Malfliet,
   Phys. Lett. {\bf B215}, 617 (1988);
   Phys. Rep. {\bf 198}, 115 (1990).
\bibitem{Bro89}
   G.E.~Brown, E.~Oset, M.~Vicente~Vacas, and W.~Weise,
   Nucl. Phys. {\bf A505}, 823 (1989).
\bibitem{Kad62}
   L.P.~Kadanoff, G.~Baym,
   {\em Quantum Statistical Mechanics} (Benjamin, New York, 1962).
\bibitem{Mro90}
   S.~Mr\'{o}wczy\'{n}ski and P.~Danielewicz,
   Nucl. Phys. {\bf B342}, 345 (1990).
\bibitem{Bey97}
   M.~Beyer and G. R\"{o}pke,
   Phys. Rev. {\bf C56}, 2636 (1997).
\bibitem{Hor83}
   C.J.~Horowitz, and B.D.~Serot,
   Nucl. Phys. {\bf A399}, 529 (1983).
\bibitem{Wol92}
   Gy.~Wolf, W.~Cassing, and U.~Mosel,
   Nucl. Phys. {\bf A545}, 139c (1992);
   Nucl. Phys. {\bf A552}, 549 (1993).
\bibitem{Gro80}
   S.~R.~de Groot, W.~A.~van Leeuwen, and Ch.~G.~van Weert,
   {\em Relativistic Kinetic Theory} (North-Holland, Amsterdam, 1980).
\bibitem{Pil79}
   H.M.~Pilkuhn,
   {\em Relativistic Particle Physics} (Springer-Verlag, Berlin, 1979).
\bibitem{Zli97}
   Zhuxia Li, Guangjun Mao, Yizhong Zhuo and Walter Greiner,
   Phys. Rev. {\bf C56}, 1570 (1997).           
\bibitem{Mos74}
   S.A.~Moszkowski,
   Phys. Rev. {\bf D9}, 1613 (1974);
   S.I.A.~Garpman, N.K.~Glendenning and Y.J.~Karant,
   Nucl. Phys. {\bf A322}, 382 (1979).
\bibitem{Wal87}
   B.M.~Waldhauser, J.~Theis, J.A.~Maruhn, H.~St\"{o}cker and W.~Greiner,
   Phys. Rev. {\bf C36}, 1019 (1987);
   P.~L\'{e}vai, B.~Luk\'{a}cs, B.~Waldhauser and J.~Zim\'{a}nyi,
   Phys. Lett. {\bf B177}, 5 (1986).
\bibitem{Bru52}
   K.~A.~Brueckner,
   Phys. Rev. {\bf 86}, 106 (1952).
\bibitem{Car71}
   A.~A.~Carter, J.~R.~Williams, D.~V.~Bugg, P.~J.~Bussey and D.~R.~Dance,
   Nucl. Phys. {\bf B26}, 445 (1971).
\bibitem{Kit86}
   Y.~Kitazoe, M.~Sano, H.~Toki, and H.~Nagamiya,
   Phys. Lett. {\bf B166}, 35 (1986).
\bibitem{Wol90}
   Gy.~Wolf, G.~Batko, W.~Cassing, U.~Mosel, K.~Niita, and M.~Sch\"{a}fer,
   Nucl. Phys. {\bf A517}, 615 (1990).
\bibitem{BasThe}
   S.~A.~Bass,
   PhD thesis, Frankfurt university, unpublished.
\bibitem{Kim97}
   Hungchong~Kim, S.~Schramm, and Su~Houng~Lee,
   Phys. Rev. {\bf C56}, 1582 (1997).
\bibitem{Eff97}
   M.~Effenberger, A.~Hombach, S.~Teis, and U.~Mosel,
   Nucl. Phys. {\bf A613}, 353 (1997).
\bibitem{Fri98}
   It should be mentioned that recent experimental determination of the       
   effective pion mass from pionic atoms gives $m^{*}_{\pi}(\rho_{0})=167$
   MeV (E.~Friedman and A.~Gal, nucl-th/9805004). Our model predicts 
   $m^{*}_{\pi}(\rho_{0})=180$ MeV. The agreement between the theoretical
   prediction and the empirical value might be further improved through 
   including the short-range correlation effect self-consistently.
\bibitem{LiPre}
   G.Q.~Li, G.E.~Brown, C.~Gale, and C.M.~Ko,
   nucl-th/9712048.
\bibitem{BraPre}
   E.L.~Bratkovskaya, W.~Cassing, R.~Rapp, and J.~Wambach,
   nucl-th/9710043.
\bibitem{ErnPre}
   C.~Ernst, S.A.~Bass, M.~Belkacem, H.~St\"{o}cker, and W.~Greiner,
   nucl-th/9712069.




\end{thebibliography}
\end{document}